# A report on the first virtual PLDI conference

Alastair F. Donaldson
Department of Computing, Imperial College London
General Chair of PLDI 2020
alastair.donaldson@imperial.ac.uk

## Abstract

This is a report on the PLDI 2020 conference, for which I was general chair, which was held virtually for the first time as a result of the COVID-19 pandemic. The report contains: my personal reflections on the positive and negative aspects of the event; a description of the format of the event and associated logistical details; and data (with some analysis) on attendees' views on the conference format, the extent to which attendees engaged with the conference, attendees' views on virtual vs. physical conferences (with a focus on PLDI specifically) and the diversity of conference registrants. I hope that the report will be a useful resource for organizers of upcoming virtual conferences, and generally interesting to the Programming Languages community and beyond.

## Contents

# 1. Introduction

The 41st ACM SIGPLAN Conference on Programming Language Design and Implementation (PLDI 2020) was due to be held in London at the Royal Geographical Society during 15-20 June 2020.  Due to the COVID-19 pandemic, a decision was made in March 2020 to cancel the physical event and to run a virtual conference instead - the first virtual edition of PLDI.

In this report I provide an overview of the format of the virtual PLDI 2020 conference, for which I was general chair.  I also present an analysis of various data sources related to the conference, including: an extensive post-conference questionnaire; data on demographics gathered at registration time and via the post-conference questionnaire; and data extracted from YouTube and Slack (two of the services used to run the conference) that provide quantitative insights into the extent to which people engaged with the conference.  The report also includes logistical details about how my team and I ran the conference, which may be informative for organisers of future virtual conferences.

The report is long due to the amount of data that was available for presentation and discussion.  I hope that it will be interesting for members of the community to skim, and for organisers of other virtual conferences to digest in detail.

## 1.1. Report structure

I start, in Section 2, by giving my personal reflections about the positive and negative aspects of virtual PLDI, based both on my own experience of organizing and attending the event, and also on my interpretation of the data that backs the rest of this report.

In Section 3 I describe the format of the virtual conference, discussing relevant logistical details, and present survey results on what attendees thought about this format.

Section 4 studies the extent to which conference attendees and external viewers engaged with the conference.  This is based on an analysis of quantitative data taken from the tools that were used to run the conference (e.g., view time associated with the ACM SIGPLAN YouTube channel during the conference), and analysis of engagement-related questions from the post-conference survey.

Section 5 is entirely based on post-conference survey data, homing on in questions that compare virtual PLDI to physical PLDI, and that assess attitudes about virtual vs. physical conferences in general.

In Section 6 I present data related to diversity and inclusion at PLDI 2020.  This is based on demographic data (a) obtained from YouTube, (b) gathered at registration time and (c) gathered via the post-conference survey.   The section also presents the results of conference surveys related to time zones, which pose a difficult challenge for virtual conferences.

Finally, in Section 7 I acknowledge the huge amount of support I received from key people when putting virtual PLDI together, and various sources of advice about virtual conferencing that were useful.

Appendix A contains free-text answers to various questions from the post-conference survey.  I decided to include all such answers as they contain a lot of ideas that should be useful for other conference organizers. Each sub-appendix is referenced from an appropriate section of the report, except for Appendix A.13, which contains answers to a final survey question asking for any additional feedback.

In addition to this report, I have made available [a set of documents related to the logistics of running the virtual conference](), including the instructions that were given to student volunteers, authors and session chairs.

### 1.2. Other articles about virtual PLDI

I am aware of (and grateful for) the following write-ups related to PLDI 2020, and have read them in varying levels of detail:

- [Lessons learned for virtual conferencing at PLDI 2020](): Curated by Benjamin Pierce, this document summarises discussions on a channel of the PLDI Slack workspace devoted to collecting attendees' thoughts about what could be improved in future virtual conferences based on PLDI 2020 (as well as aspects of PLDI 2020 that should be kept).
- [A Summary of Discussions on Virtual Conferences](): This blog post by William Bowman summarises a lengthy Twitter conversation on the pros and cons of virtual conferences that started during PLDI.
- [PLDI 2020 Conference Report](): In this blog post, Neel Krishnaswami reports that "for me, the online conference experience was a complete waste of time", and argues that for the (currently uncertain) period of time where physical conferences cannot take place "it would be better to simply convert our PL conferences fully into journals, and look for other ways to make online networking happen".
- [What I Learned from My First PLDI Experience](): a blog post from Yunjeong Lee about their PLDI experience.
- [Day 1 of PLDI2020](): a blog post from Jianyi Cheng about the first day of PLDI, mainly delving into technical details of what was presented (rather than focusing on the specifics of the conference being virtual).

In this report I have not attempted to analyse the contents of these write-ups, focusing instead mainly on the post-conference survey results.

## 2. Personal reflections on virtual PLDI

My view is based on a mixture of things – my experience putting the conference together, my personal experience of attending the event, the direct feedback I have received from many people, the results of the post-conference questionnaire, and the numeric data extracted from the various tools that were used to run the conference.

### 2.1. Positives

**Reach.** PLDI 2020 reached a lot of people, many of whom could not have attended under normal circumstances. A typical PLDI attracts around 450 participants; in contrast more than 2,800 users joined the PLDI 2020 Slack workspace (see Section 4.1). The conference was live streamed via the ACM SIGPLAN YouTube channel. YouTube analytics report a total view time for this channel of 16,773 hours during the week of PLDI. In contrast the view time for the channel for *the entire year* preceding PLDI was 6,591 hours (see Section 4.2). Demographic information from YouTube and gathered via the post-conference survey shows that participants came from a wide range of countries. More than 59% of post-conference survey respondents stated that they would have been unlikely to have attended the event if it had been held physically under normal (non-COVID-19) circumstances, and financial cost was flagged up as the main reason for this.

**Enjoyment.** The drawbacks and challenges of virtual conferences are evident, and you can search for PLDI on Twitter or look at the survey results discussed later in the document for a clear message that virtual conferences are not for everyone and that some people would simply rather not attend them. However, I thoroughly enjoyed both organizing and attending virtual PLDI and I received numerous personal messages from attendees saying that they had enjoyed the conference. A number of these people told me they had initially been sceptical about how well a virtual event would work, but that it had substantially exceeded their expectations. Look at Appendix A.13 for a lot of evidence of enjoyment. If you are an organiser of an upcoming conference that has been forced to run virtually due to COVID-19, if you're feeling daunted by the prospect, and if you are wondering "Is this actually worth doing?", I can assure you with confidence that the answer is: "Yes!"

**Engagement.** Although a huge number of people had registered for PLDI, I was still concerned that a very small fraction would actually sign up to the conference's Slack workspace and get involved in discussions and Q&A. In this new and un-tested format, I was worried that there would not be enough questions for authors of papers or for our Ask Me Anything guests to make for a lively event. A *much* larger proportion of registered participants engaged in the event than I had expected, and there was no shortage of questions. In brief (and in full knowledge that statistics like this say nothing about *depth* of engagement):

- 4,297 people registered for PLDI and 2,851 people joined the Slack workspace
- More than 32,000 messages were sent on Slack overall, by 1,083 distinct users

**Slack-based Q&A.** As described in Section 3.1, Q&A for all conference tracks involved attendees posting questions on Slack, which were then answered live by authors and presenters. There were a good number of questions for all PLDI research papers and Ask Me Anything guests, and Q&A at the co-located events was lively too. I got the impression that many attendees felt more comfortable submitting questions in text form than they would be standing up in an auditorium and asking a question at a microphone. The ability of authors to follow up for more in-depth discussion, or to answer questions that could not be asked live due to lack of time, was great. Some physical conferences already do offer this option for Q&A, and I think it would be good for more physical conferences to follow suit. I don't have any particular preference for Slack – I think any platform that allows questions to be posted in text form and facilitates following up offline via a discussion thread would work well.

**The Ask Me Anything track.** From the feedback I have received, the Ask Me Anything (AMA) track (see Section 3.4) was one of the things people liked the most about PLDI. I found it really exciting watching the conversations between guests and hosts, and being able to see, live, the questions that the audience were queuing up as candidates to be asked next. I think AMAs could become a nice feature at physical conferences, but in my view they worked particularly well in the virtual environment of PLDI: to me they felt *intimate*, because all I could see on my screen was the guest and host chatting, and at the same time *inclusive* – the questions coming from a wide range of attendees, each of whom the host named before asking the question to the guest, led to a "whole conference" feeling.

**The #mentoring channel.** A dedicated Slack channel was used for members of the community to offer their services for mentoring (see Section 3.6). The survey results suggest that at least 100 mentoring sessions took place. I personally provided six individual mentoring sessions, and I found the process of responding to a Slack message requesting a session, setting up something quickly using a Clowdr video chat room (also see Section 3.6), and holding the session in peace without needing to find a physically isolated corner of a conference venue, to be an advantage of the virtual setting.

**Hallway track experiments.** As discussed under "negatives" below, the hallway track of PLDI did not work all that well. However, it did work to a degree, and I personally enjoyed having many impromptu conversations with old friends and new acquaintances using Gather and Clowdr.

## 2.2. Negatives

**Packed schedule.** The combination of the busy PLDI Research Papers track and the Ask Me Anything track that filled almost all gaps between research paper tracks meant that there was something technical going on almost all the time. This is in part due to me planning the Ask Me Anything track after the schedule for the Research Papers track had already been set in stone, and partly due to me wanting to ensure that the conference was an intense experience with lots of material to consume. In hindsight it was too packed and did not leave enough time for people to experiment with the hallway track features that were available. I think that having at least two separate 1-hour gaps per day (in addition to a series of shorter gaps) would have been better, even if this had required double-track sessions.

**Lack of scheduled social time.** As described in Section 3.6, the conferenced use two platforms – Gather and Clowdr – to facilitate hallway interactions. However, I decided deliberately not to schedule any specific times where it was recommended to go to Gather to meet other attendees (e.g. for virtual social events), and I did not propose a plan for Clowdr video chat rooms on particular topics. This was in part due to the packed nature of the schedule (see above), and in part because I was worried that there would not be an appetite for organized virtual social events – I thought that attendees would prefer to have these platforms available to be used in an organic fashion. I was also worried that the platforms we were using might not scale to the number of people actively participating in the conference. The post-conference survey results make it clear that many attendees would have appreciated some planned periods of time where socialising via Gather or Clowdr was recommended, as well as some more direction in general about how to use these platforms and the etiquette associated with their use.

**Lack of support for newcomers.** I was delighted that a very large number of people participated in PLDI, including many for whom this PLDI was their first. I regret not doing more to actively welcome these participants and give them opportunities to engage with the PLDI community. Something that would have been very easy to set up, for example, would have been an #i-am-new channel on which newcomers to PLDI could have been encouraged to post a "Hello, world!" message, and attendees in general encouraged to reply to newcomers and set up short Clowdr video chats get to know them. Another easy thing would have been to set up a number of ~10-person video meetings and have e.g. 8 newcomers and 2 more established members of the PLDI community sign up to join each. These meetings could have involved a round of brief introductions, after which the established community members could have given the newcomers advice on the best sessions to attend and the most relevant PLDI regulars to contact, based on mutual research interests. I encourage organisers of future conferences to brainstorm lightweight ideas such as these to ensure that their conference is as welcoming as possible.

**Time zones.** The 5am-5pm PDT time zone during which the conference ran covered a lot of bases but was pretty terrible for participants in 1/3 of the world, particularly in East Asia and Oceania. I don't have good suggestions for how to solve this. Survey feedback on time zones is the subject of Section 6.3.

**There are some things you just can't recreate at a virtual conference.** An obvious negative that isn't surprising or specific to PLDI is that many of the things that are great about physical conferences – long conversations over dinner, excursions, and other social and networking opportunities – are almost impossible to recreate in a virtual environment. On the other hand, those experiences – which I admit to missing just as much as anyone – are only available to among the most privileged people in the world. If we can learn to do a really good job of facilitating social interaction at virtual conferences, we may gain exposure to a wider and much more diverse network of peers.

## 3. Format and logistics of Virtual PLDI

Sections 3.1-3.6 give an overview of the format of the conference, including relevant logistical details. Section 3.7 provides details on additional logistical matters. Section 3.8 presents an analysis of post-conference survey data for questions related to the format of the conference.

### 3.1. Overarching format

Registration for PLDI was free of charge – the costs of the virtual conference were covered by the sponsors acknowledged in Section 7. Despite the conference being free, attendees were required to register in order to gain access to the conference's Slack workspace. A welcome email provided instructions on how to join the Slack workspace. All other information on how to access the conference content was communicated via the #announcements channel in the Slack workspace.[1]

Requiring registration and linking access to Slack to a participant's registration record ensured gave me confidence that the workspace would not be subject to inappropriate behaviour by anonymous users.[2] Restricting information about how to access conference content to the Slack workspace was deliberate: I wanted to maximise the amount of Q&A that would take place during the conference, thus I wanted to incentivise signing up for Slack, which was the tool that we used for Q&A.

The main PLDI conference tracks and every co-located event used the same overarching format:

- The track/event was run as a Zoom webinar.
- The track/event had a dedicated Slack channel on which attendees could post questions to be answered by presenters.
- The track/event organisers, presenting authors, session chairs and student volunteers joined this webinar as *panellists* (in the Zoom webinar technical sense of the word). This meant that they could share their video, audio and screens to the webinar as needed.
- The webinar was live streamed to YouTube.
- All conference participants were expected to view the content via the YouTube live stream, unless YouTube was not available in their country of residence; I believe this applied exclusively to participants in China.
- Participants for whom YouTube was not available were invited to join the webinar directly as *attendees*, with view-only privileges, giving them a similar experience to that of those watching via YouTube.

---

[1] YouTube live stream links were also advertised on Twitter, but not until after live streams had started. Slack was the place to gain early notice of where to find conference content.
[2] That said, there was nothing to stop someone registering for the conference using fake details and using this as a route to gain access to Slack.

- Both YouTube comments and in-webinar chat and Q&A were disabled, (a) to encourage the use of Slack as the single place for questions and discussion, (b) to minimise the difference in experience between participants viewing the YouTube live stream and participants joining the webinar directly, and (c) to minimise the number of communication channels that the organizing team would need to monitor for inappropriate behaviour.

If YouTube were universally available, it would not have been necessary to use Zoom webinars. Instead, each track/event could have operated via a Zoom meeting, inhabited by key participants and live streamed to YouTube.

### 3.2. Absence of keynote speakers

PLDI 2020 did not feature keynote speakers. While Emina Torlak (the Program Chair) and I were confident that regular conference tracks would work reasonably well virtually, we did not think that virtual keynotes would have anything like the same feelings of excitement and togetherness associated with physical keynotes. We gave the speakers who we had already invited when planning physical PLDI the option to present at virtual PLDI or defer to a future physical version of the conference and they chose the latter options.

In place of keynotes I put together the Ask Me Anything track, discussed in Section 3.4.

### 3.3. Research Papers track

Starting at 1:22, the "[How Virtual PLDI Will Work](#)" promotional video gives an overview of the format used for PLDI Research Papers, showing slack-based Q&A in action.

The Research Papers track for the main PLDI 2020 conference ran as a single track, 5am-5pm PDT during 17, 18 and 19 June 2020. The Research Papers track is usually double tracked at physical PLDI and compressed into a shorter number of hours per day. Emina Torlak and I opted for a single track stretched over a longer time period to make at least some live conference material available in more time zones at a reasonable hour.

The track was divided into sessions of 3-4 papers. Each paper was allocated a 20-minute slot, so that a session was 1h or 1h20m in duration.

The #pldi-research-papers channel on Slack was used to collect questions about each paper. One thread per paper was created by the organisers before the conference started.

The authors of PLDI papers prepared pre-recorded videos of their conference presentations in advance. The videos were required to be between 12 and 17 minutes in length, with 14-15 minutes recommended. At least one author per paper was required to join the Research Papers webinar as a panellist in order to answer questions live. A session chair also joined the webinar live for the duration of each session, to field Slack-posted questions to the authors. Two student volunteers were responsible broadcasting talk videos and monitoring the #pldi-research-papers channel during a session.

A session ran as follows (also see [Instructions for Authors](#) and [Session Chair Duties](#) for more intricate details):

- The session chair introduced the session and the first paper's video.
- A student volunteer broadcast the pre-recorded video for the paper.
- Another student volunteer posted a message to the #pldi-research-papers channel with a link to the current paper's thread, encouraging participants to post questions to that thread.

- After the video, the session chair used the 20-|length-of-video| minutes of remaining time to read questions from the paper's thread to the available authors for them to answer live.

After the session, the paper authors were free to continue answering questions about their paper via messages on the paper's thread.

The motivation for pre-recorded videos of talks being broadcast, as opposed to having authors present live, was to avoid the conference schedule slipping due to technical difficulties and bandwidth problems.

See [Instructions for Student Volunteers](#) for more details on what the Student Volunteer roles entailed.

### 3.4. Ask Me Anything track

Instead of keynote presentations, I put together an "Ask Me Anything" (AMA) program where 15 well-known members of the programming languages community agreed to participate in live streamed Q&A. This worked as follows (see [Notes for Ask Me Anything Guests and Hosts](#) for more details):

- Each guest's AMA session was advertised in advance in the PLDI program
- A #pldi-ask-me-anything Slack channel was populated with one thread per guest
- Attendees posted questions they would like the guests to answer on the appropriate threads, either before or during an AMA session.
- The guest and a host (also from the programming languages community) joined a Zoom webinar, during which the host introduced the guest and then read them selected questions from the associated Slack thread.

As an example, check out this [AMA with Işıl Dillig, hosted by Emina Torlak](#).

The AMAs were scheduled to take place during gaps in the PLDI Research Papers track. Because most gaps were 20 minutes in length, most AMAs were formally scheduled to be 20 minutes long. However, they operated via a separate webinar and live stream so that they could overrun if there were enough questions from the audience (which – for every guest – there were).

### 3.5. Co-located events

As usual, PLDI featured a number of co-located tutorials and workshops, the ISMM conference. Each of these events ran using the overall format described at the start of this section, with its own Slack channel and live-streamed Zoom webinar. I encouraged workshops and tutorials to run live, rather than to take the PLDI Research Papers route of broadcasting pre-recorded videos, if their schedule could tolerate some slippage or reorganisation in the event of technical problems, and I believe they did all take a live approach. ISMM – which, like PLDI, had a busy schedule, took the pre-recorded approach.

The Student Research Competition featured a poster session in the conference's Gather space (see Section 3.6), followed by live presentations during a section of the main conference's webinar.

A drawback of making registration for PLDI free of charge was that a large number of attendees expressed a wish to attend all or nearly all of the co-located events. This made it hard to estimate how many people would actually engage with each event until the conference commenced, and is in part the reason why every event used the overarching format of a live-streamed webinar described in Section 3.1  A tighter estimate on numbers would have allowed some of the smaller events to have operated more intimately as Zoom meetings.

## 3.6. Hallway track, posters and mentoring

The "Hallway Interaction at PLDI 2020" promotional video showcases the Gather and Clowdr features that are discussed below.

To facilitate socialising and informal conversations, PLDI 2020 featured:

- Gather, a virtual space where, as a user, you are represented by a 2D avatar that wanders around a virtual world. When you get close to another user's avatar you can see and hear that user via their webcam and microphone. The idea is that this allows drifting in and out of conversations. Figure 1 shows a screenshot of me and Nadia Polikarpova chatting in Gather, and is taken from this example video that we shot to promote the hallway track of the virtual conference.
- Clowdr, an environment for virtual conferencing that was in early stages of development while I was putting PLDI together. Jonathan Bell, one of the main developers of Clowdr, integrated one of Clowdr's main features – video chat rooms – into the PLDI Slack workspace. From anywhere in Slack one could type /video room-name to create a video chat room with the given name or join an existing chat room with that name. Chat rooms could be of varying sizes and could be public (visible to anyone who queried the current live chat rooms) or private (so that the room creator could explicitly invite selected users to join). A Clowdr video chat room is similar in nature to a meeting in Zoom, Microsoft Teams or Google Meet. To me, the attractive feature about Clowdr for a virtual conference was the ability to create a chat room very quickly on demand, without leaving the conference's Slack workspace.

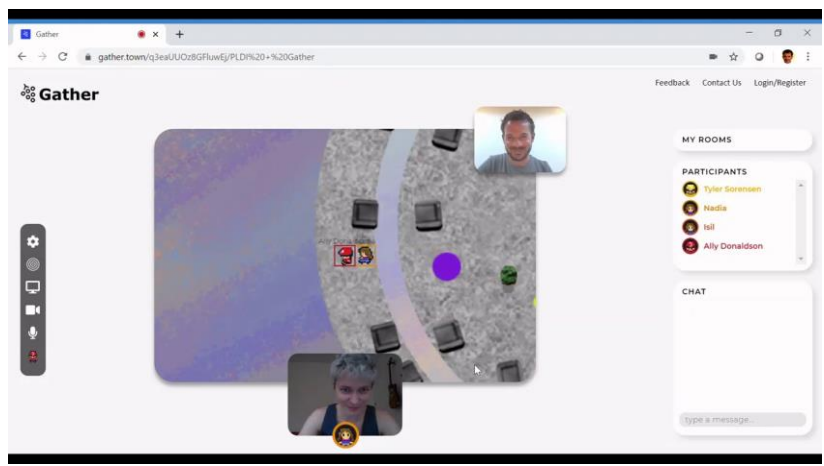

*Figure 1  A screenshot of the PLDI 2020 Gather space*

The conference's Gather space featured around 50 poster booths, with posters from some PLDI, ISMM and MAPL authors, and the Student Research Competition posters. In Gather you could walk up to a poster board and press a key to interact with the poster; this allowed you to zoom in on the poster content and scroll around the poster. I encouraged poster authors to include their Slack handles on their posters so that other attendees interested in discussing their poster with them could direct-message them on Slack to arrange a meeting.

There was a formal Student Research Competition poster session, but there was no formal session for the rest of the posters.

Via a #mentoring channel in Slack, conference attendees advertised their willingness to provide mentoring to other attendees, the idea being that someone interested in receiving a mentoring session could direct-message any of these mentors to arrange a suitable time, with Clowdr being available as a convenient option for having a video chat.  Anyone was free to put themselves forward as a mentor – I and some members of the PLMW organizing committee posted messages offering our services as mentors, and then I posted a message to the #announcements asking others to follow suit.  The etiquette was that any attendee should feel free to send a direct message to anyone who had posted on the #mentoring channel to ask for a session.  This scaled well – virtually no organization was required beyond setting up the channel.

## 3.7. Additional format and logistical details

### 3.7.1. Promoting the virtual conference

To make a strong statement that virtual PLDI was really going to take place, and our intent for it to be a proper conference, John Wickerson (the Publicity Chair) had the idea of making a promotional music video featuring several members of our conference's community.  We used the resulting song, "[This is Still PLDI](#)" – to the tune of "Always On My Mind" (by Wayne Carson, Johnny Christopher and Mark James), featuring alternative lyrics by John Wickerson – to promote the conference on social media and encourage people to register.

I don't necessarily encourage every conference to make a music video.  However, there are many different levels of intensity with which virtual conferences can be run.  If you are aiming for a relatively intense event – as I was with PLDI – then it could be worth making some kind of publicity statement to that effect as a means for increasing engagement.

To help conference attendees understand in advance how the format of the conference would work, I recorded two further promotional videos:

- A video showing how the format of live-streamed talk recordings with Slack-based questions answered live by authors would work ([link to full video](#), [link to short version of video](#))
- A video showcasing the "hallway" features of PLDI 2020 – Gather and Clowdr ([link to video](#))

My hope is that – as with the song – these videos helped to make it clear that PLDI was indeed going ahead, and that my team and I were doing our best to provide an interactive experience.  Other than YouTube views (around 2K for the short and long version of the "format" video combined, and also around 2K for the "hallway" video), I don't have a sense for how useful these videos were to people in practice.

### 3.7.2. Collecting videos from authors

PLDI Research Paper authors were asked to submit lightning videos and full talk recording videos in MP4 format with 16:9 aspect ratio.

A DropBox Professional account was purchased, providing a facility for authors to drop off video files.  The drop-off facility is anonymous (authors receive a link that they can use to drop files but cannot access files that others have dropped), and versioned (authors can update the file they have dropped off, and in the receiving folder multiple versions of a file are disambiguated via a time stamp).

Authors also had to submit a video release form, giving permission for their talk recording video to be broadcast during the conference, and optionally for their lightning/talk recording videos to be hosted on YouTube and ultimately in the ACM Digital Library.

We set two deadlines: an early deadline for receiving the lightning video and a single release form covering both lightning and talk recording videos, and a later deadline for receiving the talk recording videos. We still set the latter deadline to be two weeks before the start of the conference, to allow plenty of time for videos to be checked in case of technical problems. Requesting release forms at the earlier deadline worked well as it allowed plenty of time for late form submissions to be chased.

Our team of Student Volunteers took care of checking lightning and talk recording videos for quality, checking that release forms had been appropriately completed, and editing lighting videos together into lighting session videos that were used to promote each session. [Here are the instructions with which the Student Volunteers were provided in relation to checking and editing lighting videos](#).

### 3.7.3. Accounting for aspect ratio when broadcasting videos

During the PLDI Research Papers rehearsal (see Section 3.7.4) we found that a 16:9 video does not look right if broadcast from a machine with a display that uses a different aspect ratio. The [Student Volunteer Responsibilities](#) document has guidelines on how to work around this, including using the IINA media player on MacOS.

As a side note, we also found that not all author-submitted videos worked with all media players – e.g. there were problems with the Windows default player. VLC proved to be a very reliable option.

### 3.7.4. PLDI Research Papers rehearsal

We gave authors and session chairs the opportunity to take part in a PLDI Research Papers rehearsal one week before the conference. The main points of doing this were (a) to give our Student Volunteers a chance to practice their duties – in particular to practice the various Zoom controls they needed to use; (b) to eliminate any technical problems that authors and session chairs would face when trying Zoom and Slack for the first time, for example: not being familiar enough with Slack to identify the Slack thread for a paper; not being able to sign in to Zoom; finding that audio / video didn't work with Zoom; finding that one's Zoom client was too old; and (c) to ensure that all participants were comfortable with the conference format.

We set aside three 2-hour blocks of time on the day of the rehearsal: a Europe-friendly block, a USA-friendly block, and another block that was relatively friendly to East Asia and Australasia.

We used a Google form to ask authors and session chairs whether they wanted to take part in the rehearsal and to specify which 2-hour block they would prefer, making it clear that they would not be required to participate for the entire block.

We then put together a dummy schedule based on author and session chair availability – each session chair was assigned to a fake session consisting of three random papers; the only tie between the session chair and papers was that they had all specified a mutually convenient time block. We gave each paper around 6 minutes of time during the rehearsal, allowing the session chair to practice making an introduction to a presentation, the Student Volunteers a chance to try broadcasting a few minutes of a talk recording video, the session chair a chance to try reading out some questions from Slack (dummy questions posted by myself or the Student Volunteers), and the authors a chance to check that their audio and video was working when answering these questions.

The rehearsal was very chaotic: it successfully identified many technical problems on the parts of authors, session chairs and Student Volunteers that they could address in time for the real conference. It made it clear that not all authors and session chairs had yet understood the intended format of the conference. It also informed me that many authors had not received Zoom webinar invitations, I think due to institutional spam filtering.

The real conference ran smoothly, and I think that this was in part due to the problems solved via the rehearsal and the double-checking I put in place (see Section 3.7.5).

### 3.7.5. Sheets for author and session chair sign-off

During the rehearsal it became clear that multiple authors had not received the Zoom webinar invitations that had been sent for the rehearsal sessions, and that several authors and session chairs were not that familiar with the planned format for how the conference would work. For the duration of the rehearsals myself and Dan Iorga (Student Volunteer Captaion) were bombarded with emails and Slack messages related to problems faced by authors and session chairs, and I was concerned that the real conference would also be dominated by this.

To guard against these problems I put in place a [session chair confirmation sheet](#) (real session chair names removed in this example) and an [author confirmation sheet](#). I was then able to chase up those session chairs and authors who did not check off their paper / session, and help them resolve any questions or technical problems – e.g. in some cases it was necessary to send Zoom webinar invitations manually to authors due to institutional spam-filtering of emails sent directly by Zoom.

### 3.7.6. Onboarding registered users to Slack and authenticating users to Gather

Slack (or at least, not the free Slack plan), does not provide an "import from CSV" option to mass-register users. To work around this, Dan Iorga (Student Volunteer Captain), the Elmer van Chastelet and Danny Groenewegen from researchr and Mike Moshell from Registration Systems Lab worked together to put in place a system whereby anyone could request an invitation to the PLDI Slack workspace via the PLDI website, but such that an invitation would only be sent if the requester's email address was known to the PLDI registration database.

Similarly, anyone could attempt to create an account with Gather to access the PLDI Gather space, but this would only succeed if the email address used for the account was known to the PLDI registration database, and an email confirmation was required for account creation to succeed.

### 3.7.7. Business meeting and awards

We held the PLDI Business Meeting as a live-streamed Zoom webinar – temporarily hijacking the Ask Me Anything track's webinar. A number of awards are traditionally presented over lunch at PLDI, and we decided to have these awards presented during the Business Meeting. This worked reasonably well but was a little underwhelming as I believe many attendees did not realise that the Business Meeting would feature awards. For future conferences I recommend thinking creatively about award sessions – e.g. having award winners take part in short interviews on the topic related to the achievement being recognised by the award, perhaps with Q&A from the audience.

## 3.8. Survey feedback related to the format of virtu0al PLDI

I now present results from questions in the post-conference survey that relate to the format of the event.

**Question:** *Which meeting mode do you think would have been best for the PLDI Research Papers track?*

The options were:

- [The PLDI 2020 approach] Broadcast mode: Attendees watch via a broadcast (e.g. YouTube live stream or webinar) and do not have video or audio, but can type questions for Q&A
- Meeting mode: Attendees join a large meeting and have video and audio for Q&A. The meeting size is capped at 300 concurrent participants, admitted in a first-come-first-served manner. A broadcast is available to attendees for whom there is no space in the meeting, and those attendees can type questions for Q&A.
- Asynchronous mode: pre-recorded videos are available, and a text-based platform is available for Q&A. There are no live sessions.
- Other (please describe)

**Results (458 responses):**

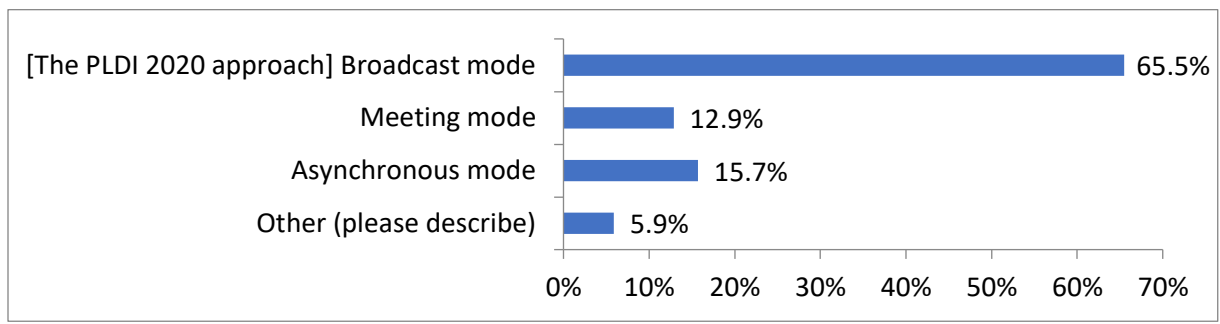

The overwhelming feedback is that attendees favoured the format used by PLDI compared with the asynchronous format that some recent conferences have used, or idea of using a large interactive meeting.

Several respondents provided alternative suggestions, which are presented in Appendix A.1. Multiple respondents make the point that providing pre-recorded videos ahead of time could alleviate the problem of time zones to some degree – members in incompatible time zones could watch a presentation in advance and post questions via Slack for the authors to answer during the live session.

**Question:** *Which meeting mode do you think would have been best for the PLDI Ask Me Anything track?*

The options were:

- [The PLDI 2020 approach] Broadcast mode: Attendees watch via a broadcast (e.g. YouTube live stream or webinar) and do not have video or audio, but can type questions for Q&A
- Meeting mode: Attendees join a large meeting and have video and audio for Q&A. The meeting size is capped at 300 concurrent participants, admitted in a first-come-first-served manner. A broadcast is available to attendees for whom there is no space in the meeting, and those attendees can type questions for Q&A.
- Asynchronous mode: guests interact with the audience via a text-based platform. There are no live sessions.
- Other (please describe)

**Results (437 responses):**

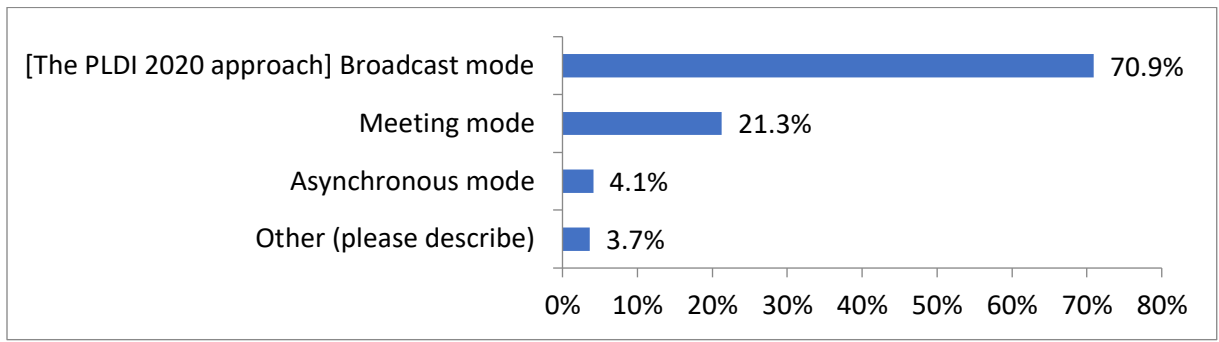

Again, the overwhelming response is that the PLDI format was appropriate for AMA sessions. The asynchronous option received even less support than in the above question about the Research Papers track, while the "large meeting" mode received more support.

Free text responses providing alternative options are presented in Appendix A.2.

**Question:** *What, in your view, is the ideal length of a paper presentation at a virtual conference (excluding time for Q&A)?*

**Results (465 responses):**

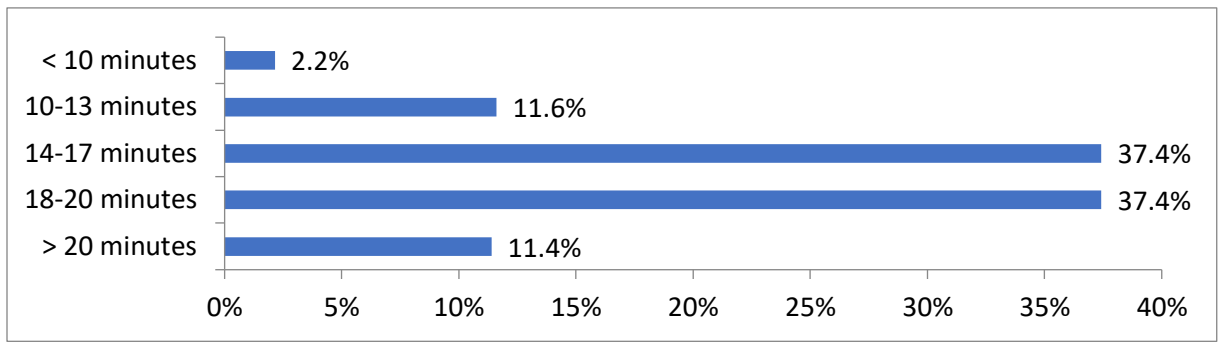

Paper presentations at PLDI were between 12 and 17 minutes in length. The results suggest that this is acceptable, but that slightly longer presentations would also be welcome.

**Question:** *If you were an author of a paper at another virtual conference, would you prefer to present the paper live, or to submit a pre-recorded talk?*

**Results (454 responses):**

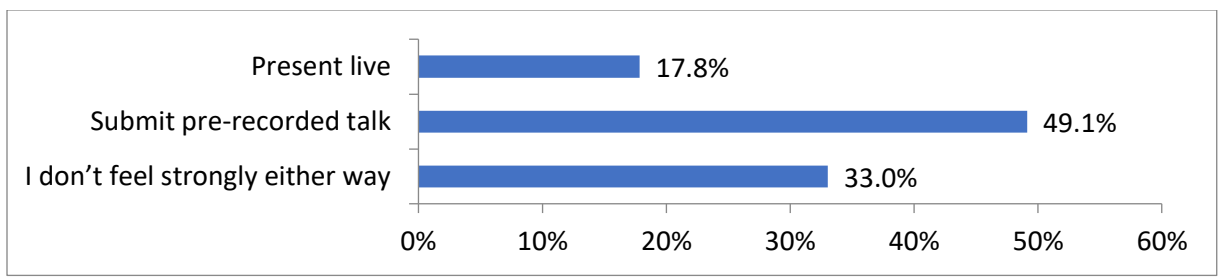

The above results are across *all* conference attendees. I thought it would be interesting to see the results restricted to respondents who answered "Yes" to the question "Are you an author of any material that was presented at PLDI or one of the co-located meetings?". The responses have a similar shape.

**Results restricted to authors of material at PLDI or a co-located meeting (110 responses):**

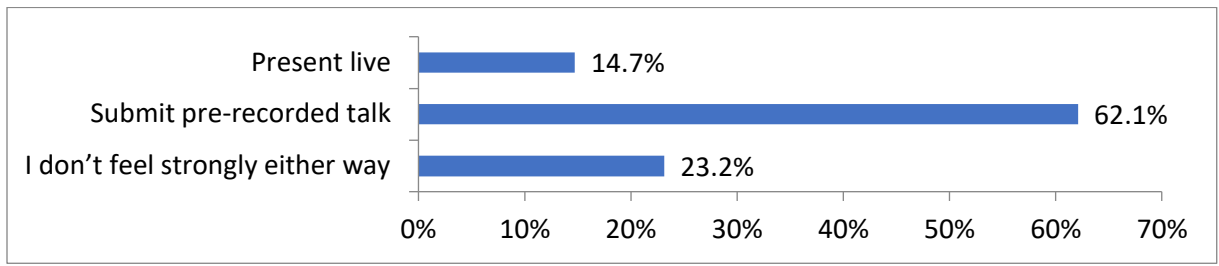

- Present live — 14.7%
- Submit pre-recorded talk — 62.1%
- I don't feel strongly either way — 23.2%

**Question:** *Please share your thoughts on how well the Q&A process of Slack-based questions answered live by authors / guests worked.*

This open question received a large number of responses which are given in Appendix A.3. The consensus appears to be that this approach worked well. There are some remarks that the Slack specifically has shortcomings that alternatives may have, and some objection to Slack being non-free software. Here are a few responses that I found insightful:

- *"It was pretty good. I find the fact that co-authors can answer live on Slack a benefit, which I've seen work in other, physical, conferences too. I also like having questions moderated and relayed by session chairs. I think that really helps avoid grandstanding by big names."*
- [This respondent is saying that they preferred the Slido-based Q&A process used by PLMW to the Slack-based process of PLDI.]  *"I liked the Q&A process of PLMW, where questions could be upvoted. In this way, the questions got answered that most people were interested in, and not the questions that came in first."*
- *"I liked it better than live conferences. Far more effective medium for asking questions (no waiting in line, no worrying about how loud to speak, confusion over accents, etc.) And, easy to spawn followup discussion."*
- *"I liked the Q&A process a lot. I didn't ask questions, but I found the interaction mode, where moderators fielded the questions, simple and efficient. The one issue I saw was that some moderators are better than others at this process. Some just went from the top to the bottom, which is not great because it advantages people who ask questions many hours before the talk even happens and takes neither continuity nor general interest in a question into account. Others tried to ask the most popular questions and perhaps group those with a similar theme, both of which are nice approaches."*
- *"It worked really, really well. In some cases, for example, clarifying questions or statements were provided in Slack, based on answers, and were picked up by the moderator. It yielded a really effective back & forth. In some cases, moderators simply read questions from the top, which was a little less interesting. Effective moderators grouped questions together to forms themes that directed the arc of the Q&A or AMA. Some clarity around voting for questions may have helped weaker moderators pick more interesting questions."*

**Question:** *Please indicate how strongly you agree or disagree with the following statement: "I recommend that future virtual conferences make Gather available for social interaction"*

**Results (494 responses):**

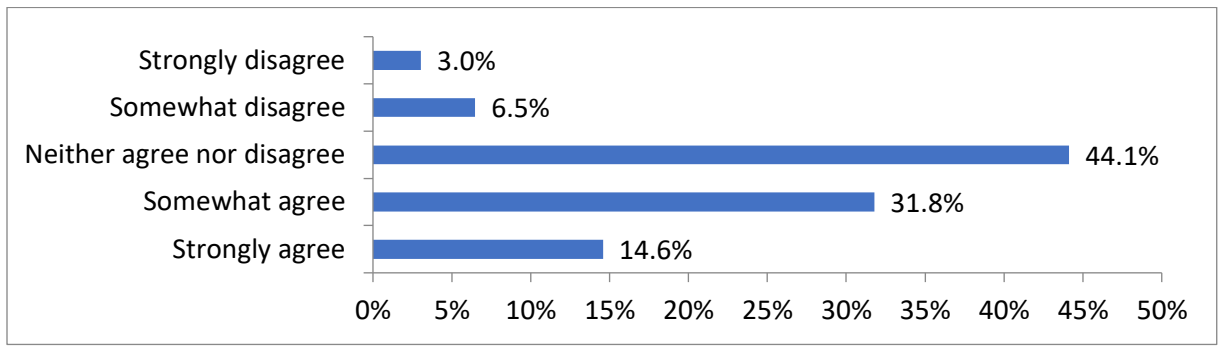

Overall, respondents seem ambivalent about Gather, leaning towards being positive about it. As discussed in Section 4.4, a lot of survey respondents did not actually try Gather.

**Question:** *Please indicate how strongly you agree or disagree with the following statement: "I recommend that future virtual conferences make Clowdr video chat rooms available for social interaction"*

**Results (447 responses):**

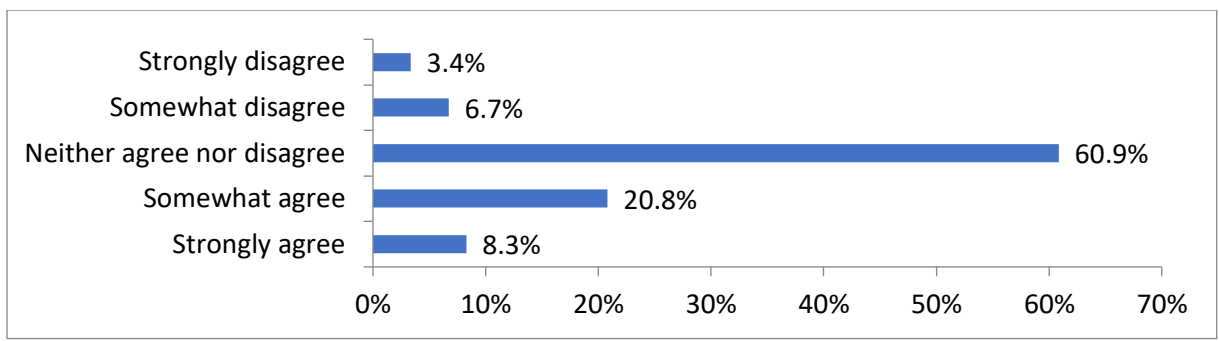

Ambivalence towards Clowdr is higher than for Gather, but there is still a positive lean overall. As discussed in Section 4.4, relatively few survey respondents tried Clowdr.

**Question:** *Do you have suggestions for ways to provide more social interactions, networking or mentoring opportunities at a virtual conference?*

There were a very large number of insightful responses to this free-text question. They are presented in Appendix A.4 and I strongly encourage organizers of future virtual conferences to read them. Here is a selection of responses that I found noteworthy and that are illustrative of the variety of responses to this question:

- *"Offer routine group video meetings for the audience following a session, as an automatic follow-on from the presentation platform."*
- [This respondent is referring to a series of "coffee hour" Zoom meetings that conference participants set up.] *"More meetings of the Zoom coffee hour style, where conversations are small (roughly 5 people), time-limited, and randomized. In my experience, these events were orders of magnitude better at encouraging participation and interaction than any of the other platforms"*
- *"I think PLDI did tremendously well with trying to establish social interactions. However, I think if you are not already in the "in-group" (I'm more in the SE community, ICSE/FSE/OOPSLA), it's very hard to initiate the same levels of interaction."*

- *"The mentoring channel was fantastic and I think that worked probably even better than it would have in person (I had 5 mentoring conversations that I think would not have happened in London)."*
- *"During IEEE S&P there was a slack bot called Donuts that was starting a private chat with a random member of the audience twice a day. This helped (and strongly invited) to connect with people as there was no need for initiating the chat (as was necessary with gather)"*
- *"Arrange for small groups of people to meet on Gather. At POPL 2020, there were mentoring breakfasts with about 5-6 people per table. Similar events can be held over Gather, we just need some rules on how to make them work."*
- *"Accessibility is important. Not having meals together was a major bummer. That is usually when I meet most people at conferences. I think some kind of shared online activity that is not a meal would help with that, like an online game."*
- *"One idea I had was to have some sort of [speed-networking] online for new students/researchers. I found it a little intimidating to use the software for social interaction given that I am a young researcher who does not know many people. It would be nice for social interactions to be quickly formed in this way (preferably near the beginning of the conference)"*

The post-conference survey also featured open questions on suggestions for specifically improving Gather and Clowdr which I have passed on to the developers of those platforms.

## 4. Engagement with the virtual conference

Several sources of data on engagement were available. I discuss the number of registrants for the conference and for Slack (Section 4.1), statistics provided by YouTube (Section 4.2), statistics on usage provided by Slack (Section 4.3), and results from questions in the post-conference survey related to engagement (Section 4.4).

### 4.1. Registrations, Slack sign-ups and active Slack users

Table 1 summarises the number of people who registered for PLDI, the number of Slack workspace users and the number of those users who posted at least one message.

*Table 1 Engagement with PLDI measured by number of registrations, Slack sign-ups, and active Slack users*

| Number of registrations | Number of registrants who joined the Slack workspace | Number of additional Slack users | Total Slack users | Number of Slack users who posted at least once |
|---|---|---|---|---|
| 4,297 | 2,685 | 166 | 2,851 | 1,083 |

Of the 4,297 people who registered for PLDI, 2,685 joined the PLDI Slack workspace using the same email address used for registration. A further 166 users were added manually to the workspace using an email address that did not match any registered participant. This was done in order to add our student volunteers (who needed access to the workspace before it had been opened to registered participants), representatives from our Gold and Platinum sponsors, and various authors who had forgotten to register in advance.

The number of registrants for a typical edition of physical PLDI is around 450 – substantially lower than the number of PLDI 2020 participants who engaged with the conference by posting at least one Slack message. Thus, if measured by number of active Slack users, engagement with PLDI was higher than usual. But, of course, sending even a moderate number of Slack messages does not imply that a participant engaged with PLDI in a *deep* manner. Furthermore, many participants *may* have engaged deeply with the PLDI content despite sending no messages: we have no way of knowing how many of the remaining Slack users participated as silent observers of the conference live streams and Slack activity vs.; likewise, we cannot know how many of the registrants who did not join the Slack workspace lost interest in attending, vs. found out that the conference was being live streamed and decided to watch the conference content via that route without feeling the need to interact via Slack.[3]

### 4.2. YouTube statistics

The live streams for PLDI and co-located events were made available via the ACM SIGPLAN YouTube channel. Table 2 presents overall analytics for the channel for the year immediately preceding the conference, the week immediately preceding the conference, the week of the conference, and the week after the conference.

*Table 2 YouTube analytics for the SIGPLAN YouTube channel for the weeks associated with several recent ACM SIGPLAN conferences, the week prior to PLDI 2020, the week of PLDI 2020 and the week following PLDI 2020*

| Significance of week | Dates | Views | Watch time (hours) | Unique viewers | Average views per viewer | Average view duration (mm:ss) |
|---|---|---|---|---|---|---|
| Year before PLDI 2020 | 15 June 2019 – 14 June 2020 | 86,032 | 6,591 | Not available | Not available | 4:35 |
| Week before PLDI 2020 | 8-14 Jun 2020 | 4,179 | 207 | 1,982 | 2.1 | 2:58 |
| Week of PLDI 2020 | 15-21 Jun 2020 | 60,947 | 16,773 | 23,236 | 2.6 | 16:30 |
| Week after PLDI 2020 | 22-28 June 2020 | 6,917 | 747 | 3,336 | 2.1 | 6:28 |

I find the figures for watch time particularly striking: the watch time for the SIGPLAN YouTube channel during the *week* of PLDI was more than 16K hours, which is more than 2.5 times the watch time for the preceding *year*. The number of unique viewers during PLDI – more than 20K – shows that the virtual conference material reached a very large audience, but that on average an individual viewer watched conference material for less than 1 hour in total during the event. This supports the hypothesis that the virtual conference reached a wide audience, but that engagement was not as deep as with a physical event.

The average view time during PLDI of 16:30 minutes is also much higher than the view times associated with the year and week preceding the conference, and a lot higher than the view time for the week following.

Table 3 shows the watch time during the week of PLDI associated with live stream videos for the Research Papers and Ask Me Anything tracks of PLDI, for the Programming Languages Mentoring Workshop (PLMW) – the co-located event with the highest watch time – and for other co-located events. The "Total" column shows that the watch time associated with PLDI-related live streams accounts for nearly all of the watch time for the SIGPLAN YouTube channel for the week of PLDI.

---

[3] Two registered participants did get in touch to tell me that they specifically would *not* join Slack, in one case because they do not like Slack, and in the other case because they are not willing to use software that is not free and open-source.

*Table 3  Total watch time for the three most-watched PLDI tracks, and for other co-located events.*

| Track | Research Papers | AMAs | PLMW | Other co-located events | **Total** |
|---|---:|---:|---:|---:|---:|
| Watch time (hours) | 5,466 | 2,935 | 2,136 | 6,416 | **16,496** |

## 4.3. Slack activity

Reports from Slack downloaded at the end of the conference state that 32,618 messages were posted in total (this includes some messages posted in the run-up to the conference, including during a rehearsal with authors).

Table 4 shows details of activity on the conference's public Slack channels, showing the number of messages posted in total, the number of members subscribed to each channel, and the number of those members who posted messages.  The sum total of these message is much less than the total number of messages posted during the conference – i.e., the vast majority of messages exchanged were private.

*Table 4  Data from Slack about activity on each of the conference's public channels*

| Channel name | Messages | Total members | Members who posted |
|---|---:|---:|---:|
| pldi-research-papers | 1,157 | 1,223 | 252 |
| mapl | 424 | 409 | 99 |
| pldi-ask-me-anything | 370 | 867 | 130 |
| soap | 356 | 659 | 89 |
| discussion | 349 | 726 | 98 |
| lessons-learned-for-virtual-conferencing | 344 | 299 | 72 |
| ismm | 328 | 323 | 69 |
| spoofax | 324 | 133 | 37 |
| rems-deepspec | 317 | 438 | 67 |
| infer | 224 | 302 | 49 |
| pldi-business-meeting | 132 | 100 | 35 |
| technical-support | 111 | 64 | 23 |
| sponsor-github | 88 | 100 | 18 |
| mentoring | 77 | 327 | 51 |
| quantum-tutorial | 67 | 256 | 23 |
| dse-tutorial | 60 | 171 | 13 |
| student-research-competition | 45 | 138 | 13 |
| announcements | 41 | 2,830 | 5 |
| lctes | 33 | 29 | 10 |
| programming-for-autonomy-tutorial | 27 | 77 | 3 |
| job-adverts | 24 | 160 | 18 |
| soap-informal | 19 | 204 | 9 |
| cares | 11 | 43 | 5 |

Figure 2 illustrates the distribution of active Slack users according to the number of messages they posted.  Inactive users – who posted zero messages – are not included.  To make the figure readable, data for myself and the Student Volunteer Captain (Dan Iorga) are also excluded, as we each posted in excess of 1000 messages.  The figure shows that most active participants posted a small number of messages, with a long tail of more active participants.

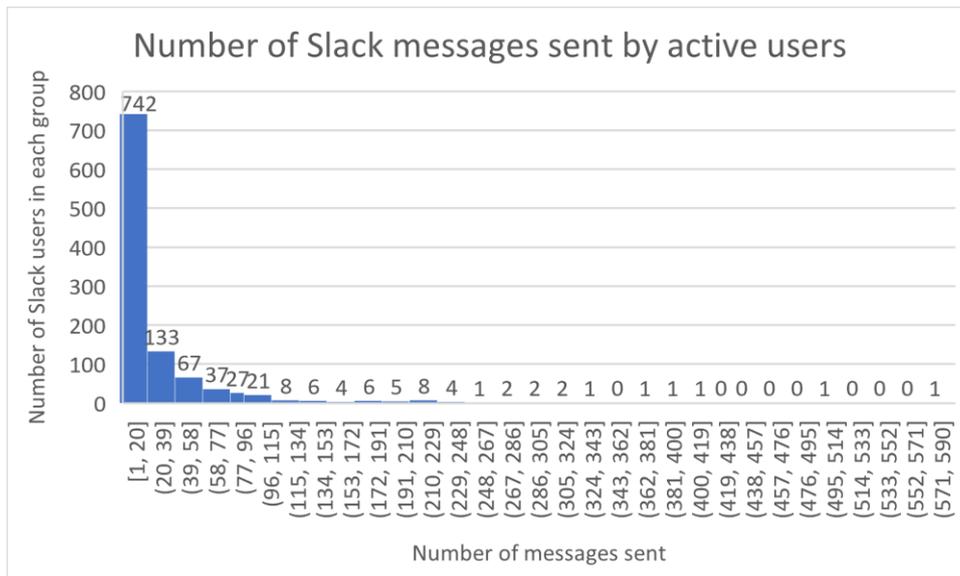

*Figure 2 Histogram showing the distribution of active Slack users according to the number of messages they posted, excluding the General Chair and Student Volunteer Captain*

### 4.4. Survey feedback related to engagement

Results for engagement-related post-conference survey questions now follow, covering engagement with the main conference tracks, co-located events, mentoring, Gather and Clowdr, and use of Slack. I have included some personal thoughts on how to interpret the results for some of the questions.

**Question:** *During the week of PLDI events, how many hours in total did you spend watching conference material?*

**Results (461 responses):**

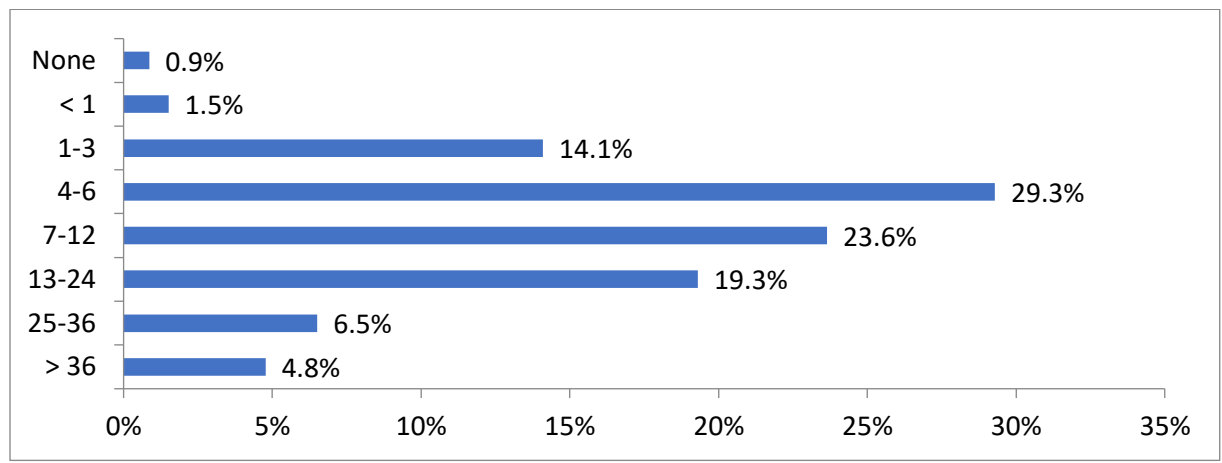

**Question:** *How many PLDI research paper presentations did you watch a substantial part of, either live or later?*

**Results (466 responses):**

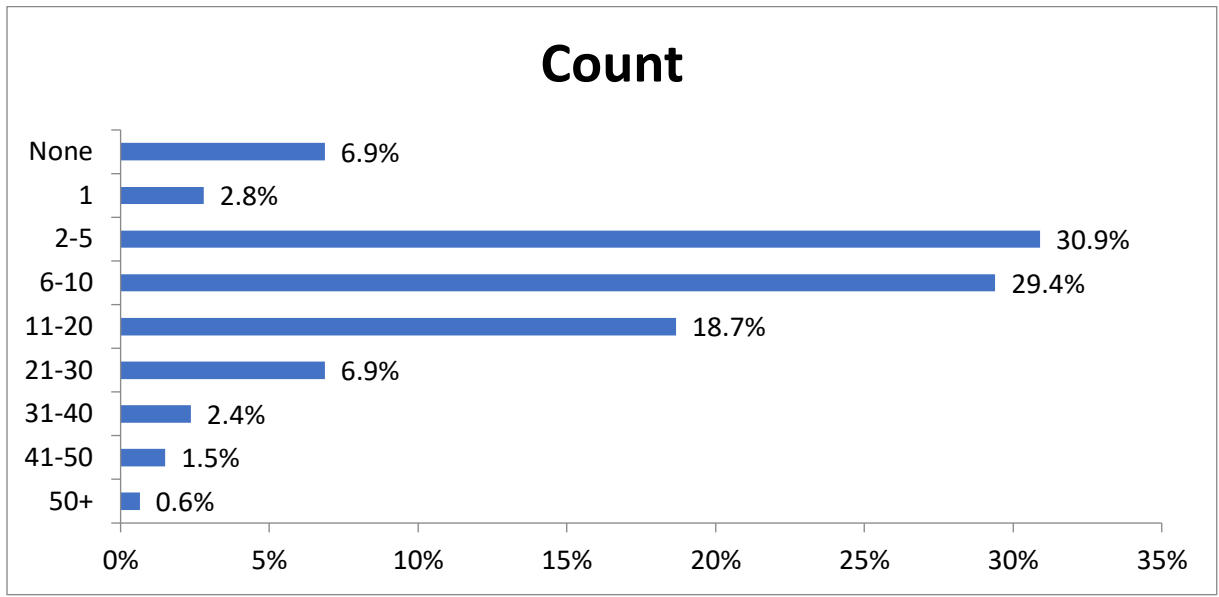

The results for this and the previous question appear to support the hypothesis that while the virtual conference reached a very large audience, engagement with conference material on a per-attendee basis was lower than it would be at a physical conference.

**Question:** *Which Ask Me Anything guests' sessions did you watch a substantial part of, either live or later?* (Respondents were invited to select multiple options if applicable.)

**Results (339 responses) –** the names of guests have been anonymised, the results are shown in decreasing order, and absolute counts are shown, not percentages, since respondents could select multiple options:

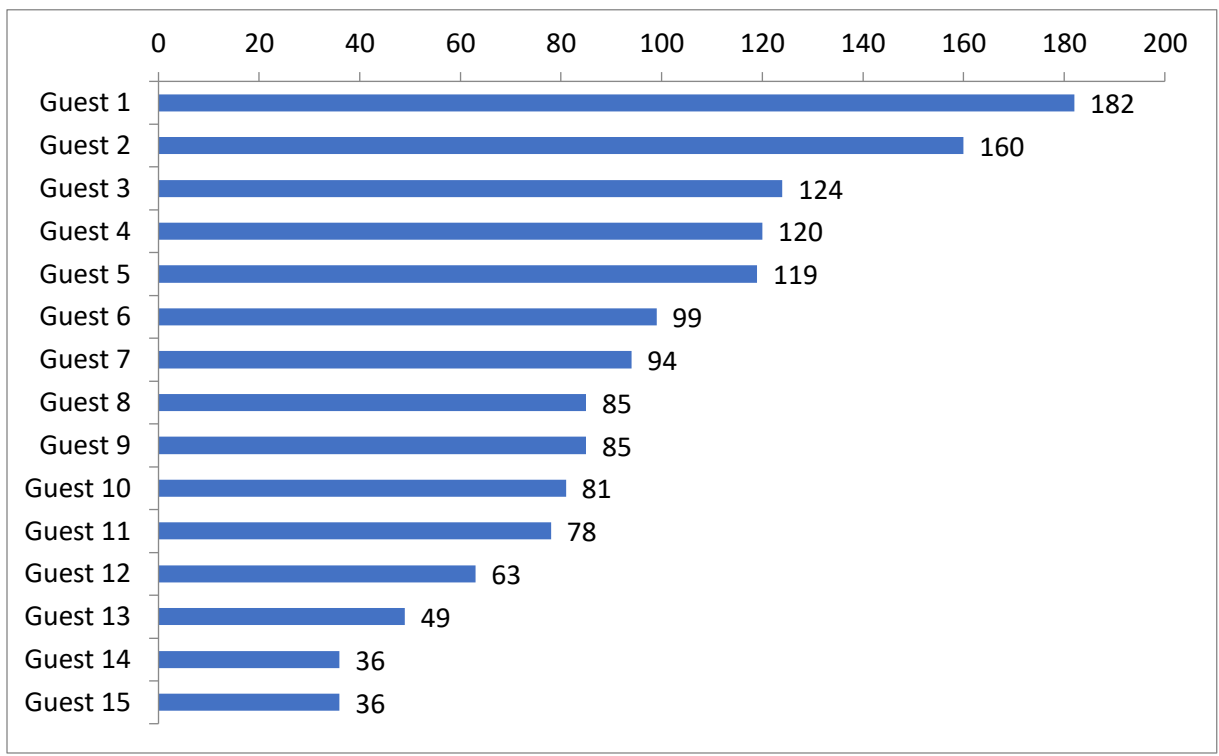

**Question:** *Which co-located events did you watch material from, either live or after-the-event?* (Respondents were invited to select multiple options if applicable.)

**Results (414 responses)** – absolute counts are shown, not percentages, since respondents could select multiple options:

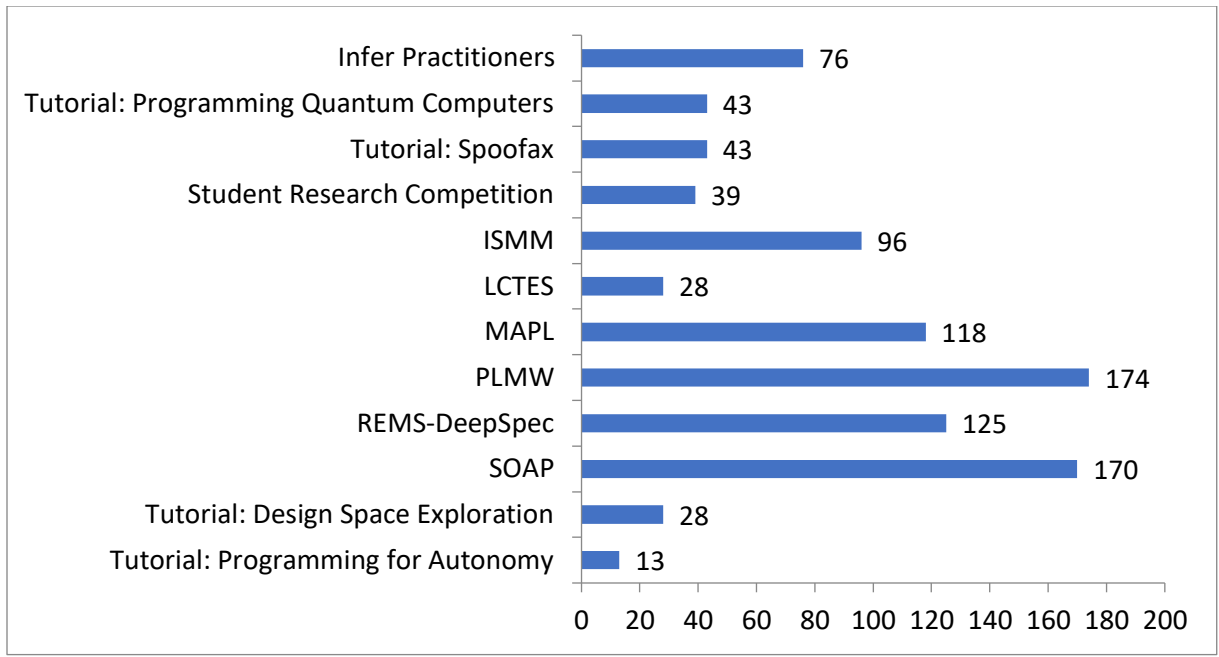

**Question:** *Please estimate how many questions you asked on the PLDI Slack, in total (i.e., questions about papers in any event or questions to Ask Me Anything guests)?*

**Results (469 responses):**

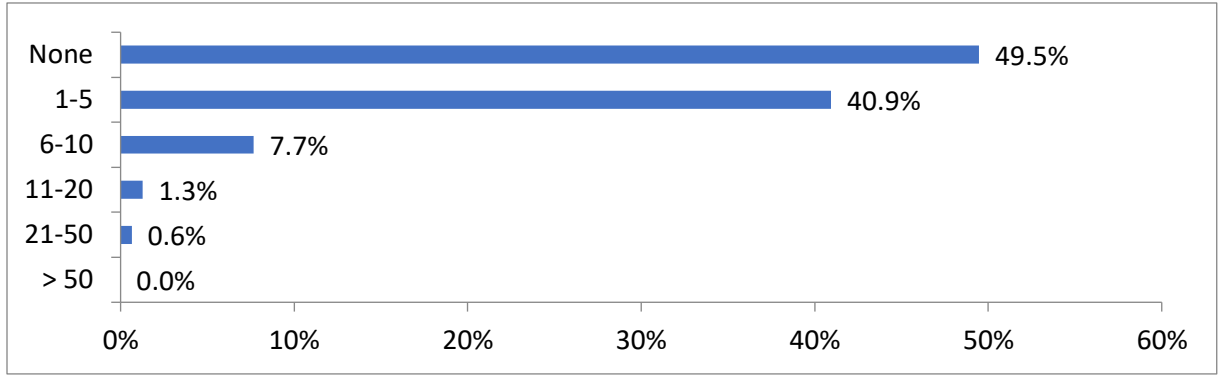

These results match the trend shown in Figure 2 – just as that figure showed that most active Slack users posted a small number of messages with a small number being significantly more active, we see that most users who participated in Q&A asked a small number of questions, with some users being much more active in Q&A.

**Question:** *Did you offer your services as a mentor via the #mentoring channel?*

**Results (513 responses):**

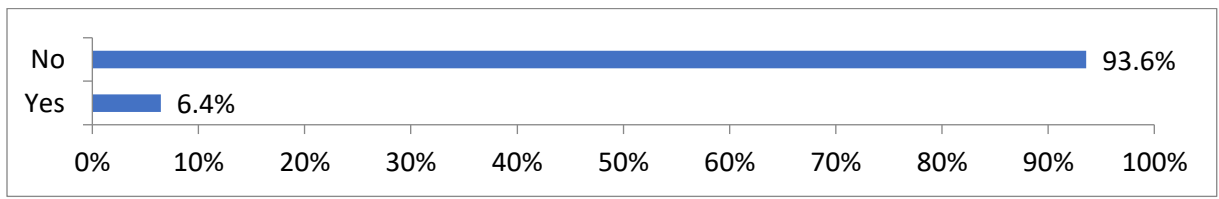

**Question:** *If you answered "Yes", how many mentoring sessions did you provide?*

**Results (31 responses)**, presented as raw numbers rather than percentages to facilitate estimating a lower bound on sessions delivered by mentors, and filtered to only include responses where the respondent had answered "Yes" to the previous question (a lot of respondents selected "No" to the previous question and opted to answer this question with "None" rather than leaving it un-answered):

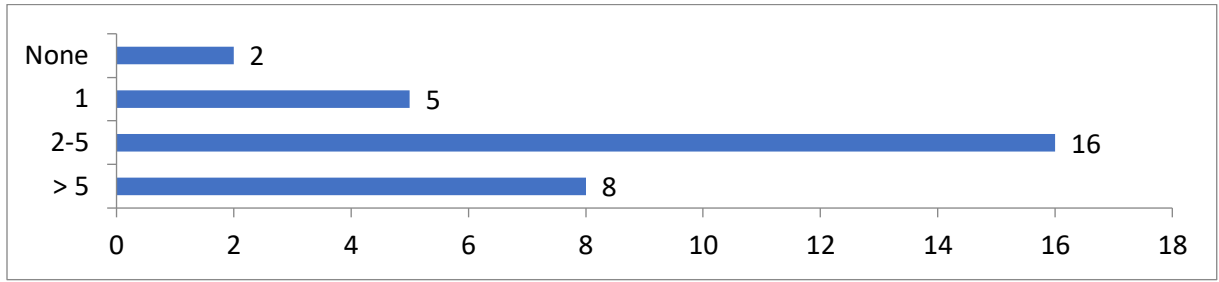

| | |
|---|---|
| None | 2 |
| 1 | 5 |
| 2-5 | 16 |
| > 5 | 8 |

Counting each "2-5" entry as 3.5 and each ">5" entry as 6, 109 seems like a reasonable estimate for a lower bound on the number of mentoring sessions that these respondents provided.

**Question:** *Did you approach a mentor who had offered their services via the #mentoring channel to request a session?*

**Results (481 responses):**

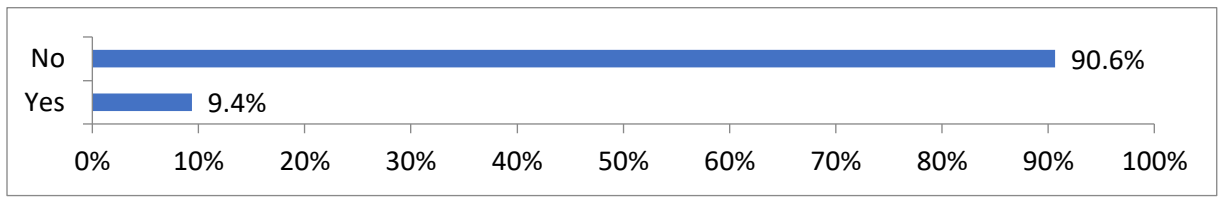

| | |
|---|---|
| No | 90.6% |
| Yes | 9.4% |

**Question:** *If you answered "Yes", how many mentoring sessions did you receive?*

**Results (45 responses)**, presented as raw numbers rather than percentages to facilitate estimating a lower bound on sessions received by mentees, and filtered to only include responses where the respondent had answered "Yes" to the previous question (a lot of respondents selected "No" to the previous question and opted to answer this question with "None" rather than leaving it un-answered)**:**

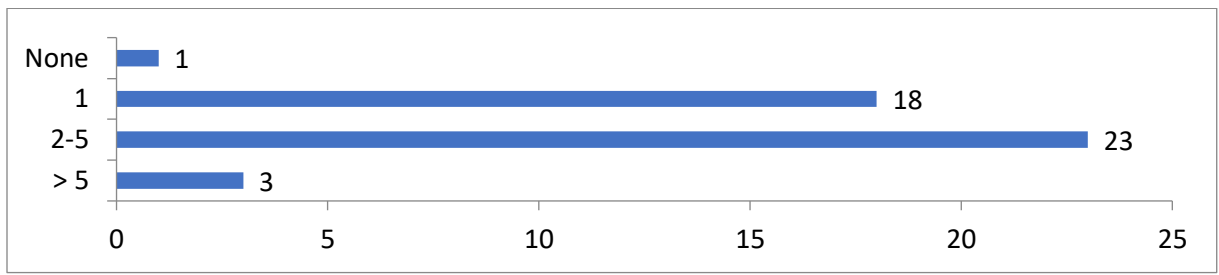

| | |
|---|---|
| None | 1 |
| 1 | 18 |
| 2-5 | 23 |
| > 5 | 3 |

Again, counting each "2-5" entry as 3.5 and each ">5" entry as 6, 116 seems like a reasonable estimate for a lower bound on the number of mentoring sessions that these respondents received.

There appears to be little overlap between mentors and mentees. Three respondents report having offered their services as a mentor and also approached other attendees about receiving mentorship; in all cases the respondents reported taking part in mentoring sessions both as mentor and mentee.

**Question:** *How many hours total did you spend in the PLDI 2020 Gather space?*

**Results (524 responses):**[4]

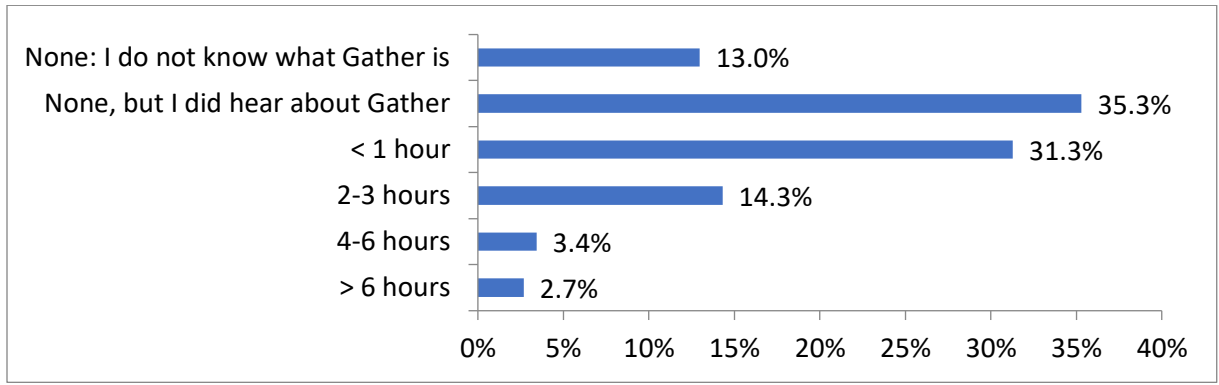

| Category | % |
|---|---|
| None: I do not know what Gather is | 13.0% |
| None, but I did hear about Gather | 35.3% |
| < 1 hour | 31.3% |
| 2-3 hours | 14.3% |
| 4-6 hours | 3.4% |
| > 6 hours | 2.7% |

The majority of respondents did not try Gather at all. This tallies with the somewhat ambivalent response to whether future conferences should use Gather (see Section 3.7), and I think reinforces a negative about PLDI: that not enough time was specifically earmarked for social interaction using this platform (see "Negatives about virtual PLDI" in Section 2).

**Question:** *How many hours total did you spend in Clowdr video chat rooms (accessed via the /video command from Slack)?*

**Results (517 responses):**

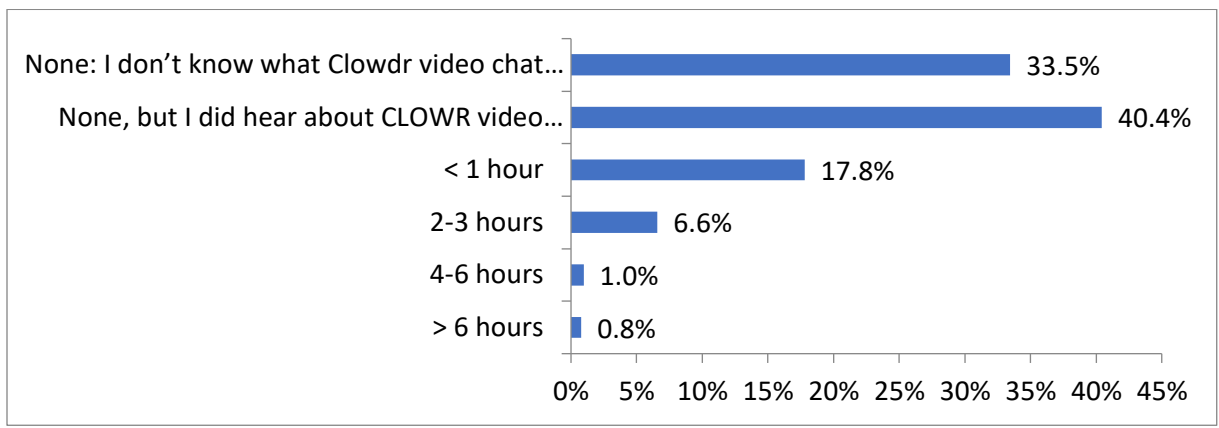

| Category | % |
|---|---|
| None: I don't know what Clowdr video chat… | 33.5% |
| None, but I did hear about CLOWR video… | 40.4% |
| < 1 hour | 17.8% |
| 2-3 hours | 6.6% |
| 4-6 hours | 1.0% |
| > 6 hours | 0.8% |

Only 26% of respondents tried Clowdr, tallying with the ambivalent response to whether future conferences should use Clowdr (see Section 3.7). Again, this reinforces the fact that attendees were not sufficiently incentivised to use this feature; I believe a carefully crafted and well-advertised series of chat rooms would have been useful here.

**Question:** *How did you use Clowdr video chat rooms?* (Respondents were invited to select multiple options if applicable.)

**Results (137 responses)** – absolute counts are shown, not percentages, since respondents could select multiple options:

---

[4] Guy Steele pointed out a bug in the responses offered for this question and the similar question for Clowdr: they do not allow one to specify that one spend 1 hour in Gather.

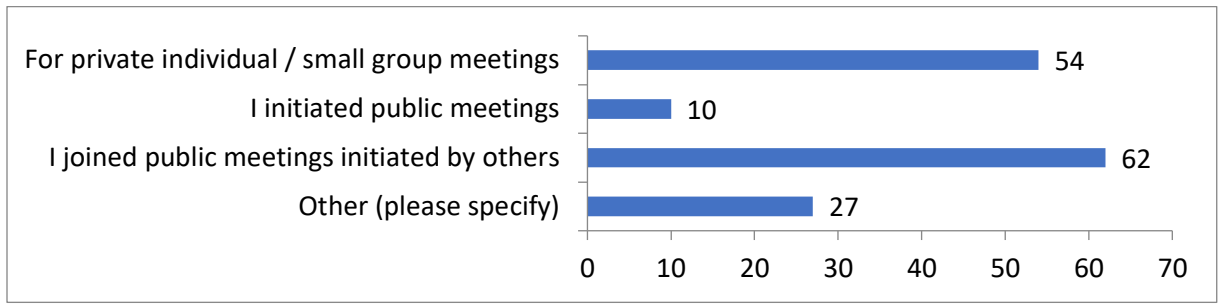

Free text responses to the "Other (please specify)" option highlighted some problems attendees had working out how to use Clowdr but did not specify any additional usage scenarios to the options offered in the question.

## 5. Survey feedback related to virtual vs. physical PLDI

This section is purely based on post-conference survey results, focusing on those questions that relate to comparing virtual and physical conferences, with a focus on PLDI.

**Question:** *In what ways did PLDI 2020 fall short of a physical conference?*

This free-text question received a large number of responses which are presented in Appendix A.5. Unsurprisingly, the dominant answer relates to lack of effective substitutes for physical interaction. There are also many comments relating to the difficulties posted by time zones, and the challenge of attending a virtual conference while in one's regular environment, with normal life going on at the same time. Here are some illustrative examples:

- *"No easy way to have serendipitous interactions in the hallway. I just wasn't interested in pushing an icon around on a screen like in a video game so I didn't. I just didn't spend much time at PLDI so I didn't get nearly as much out of it as if I had travelled to London."*
- *"People not being in the same time zone. Many "social" things in the late evening for Europe, where the conference was expected to be located."*
- *"The physical conference provides a way to stay on the periphery and still make new contacts. Further, the multiple tracks of the physical conference forced me to prioritize the talks I wanted to see in person (this PLDI was intense when I tried to go to all the talks, and I entered brain-dead state a little earlier)."*
- *"Gathering enough momentum to be able to block hours of online PLDI time for attendees. It is very hard to spend anywhere near the same amount of time connected with fellow participants when sitting or standing at a computer desk anyway (or any connected environment, really)."*
- *"It felt like there is a bunch of high-profile people+cohorts that all know each others and enjoy connecting with their buddies, and newcomers/outsiders/introverts may have a hard time finding a way in or feeling to belong in the community/event. This is probably true for physical conferences too, but online makes it much more easy to lurk and to ignore who you*

*don't know (they are just absorbed by the huge contact list and you end up searching for who you know)"*
- *"Colleagues and family did not understand that I was at a conference and thus do not reduce the expectations during this period. It makes it difficult to have time to read the articles and follow the talks. Social interaction and technical discussions are also much more difficult."*

**Question:** *In what ways did PLDI 2020 exceed your expectations?*

This free-text question received a large number of responses which are presented in Appendix A.6. Common positive theme is the Slack-based Q&A model, the ability to catch up on material after the event at 1.5x speed, and the clarity associated with pre-recorded talks. Here are some illustrative examples:

- *"I think combining a core of proven technology (Youtube, Slack, and to some extent Zoom) with younger experimental services like gather and clowdr was bold and effective - certainly much better than another virtual conference I attended. The organisers clearly put a lot of work and passion into pulling off the conference in difficult circumstances which I think was key to making it work this well."*
- *"The Zoom coffee hour, mentoring channel, and AMAs really exceeded my expectations. The most significant value of attending PLDI seemed to be the encouraged interaction with other researchers in the community, since the majority of papers and talks now are published after the conference anyway (not to say these aren't valuable, but the social interaction is hard to obtain any other way), and all three of these provided a fantastic and welcoming platform to meet and learn from both other students and senior members of the community."*
- *"The talks are extremely clear. Asking a question is now very easy as well as following it up in the slack thread is very very helpful/productive."*
- *"The Slack seemed more bubbling and engaged than the virtual ASPLOS Slack. AMA sessions were great and super interesting to watch as a junior researcher. Moderated Q & A seemed to improve question quality, fewer "it's more of a comment than a question" type responses. Mostly single-track made it easier to watch all the talks I wanted to see live. Video abstracts were awesome, great to be able to get high-level overviews of lots of talks before seeing the whole thing."*
- *"Probably better attendance than a physical conference, and probably slightly more polished talks. Talk Q/A also worked well, maybe better than a physical conference, since overflow questions could easily be answered electronically. Easier to attend the subset of talks I'm really interested in."*
- *"The asynchronous question mechanism was amazing, I got a lot more discussion on my paper than in many regular conferences. It was also great to be able to rewind/pause the talks on youtube, and even to cook dinner while watching talks, etc. The slack workspace was very active, and the ask-me-anything sessions were particularly valuable."*

**Question:** How does the amount of time that you spent watching the virtual conference compare with what you might have spent, had the conference been physical, occurring under normal (non-COVID) circumstances:

**Results (463 responses):**

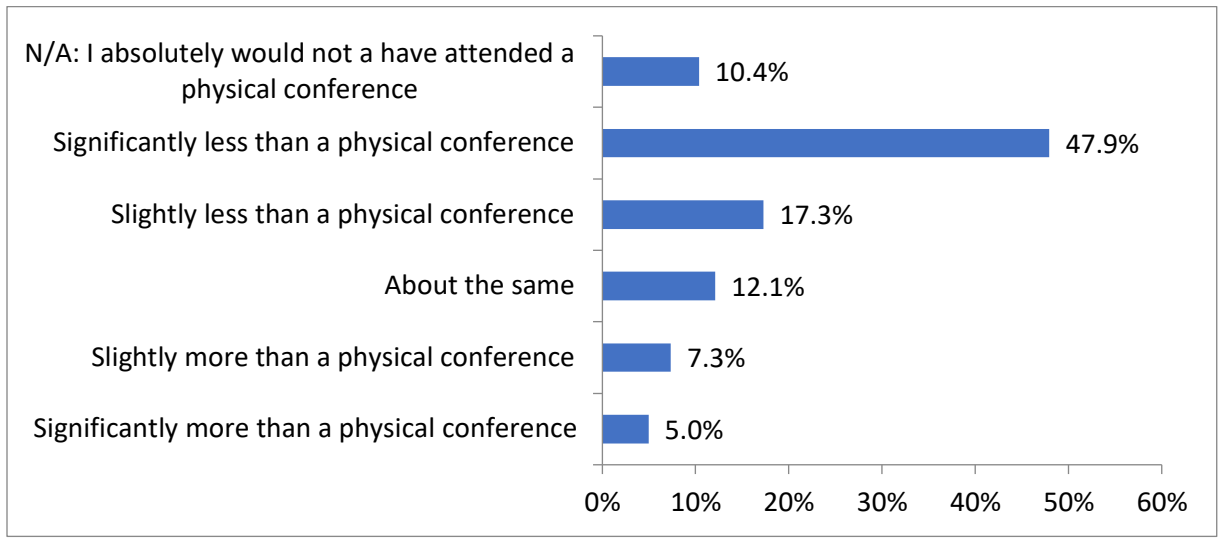

These results support the hypothesis that while the virtual conference may have reached a significantly larger audience than a physical conference would have, the depth of engagement per participant is lower.

**Question:** *Please indicate how strongly you agree or disagree with the following statement: "I enjoyed attending PLDI 2020 more than previous PLDIs that I have attended"*

**Results (480 responses):**

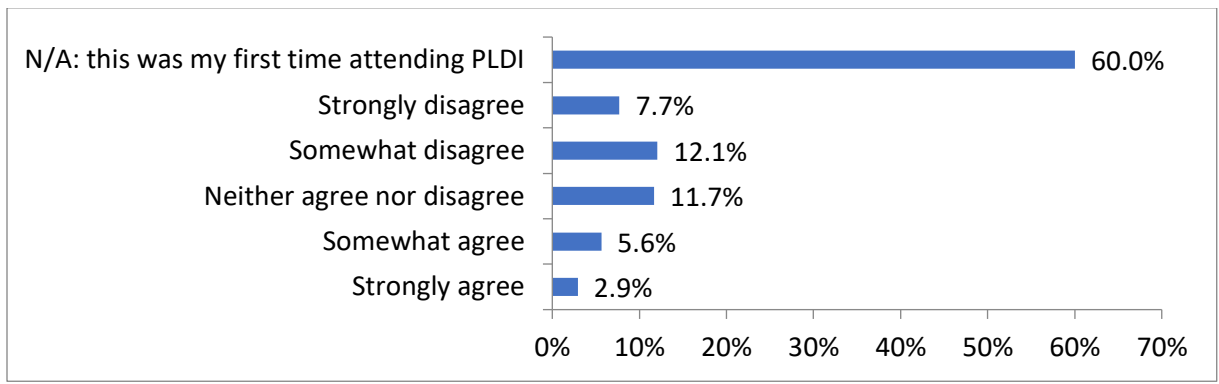

I had been expecting more disagreement here from respondents who had attended PLDI before. The large number of first-time PLDI attendees again demonstrates the extended reach of the virtual event.

**Question:** *Compared to a physical conference, we did not have a hallway for informal conversations, social interactions, and mentoring. Instead we had Gather, Clowdr, and Slack, including channels for discussion and mentoring. For the purpose of social interaction, how did these virtual options compare to a physical conference?*

**Results (484 responses):**

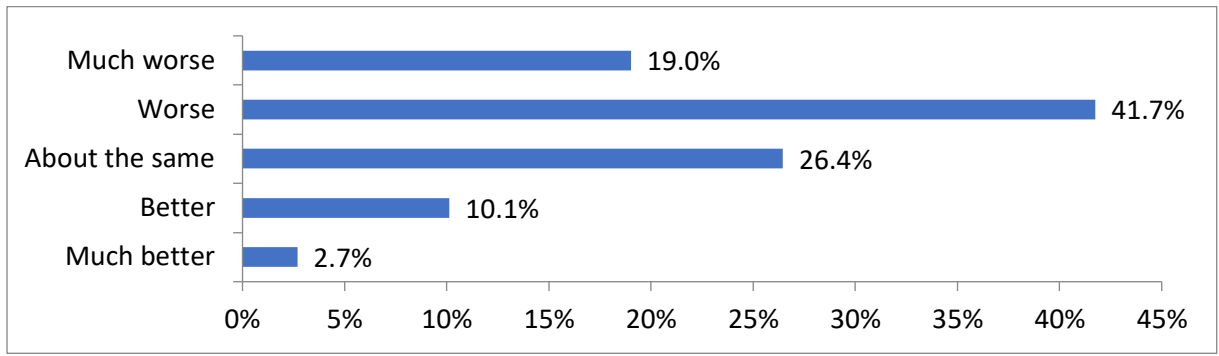

It is unsurprising that the majority of participants found the options for interaction compared unfavourably to the interaction that is possible at a physical conference.

**Question:** *Please indicate how strongly you agree or disagree with the following statement: "The quality of online video presentations at PLDI was at least as high as the quality of physical conference talks"*

**Results, shown with respect to:**
- Respondents who claim this was their first time attending PLDI (292 responses)
- Respondents who claim to have attended a previous PLDI (177 responses)
- All respondents (469 responses)

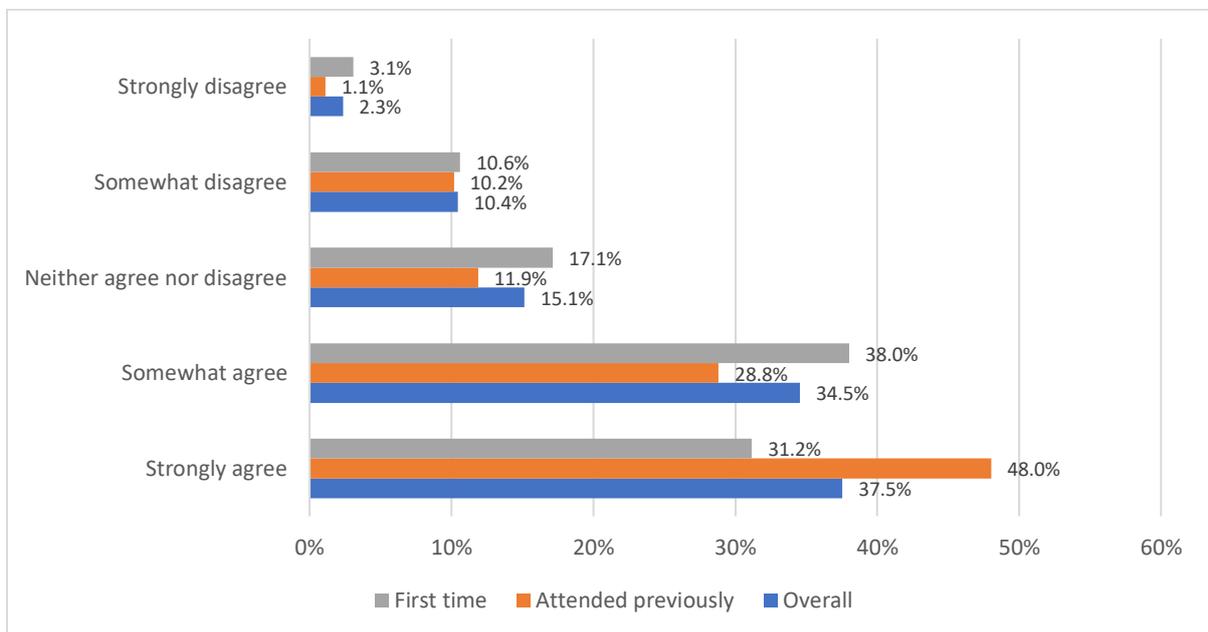

**Question:** *Please indicate how strongly you agree or disagree with the following statement: "If the COVID-19 pandemic had not happened and PLDI 2020 had been held as normal, I would have attended the meeting physically"*

**Results, shown with respect to:**
- Respondents who claim this was their first time attending PLDI (320 responses)
- Respondents who claim to have attended a previous PLDI (181 responses)
- All respondents (501 responses)

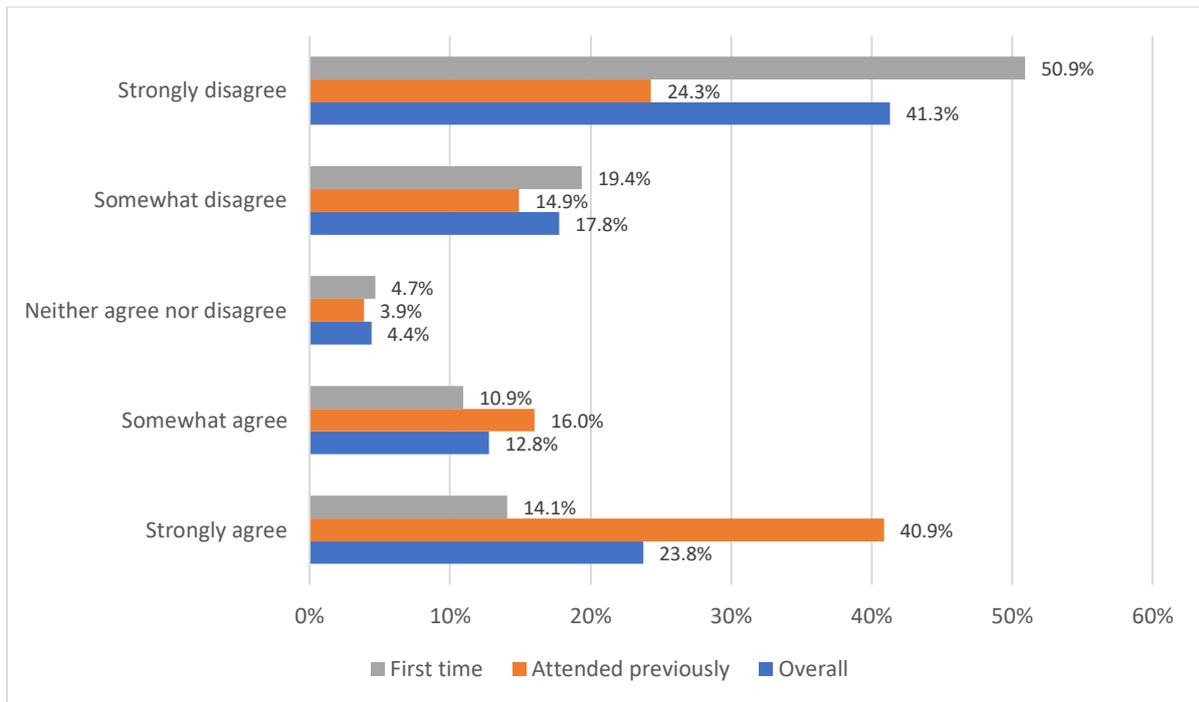

The results show that a significant number of respondents would not have attended PLDI if it had been held physically under normal circumstances, but that most previous attendees of PLDI would have been likely to have attended physically.

**Question:** *If you would have been unlikely to attend, which factors would have influenced this?* (Respondents were invited to select multiple options if applicable.)

**Results (384 responses)** – absolute counts are shown, not percentages, since respondents could select multiple options:

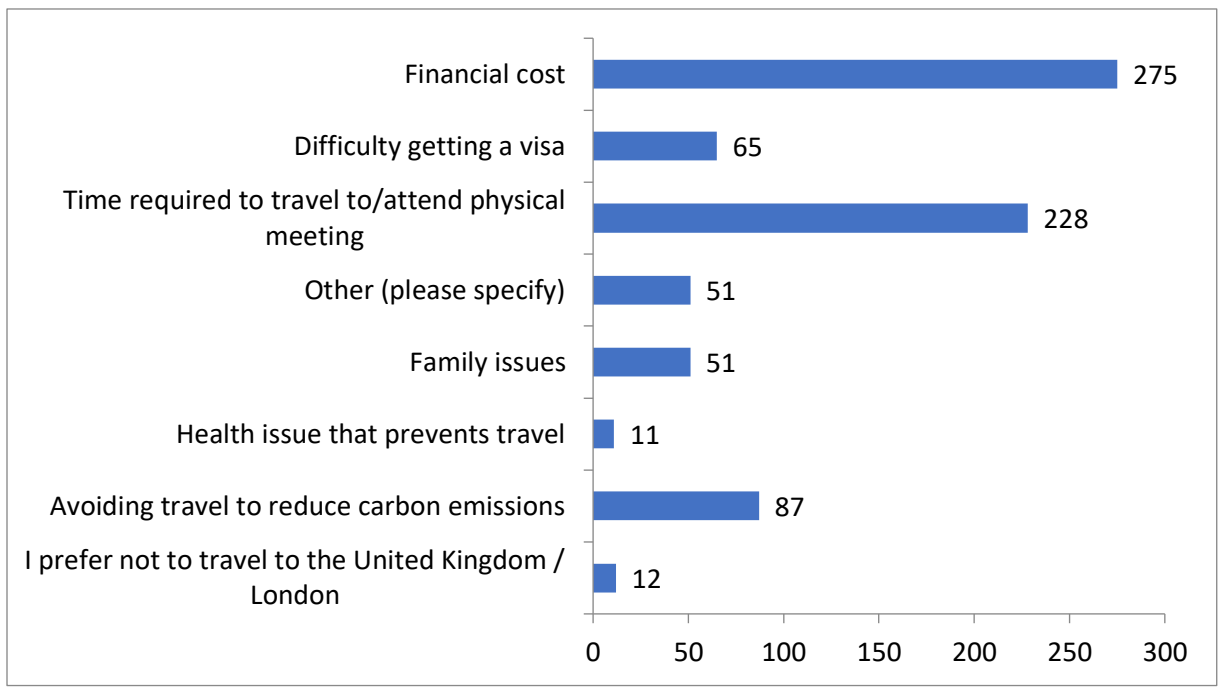

A number of additional factors were provided as free-text responses and are presented in Appendix A.7. The fact that financial cost is the dominating factor here illustrates what in my view is the biggest advantage of virtual conferences: that they are more inclusive.

**Question:** *Please indicate how strongly you agree or disagree with the following statement: "Based on my experience of virtual PLDI 2020, I would attend a future virtual PLDI"*

**Results, shown with respect to:**
- Respondents who claim this was their first time attending PLDI (315 responses)
- Respondents who claim to have attended a previous PLDI (179 responses)
- All respondents (494 responses)

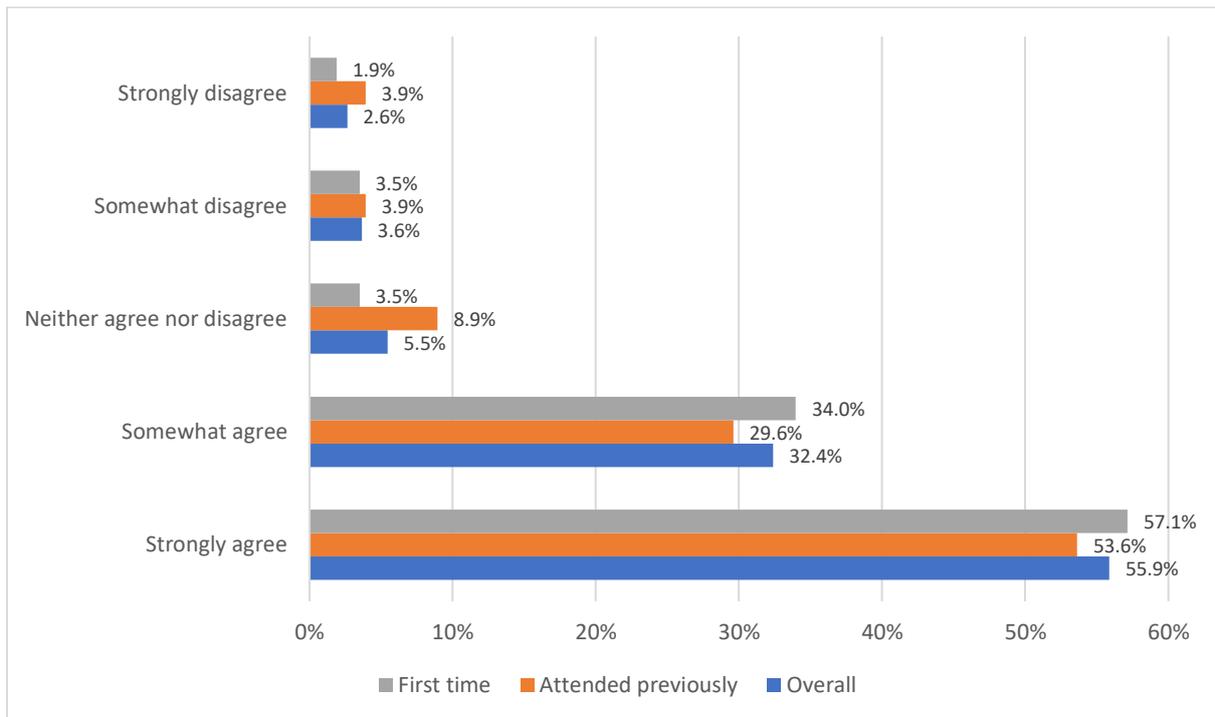

While the survey results overall show a mixed appetite for virtual conferences in general, these results show that the vast majority of respondents would attend another virtual PLDI, regardless of whether PLDI 2020 was their first time attending the conference.

**Question:** *If the COVID-19 pandemic ends, how keen would you be for future PLDI conferences to be virtual?*

**Results, shown with respect to:**
- Respondents who claim this was their first time attending PLDI (303 responses)
- Respondents who claim to have attended a previous PLDI (176 responses)
- All respondents (479 responses)

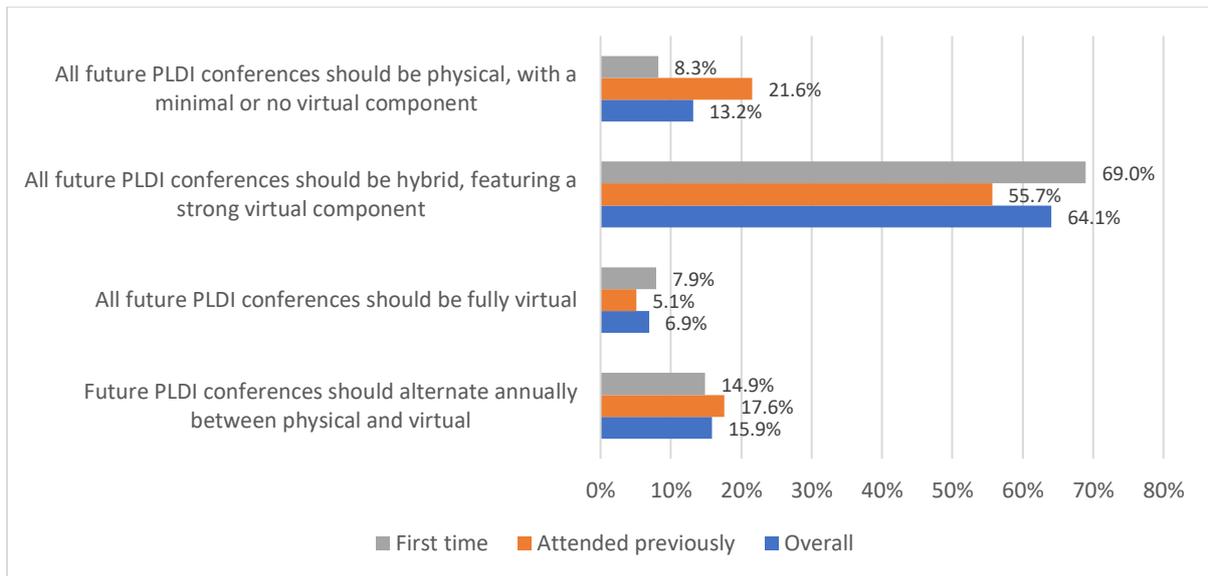

The results make it clear that there is little appetite from the respondents for virtual conferences as the default format for PLDI, but a lot of enthusiasm for hybrid events that combine a physical meeting with a significant virtual component. My concern about this is that I believe a hybrid conference may be much harder to run successfully than a fully physical or fully virtual conference, due to the disparity between attendees who are / are not physically present.

**Question:** *What is the maximum registration fee you estimate that you, or your employer or institution, would have been willing to pay in order for you to attend virtual PLDI 2020, if you had known in advance what the conference would offer?*

In the results, "Nothing" is short for "Nothing: I would only have attended if there was no fee", and "Typical SIGPLAN" is short for "I think it would be reasonable to charge the fees associated with a typical SIGPLAN conference".

**Results, shown with respect to:**
- Respondents who claim this was their first time attending PLDI (308 responses)
- Respondents who claim to have attended a previous PLDI (180 responses)
- All respondents (488 responses)

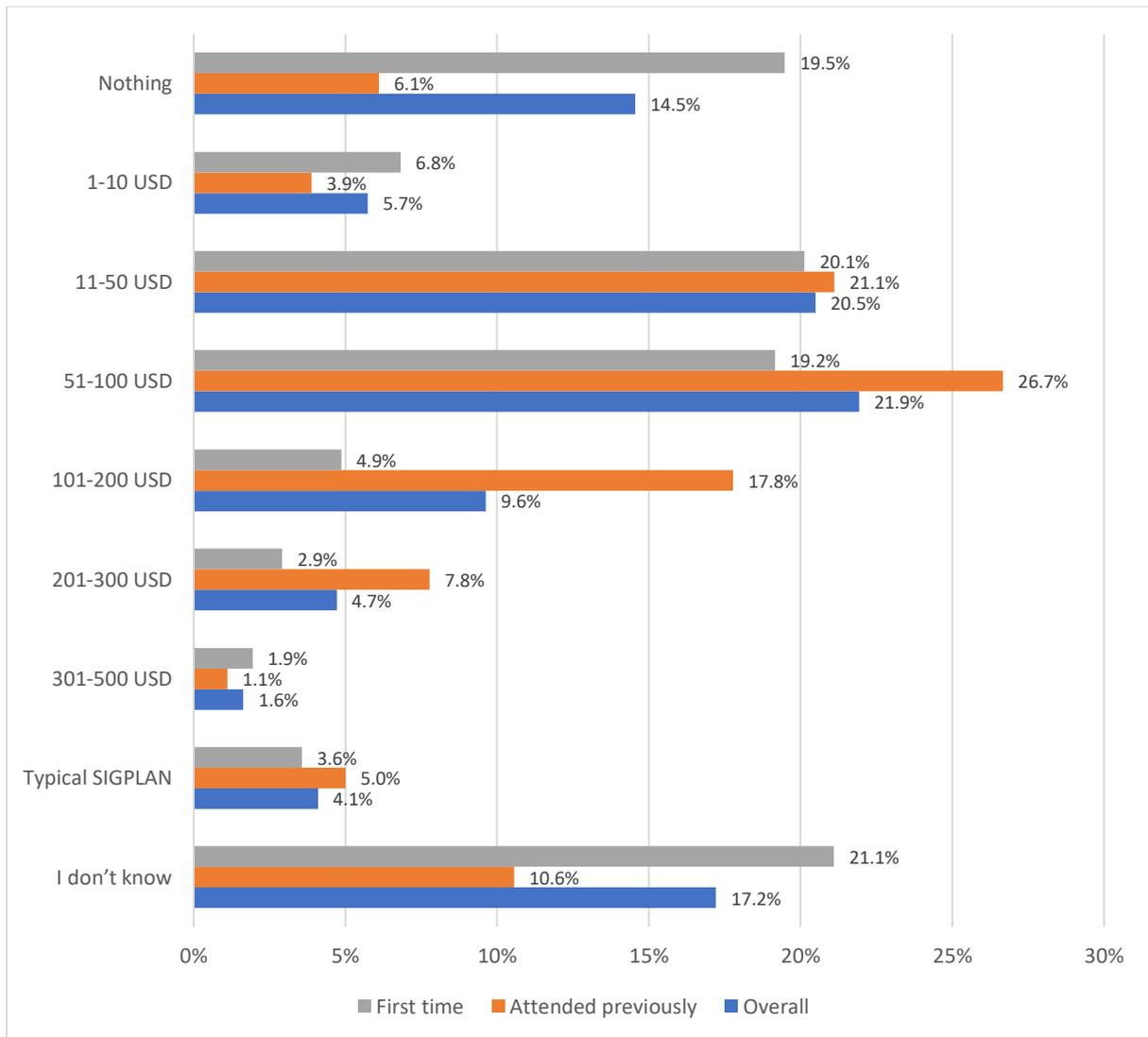

The results suggest that charging up to 50 USD for registration would deter a substantial, but perhaps not insurmountable, number of attendees. Most attendees who would be unwilling to pay to attend PLDI were first-time attendees.

## 6. Diversity and inclusion

I present demographic data obtained from YouTube related to viewers of the conference material (Section 6.1), and demographic data obtained during conference registration and via the post-conference survey (Section 6.2). I present the data as is and have refrained from discussing it in depth; I believe it is valuable to make this data available to the community to inform initiatives that aim to increase diversity, yet analysis of demographic data is not within my area of expertise. I then give details of post-conference survey results related to time-zones (Section 6.3).

### 6.1. Demographic data from YouTube

Recall from Section 4.2 that, during the week of PLDI 2020, the ACM SIGPLAN YouTube channel saw a great deal of activity, principally due to the live streaming of PLDI events.

YouTube provides analytics data on the age, gender and country associated with viewers. According to the "Audience" section of this page on YouTube Studio analytics basics, age and gender *"is based on signed in viewers across all devices"*, while country *"is based on IP address"*. The page also states the following caveat, which should be borne in mind when interpreting these results: *"Note that you may only see demographics data for a subset of your viewers. The data may not represent the overall composition of your traffic."*

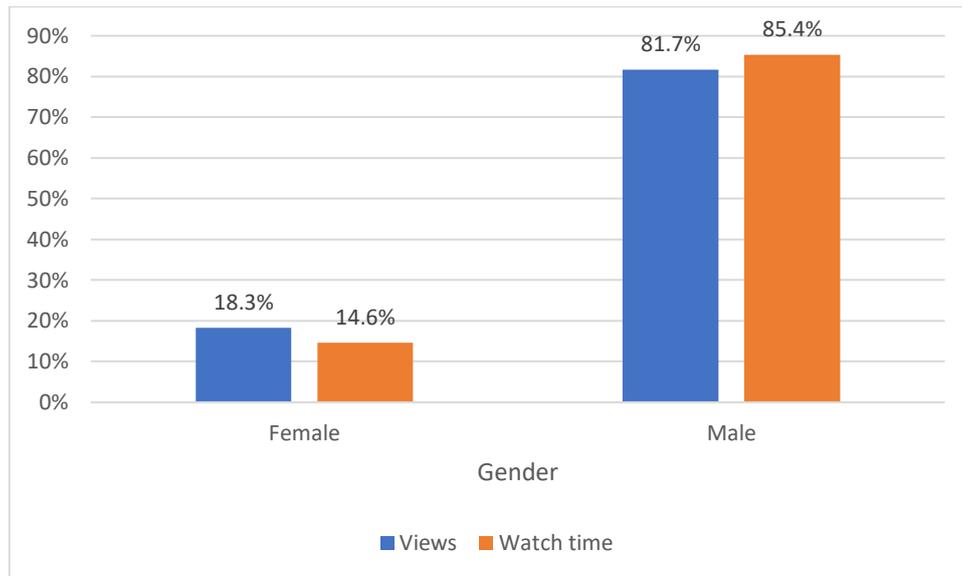

*Figure 3 Percentage of total views and total watch time for the ACM SIGPLAN YouTube channel during the period 15-21 June 2020 – the week of PLDI – grouped according to gender. Data provided by YouTube analytics.*

Figure 3 shows a percentage breakdown of total views and total view time according to gender. The percentages associated with Female and Male in Figure 3 do correspond roughly to the percentages associated with identifying as Female and identifying as Male in the data on gender identity collected during conference registration and via the post-conference survey (see Section 6.2). However, the data cannot be directly compared because gender and gender identity are not the same thing.

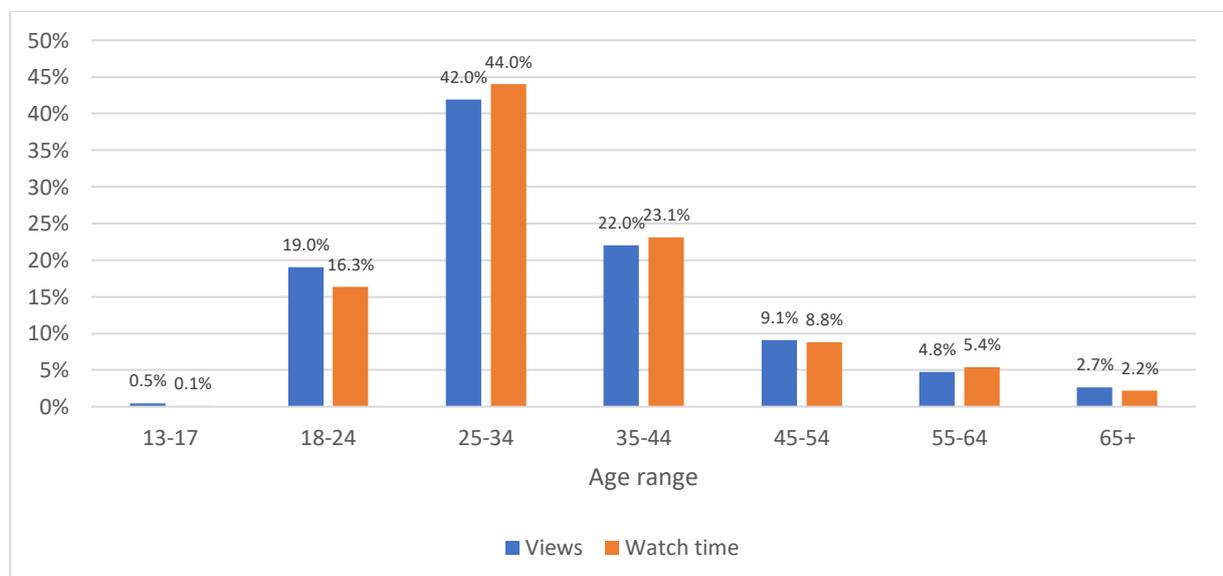

*Figure 4 Percentage of total views and total watch time for the ACM SIGPLAN YouTube channel during the period 15-21 June 2020 – the week of PLDI – grouped according to age. Data provided by YouTube analytics.*

Figure 4 shows a percentage breakdown of total views and total view time according to age range. The shape of the figure broadly corresponds to the shape of the figure associated with age-related demographic information gathered during registration and via the post-conference survey, discussed in Section 6.2.

Figure 5 shows percentage view time individually for the 20 countries with the highest view time, and in aggregate for the remaining countries. Figure 6 shows percentage view time grouped by geographical continent.

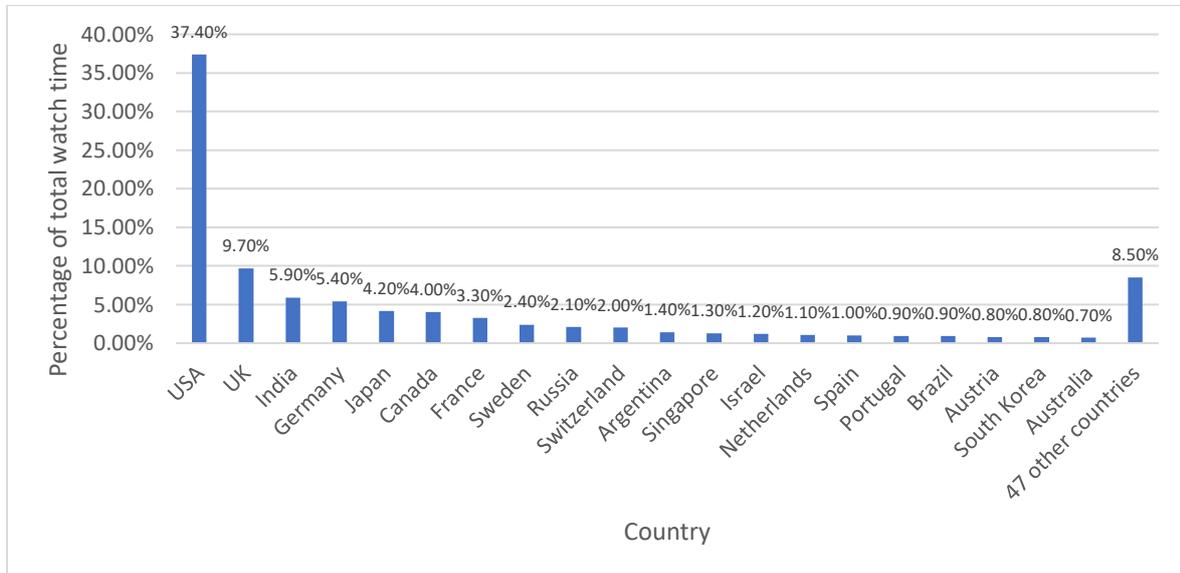

*Figure 5 Percentage of total watch time for the ACM SIGPLAN YouTube channel during 15-21 June – the week of PLDI – grouped by country. Results for the 20 countries with the highest watch time are shown individually. Results for the remaining countries shown in the last bar of the chart.*

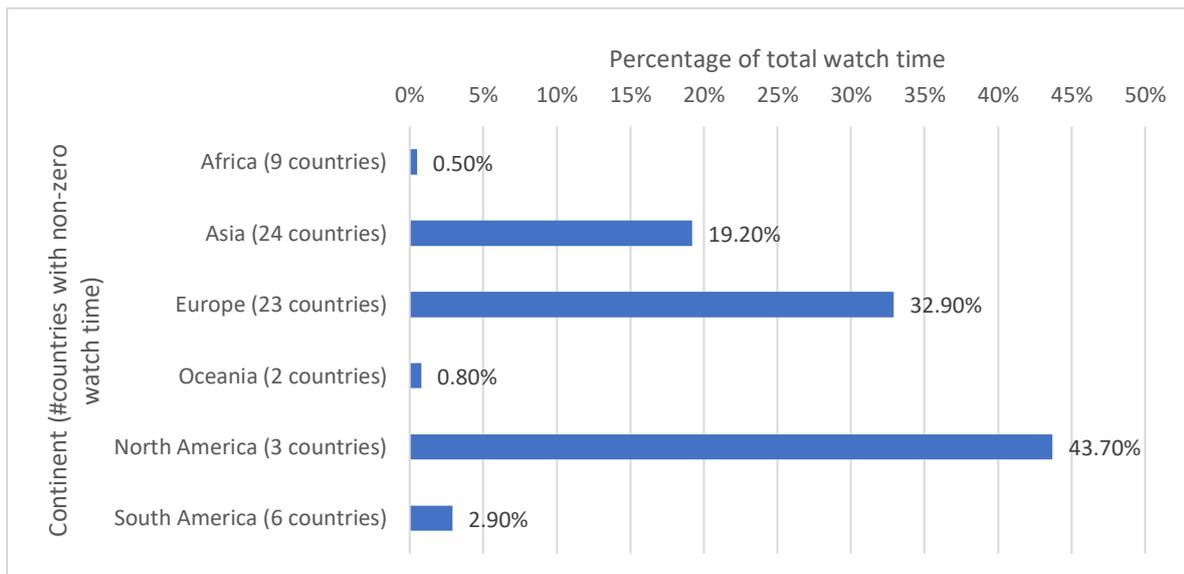

*Figure 6 Percentage of total watch time for the ACM SIGPLAN YouTube channel during 15-21 June – the week of PLDI – grouped by geographical continent. The number of distinct countries in each geographical continent for which non-zero watch time is shown in parentheses.*

## 6.2. Demographic data collected at registration and via the post-conference survey

The PLDI 2020 registration form featured a set of demographic-related questions that were used during POPL 2020 registration, and that I presume evolved via their use during registration for other ACM SIGPLAN conferences. I used the POPL 2020 questions except that I added more options to the question about ethnicity. I repeated these questions verbatim in the post-conference questionnaire to allow comparing registration results with post-conference results.

I present results for registration-gathered data over: the whole data set (called "Registered" in the charts below); the data set restricted to those registrants who subsequently joined the conference's Slack workspace (called "On Slack" below); and the data set restricted to those registrants who posted at least one message on the Slack workspace (called "Active on Slack" below). I felt this was important because the very large number of overall registrants might give an inaccurate impression of diversity at PLDI given that we do not know whether a large proportion of those registrants – the ones who did not join Slack – engaged further.

Results for respondents of the post-conference survey are labelled "Post-conference" in the charts below.

**Question:** *To which gender identity do you most identify?*

**Results, with numbers of responses as follows:**
- Registered: 3817 responses
- On Slack: 2431 responses
- Active on Slack: 901 responses
- Post-conference: 562 responses

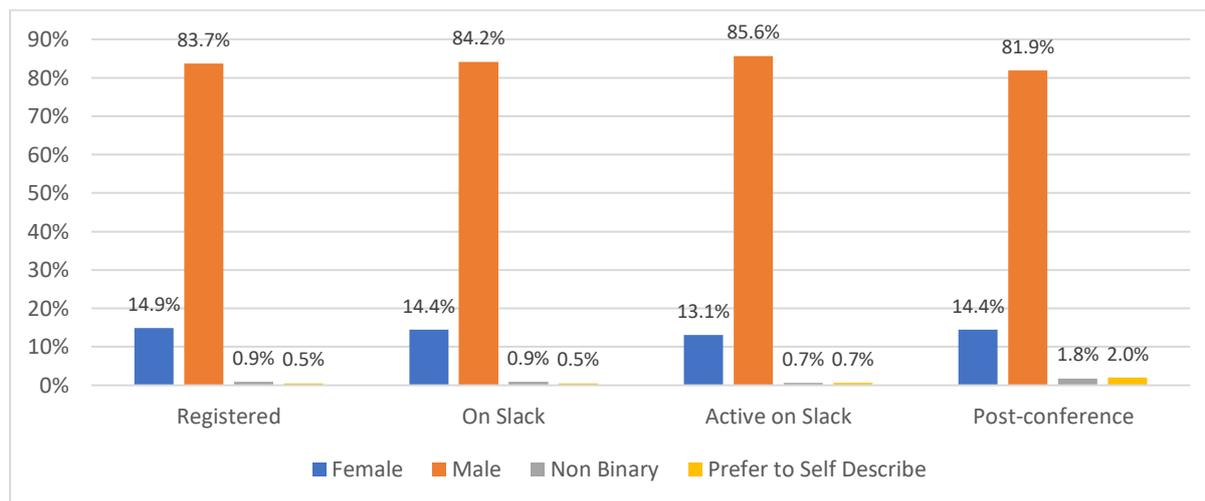

Free-text responses associated with "I prefer to self-describe" from the post-conference survey are provided in Appendix A.8.

I thought it would also be interesting to view these results restricted to those respondents who answered "Yes" to the question "*Are you an author of any material that was presented at PLDI or one of the co-located meetings?*"

**Results, restricted to authors, with numbers of responses as follows:**
- Registered: 356 responses
- On Slack: 279 responses
- Active on Slack: 183 responses

- Post-conference: 110 responses

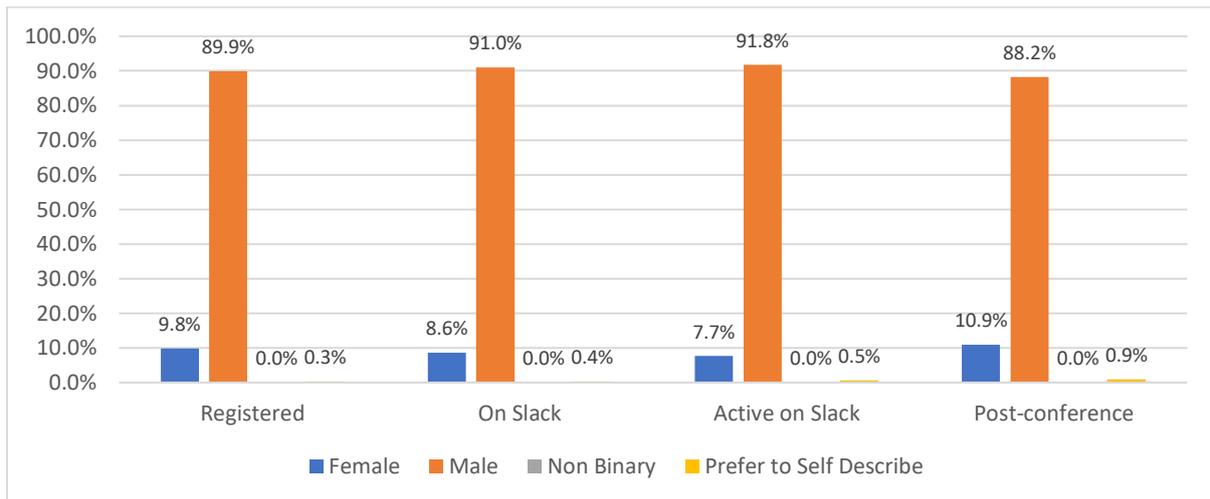

Notice that the female:male ratio decreases significantly when we restrict results to those registrants/respondents who authored material.

**Question:** *What is your age group?*

**Results, with numbers of responses as follows:**
- Registered: 3847 responses
- On Slack: 2447 responses
- Active on Slack: 913 responses
- Post-conference: 564 responses

Data labels above bars are omitted as they were unreadable.

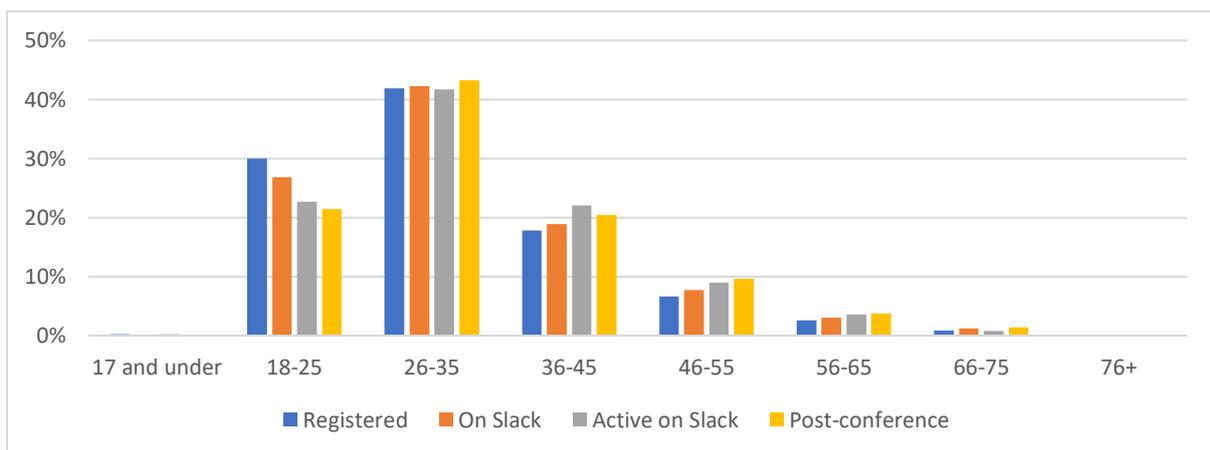

**Question:** *What is your ethnic group?* (Respondents were invited to select multiple options if applicable.)

**Results, with numbers of responses as follows:**
- Registered: 3648 responses
- On Slack: 2315 responses
- Active on Slack: 855 responses
- Post-conference: 568 responses

Because respondents could select more than one option, the percentages associated with each data set sum to slightly more than 100%. I still felt that it made sense to present percentages rather than raw numbers because I wanted to compare the "Registered", "On Slack", "Active on Slack" and "Post-conference" data sets, which have different sizes. Data labels next to bars are omitted as they were unreadable.

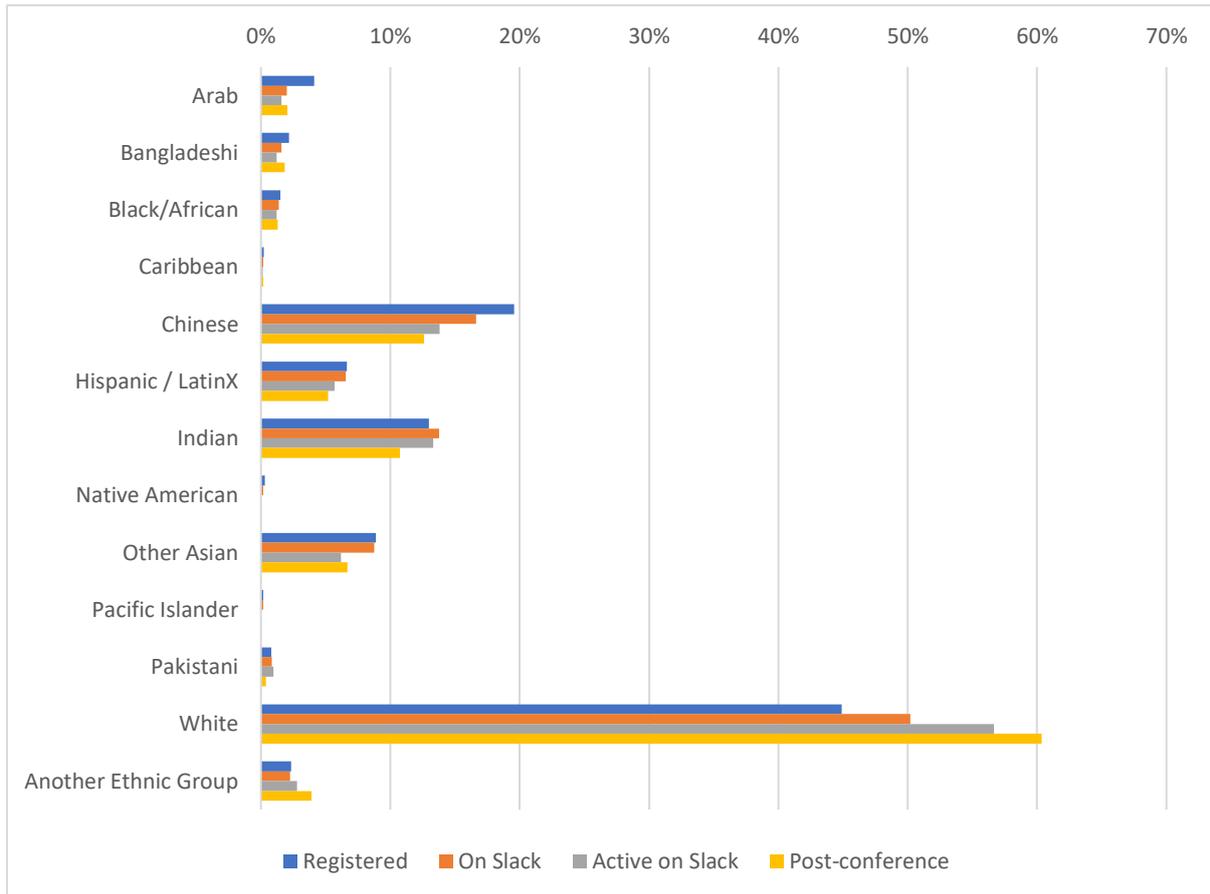

Free-text responses associated with "Another Ethnic Group" from the post-conference survey are provided in Appendix A.9.

**Question:** *Do you have a disability or special need that impacts your access to ACM conferences, special interest groups, publications, or digital resources?*

**Results, with numbers of responses as follows:**
- Registered: 3675 responses
- On Slack: 2367 responses
- Active on Slack: 889 responses
- Post-conference: 559 responses

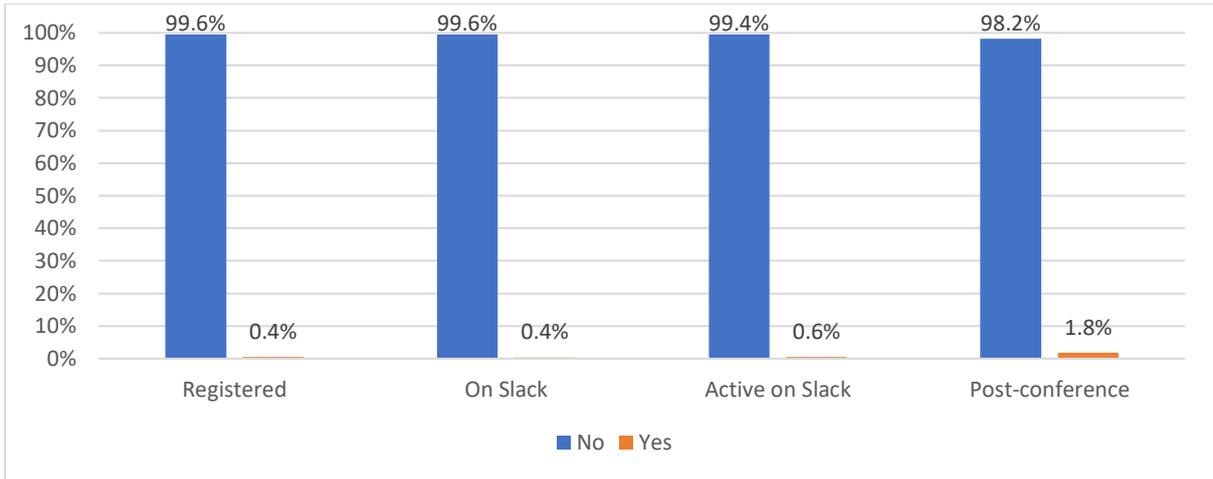

Free-text responses associated with "Yes" from the post-conference survey are provided in Appendix A.10.

**Question:** *Do you identify as being a member of an underrepresented group?*

**Results, with numbers of responses as follows:**
- Registered: 3288 responses
- On Slack: 2117 responses
- Active on Slack: 812 responses
- Post-conference: 554 responses

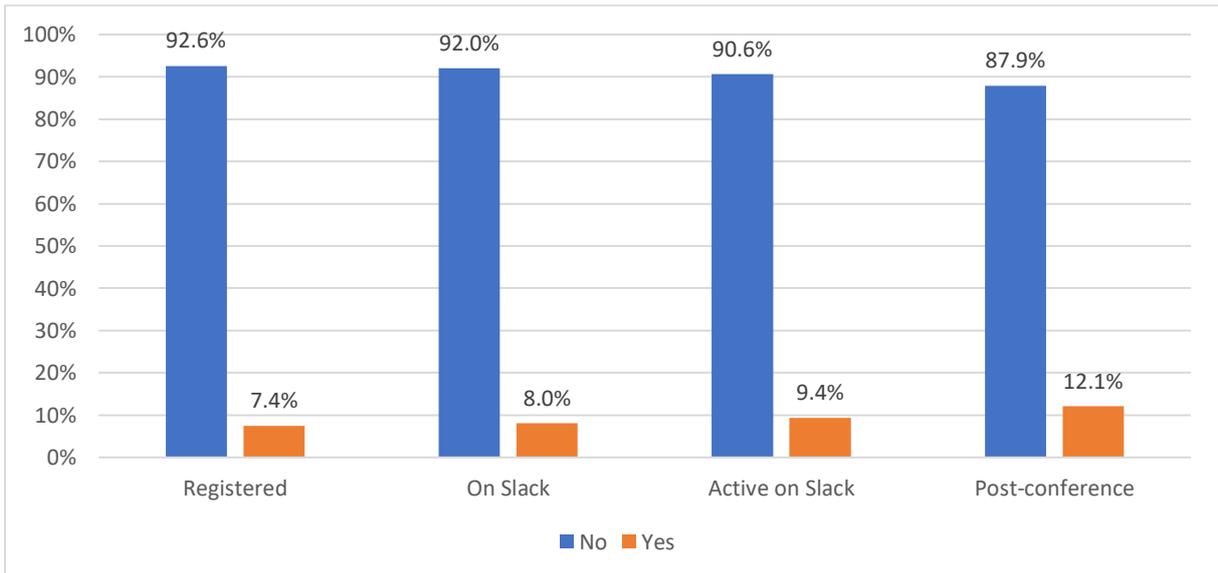

Free-text responses associated with "Yes" from the post-conference survey are provided in Appendix A.11.

**Question:** *What is your current role*

**Results, with numbers of responses as follows:**
- Registered: 3887 responses
- On Slack: 2473 responses
- Active on Slack: 925 responses
- Post-conference: 564 responses

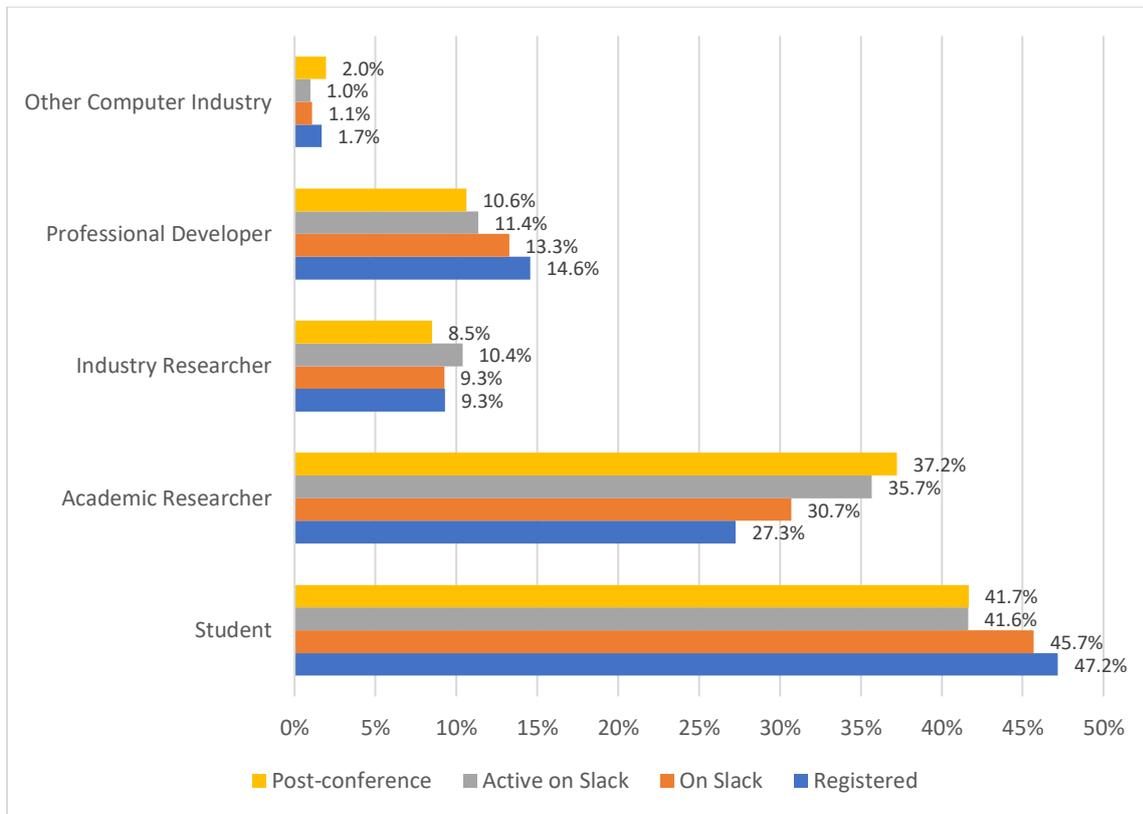

**Question:** *How many PLDIs had you previously attended?*

**Results, with numbers of responses as follows:**
- Registered: 3875 responses
- On Slack: 2480 responses
- Active on Slack: 926 responses
- Post-conference: 566 responses

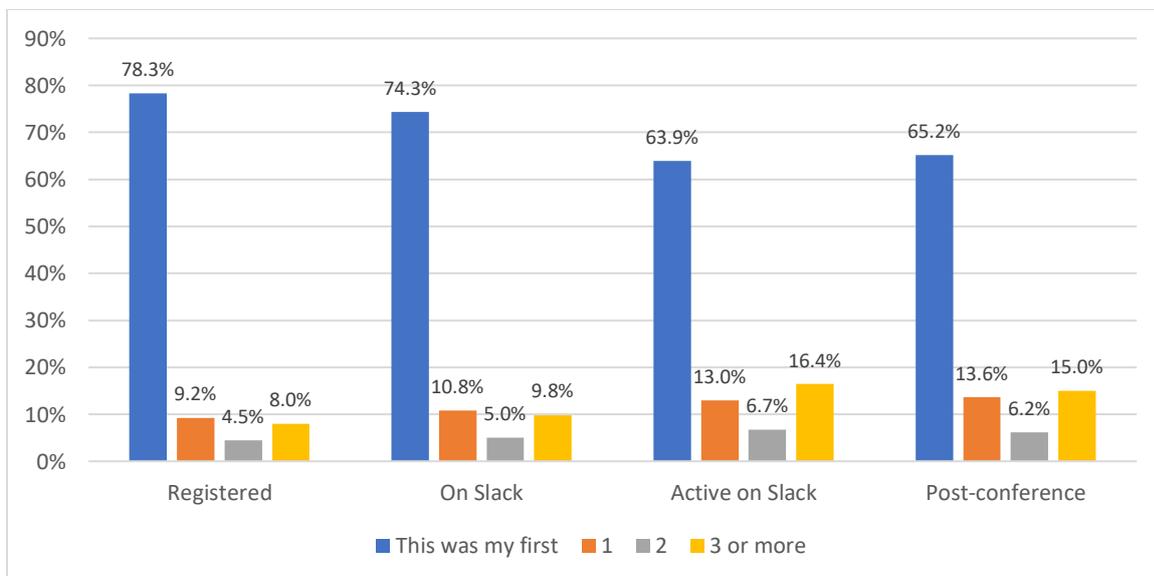

**Question:** *Are you an author of any material that was presented at PLDI or one of the co-located meetings?*

**Results, with numbers of responses as follows:**

- Registered: 4297 responses
- On Slack: 2688 responses
- Active on Slack: 1001 responses
- Post-conference: 563 responses

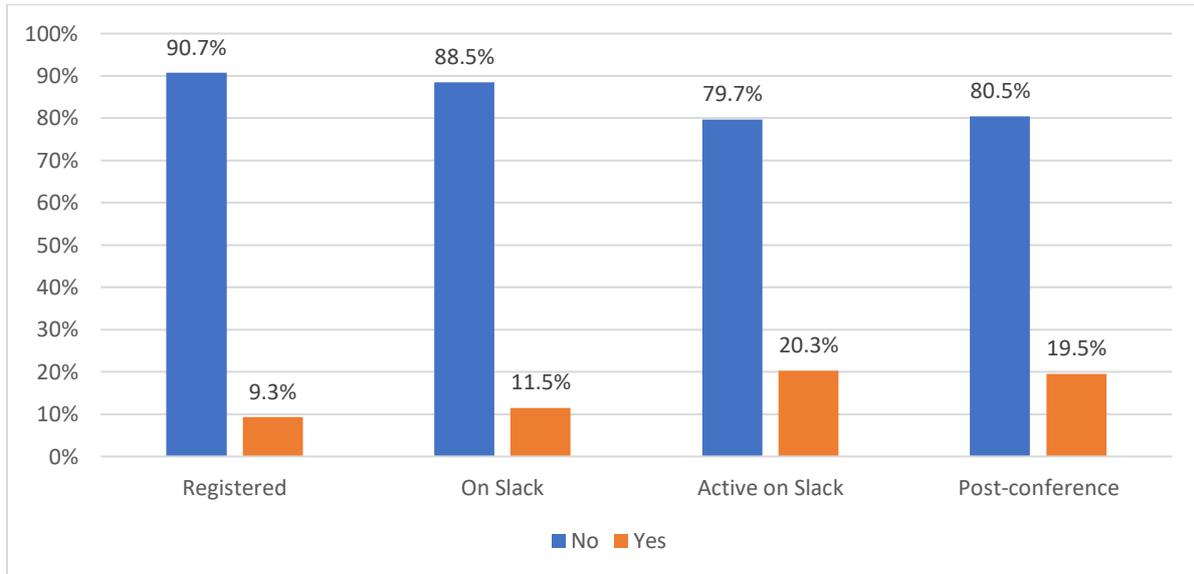

**Country of residence at registration time / during the conference.** During registration, registrants were asked for their country of residence. In the post-conference questionnaire, respondents were asked: *Which country were you in during the conference?*

While these are not identical questions, I think it is reasonable to present the response data for them side by side. Here are the results grouped by geographical continent, with numbers of responses as follows:

- Registered: 4293 responses
- On Slack: 2684 responses
- Active on Slack: 999 responses
- Post-conference: 561 responses

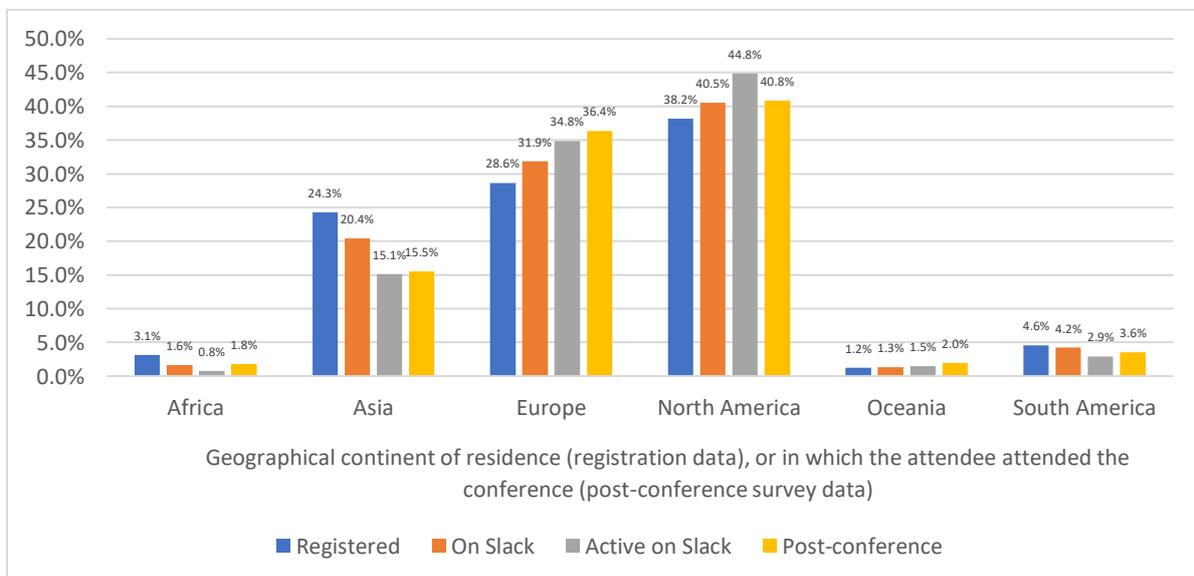

The drop-off in numbers from Asia when comparing numbers at registration to those who signed up for / were active on Slack, and who filled out the post-conference questionnaire, can mostly be attributed to a drop-off in numbers from China.

### 6.3. Time zones

The survey included questions to understand the distribution of time zones inhabited by attendees, attendees' attitudes towards taking time zones into account to make virtual conferences more inclusive, and attendees' views on a number of potential approaches.

**Question:** *To the nearest hour, which time zone were you in during the conference?*

**From post-survey questionnaire (561 responses):**

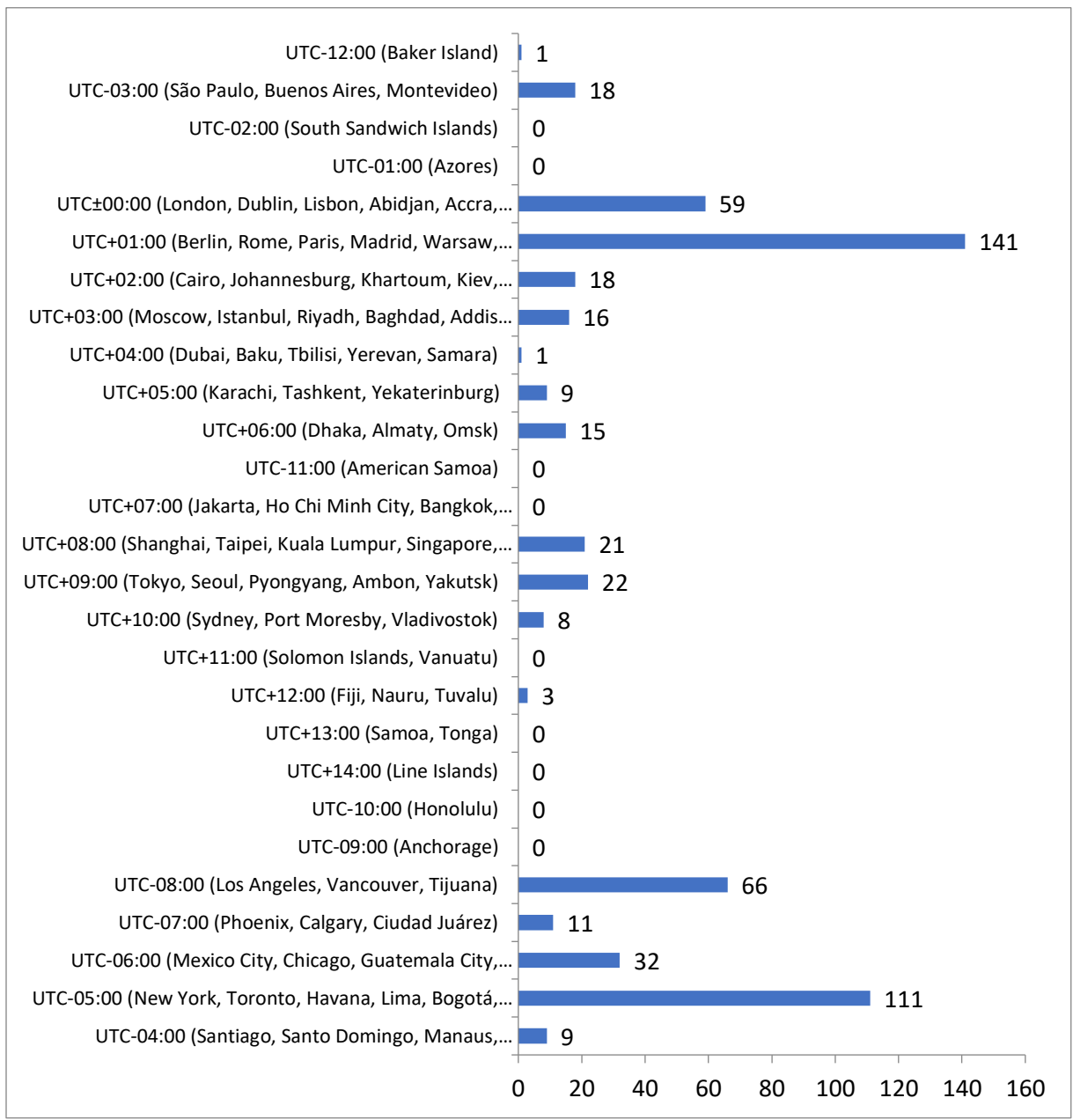

| Time zone | Responses |
|---|---|
| UTC-12:00 (Baker Island) | 1 |
| UTC-03:00 (São Paulo, Buenos Aires, Montevideo) | 18 |
| UTC-02:00 (South Sandwich Islands) | 0 |
| UTC-01:00 (Azores) | 0 |
| UTC±00:00 (London, Dublin, Lisbon, Abidjan, Accra,… | 59 |
| UTC+01:00 (Berlin, Rome, Paris, Madrid, Warsaw,… | 141 |
| UTC+02:00 (Cairo, Johannesburg, Khartoum, Kiev,… | 18 |
| UTC+03:00 (Moscow, Istanbul, Riyadh, Baghdad, Addis… | 16 |
| UTC+04:00 (Dubai, Baku, Tbilisi, Yerevan, Samara) | 1 |
| UTC+05:00 (Karachi, Tashkent, Yekaterinburg) | 9 |
| UTC+06:00 (Dhaka, Almaty, Omsk) | 15 |
| UTC-11:00 (American Samoa) | 0 |
| UTC+07:00 (Jakarta, Ho Chi Minh City, Bangkok,… | 0 |
| UTC+08:00 (Shanghai, Taipei, Kuala Lumpur, Singapore,… | 21 |
| UTC+09:00 (Tokyo, Seoul, Pyongyang, Ambon, Yakutsk) | 22 |
| UTC+10:00 (Sydney, Port Moresby, Vladivostok) | 8 |
| UTC+11:00 (Solomon Islands, Vanuatu) | 0 |
| UTC+12:00 (Fiji, Nauru, Tuvalu) | 3 |
| UTC+13:00 (Samoa, Tonga) | 0 |
| UTC+14:00 (Line Islands) | 0 |
| UTC-10:00 (Honolulu) | 0 |
| UTC-09:00 (Anchorage) | 0 |
| UTC-08:00 (Los Angeles, Vancouver, Tijuana) | 66 |
| UTC-07:00 (Phoenix, Calgary, Ciudad Juárez) | 11 |
| UTC-06:00 (Mexico City, Chicago, Guatemala City,… | 32 |
| UTC-05:00 (New York, Toronto, Havana, Lima, Bogotá,… | 111 |
| UTC-04:00 (Santiago, Santo Domingo, Manaus,… | 9 |

**Question:** *Time zones mean that at any given time one third of the world will be unavailable. If virtual PLDI becomes a regular thing, should PLDI:*

**Results (475 responses):**

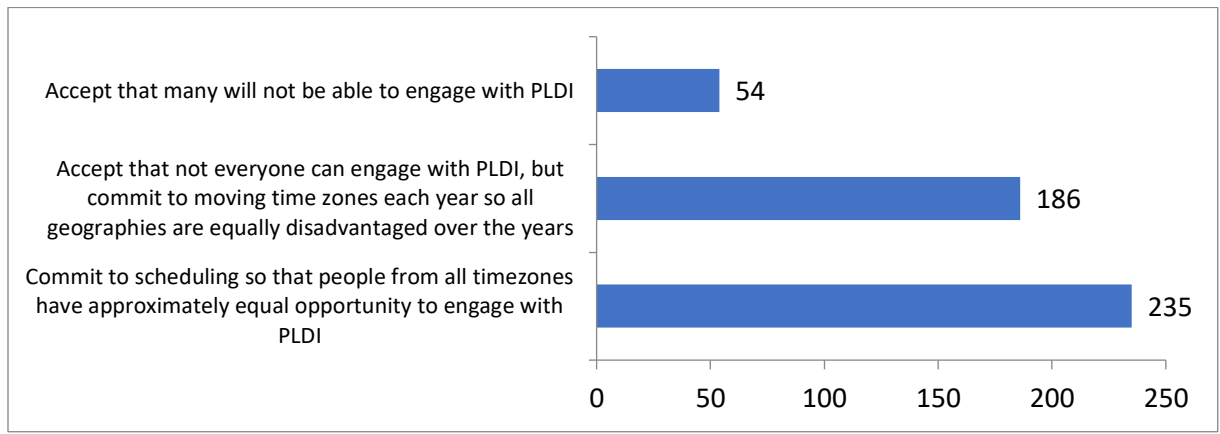

**Question:** *Ways to mitigate the effects of time zones include spreading out the schedule and mirroring key events such as Q&A 12 hours later (so that regardless of time zone, one or other edition of the event will be accessible at a reasonable time).  Assuming roughly 8 hours of content per day, which of the following approaches to handling time zones would you prefer organisers to take for future virtual conferences of a similar size to PLDI?  Respondents were allowed to select multiple options and were asked to select the options they thought were strongly preferable.*

**Results (442 responses):**

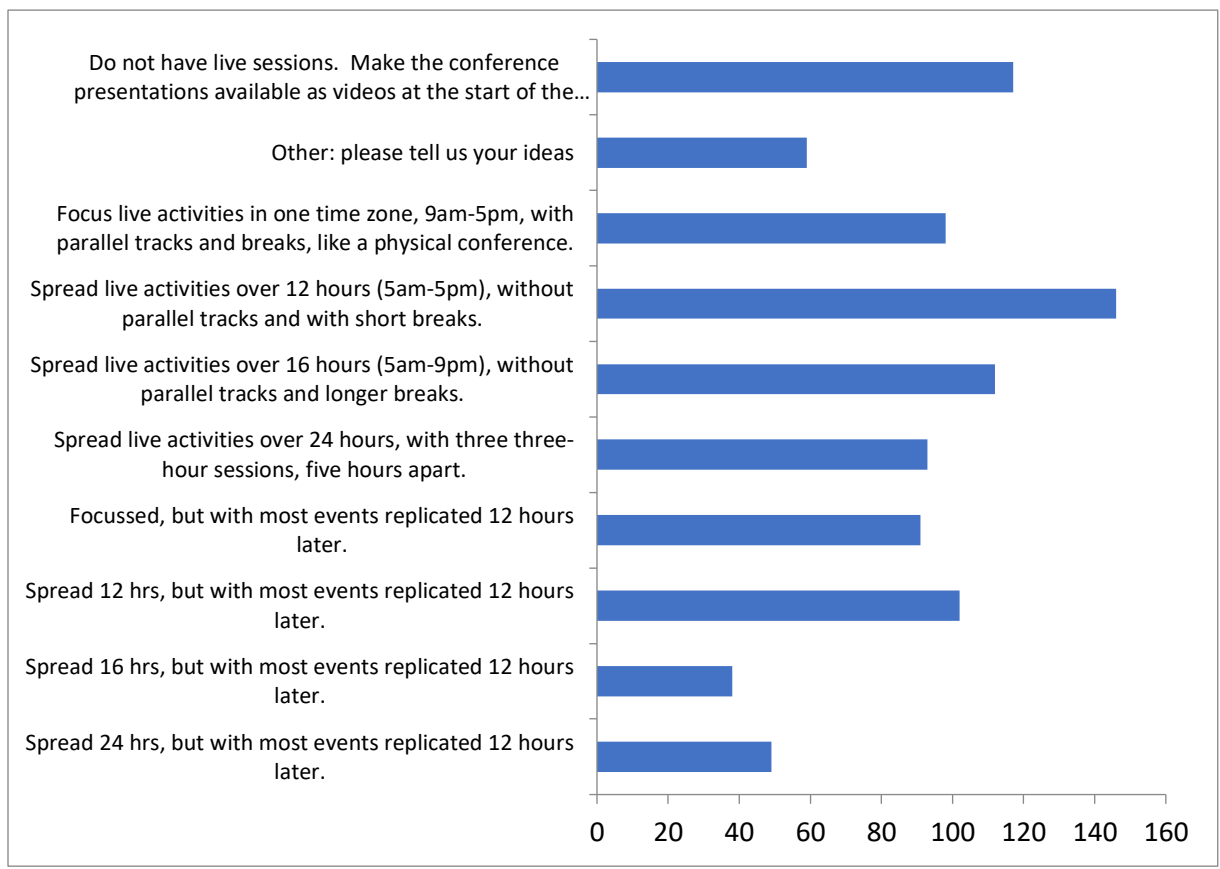

Unfortunately, the somewhat uniform distribution of responses does not give a strong steer.  The most popular option – "Spread live activities over 12 hours (5am-5pm), without parallel tracks and with short breaks." – is the model used for PLDI. It should be borne in mind, though, that most survey respondents were in time zones for which the PLDI model proved to be convenient.

Many ideas for ways to mitigate the effects of time zones were provided as free-text answers; they are included in Appendix A.12.

## 7. Acknowledgements

I am grateful to Dan Grossman and Michelle Strout for their feedback on an earlier draft of this report.

Many of the questions for the post-conference survey used for PLDI 2020 were inspired questions asked in the post-conference survey for the recent EDBT/ICDT virtual event (Bonifati, et al., 2020). Thanks to Dan Grossman, Jeff Foster, Benjamin Pierce, Emina Torlak, Stephanie Weirich and John Wickerson for providing feedback on the survey questions.

The "Guide to Best Practices for Virtual Conferences" from the ACM Presidential Task Force on What Conferences Can Do to Replace Face-to-Face Meetings was an extremely valuable resource in the design of virtual PLDI (ACM Presidential Task Force on What Conferences Can Do to Replace Face to Face Meetings, 2020), as was [this report on the ASPLOS 2020 virtual conference](#).

I received a lot of advice and encouragement from Crista Lopes and Benjamin Pierce on virtual conferences, including useful details of the pros and cons of the various tools that are available. I benefited from several discussions with Milind Kulkarni, who was setting up a virtual version of PLMW to run in the same week as PLDI, related to virtual conferencing technology. Jonathan Bell put an enormous amount of effort into getting an early version of Clowdr ready for PLDI so that we could benefit from video chatrooms. Jonathan also provided invaluable advice related to technical aspects of Slack.

I am extremely grateful to the PLDI 2020 Organizing Committee for the hard work they put in to make the conference a success. Extra special thanks are due to: Dan Iorga (Student Volunteer Captain), who helped me figure out every logistically aspect of the virtual conference, including the intricacies of Zoom, Slack and YouTube live streaming, and who worked tirelessly to make sure we had every base covered; John Wilkerson (Publicity Chair), who, as well as doing a brilliant job of all aspects of publicity, was a constant source of advice and a willing sounding board for my thoughts on how the virtual conference would work; Emina Torlak (Program Chair), who – in addition to doing an outstanding job of all the standard duties of a Program Chair – worked with me to figure out a suitable program for the virtual event, and took care of communicating the many details of how the virtual conference would work with paper authors and session chairs; Annabel Satin (Conference Manager), for support throughout the organization of PLDI, and in particular for taking care of the steps necessary to cancel the physical meeting; Stefan Marr (Workshops and Tutorials Chair), who worked with the organizers of the many co-located events to figure out a suitable format via which they could run virtually; Manu Sridharan (Sponsorship co-chair), for liaising with the corporate sponsors of PLDI 2020 to work out suitable alternative sponsor perks when we moved to a virtual conference format; and Na Meng (Web chair), for putting in a lot of extra hours in relation to the website changes that were necessitated by moving from physical to virtual PLDI.

Special thanks also to Mike Moshell at Registration Systems Labs for extra work relating to the unexpected registration needs of the virtual event; J.C. Peeples at ACM for support with setting up contract related to the virtual event at short notice; and Elmer van Chastelet and Danny Groenewegen from researchr for integrating a Slack sign-up system with the registration system.

Thanks to the wonderful PLDI 2020 student volunteers without whose hard work we could not have run the event: Noaman Ahmad, Cezar Andrici, Sabree Blackmon, Matt Bowers, Madhurima Chakraborty, Yixuan Chen, Jianyi Cheng, Julia Gabet, Momoko Hattori, Paul He, Yann Herklotz, Linnea Ingmar, Dan Iorga, Thomas Koehler, Paulette Koronkevich, Martin Kristien, Peeyush Kushwaha, Michelle Li, Paul Lietar, Amanda Liu, Daniel Marshall, Amir Naseredini, Chao Pang, Zach Patterson, Radu Prekup, Baber Rehman, John Renner, Ahmad Rezaii, Salim Salim, Klas Segeljakt, Kartik Singhal, Alexandros Tasos, Siddhant N Trivedi, Darya Verzhbinsky, Irene Vlassi Pandi, Daniel Wang, Dominik Winterer, Tiancheng Xu, Irene Yoon, Stefan Zetzsche.

Thanks to SIGPLAN Cares for providing representatives to be available throughout the week of PLDI events.

Thanks to the sponsors of PLDI 2020, whose generous support allowed us to make the event free to attend: Apple, Microsoft, VeTTS, Facebook, Novi, GitHub, JetBrains, Oracle, Stripe, Googe, Huawei and Uber.

Finally, thanks to all the authors of papers presented at PLDI and its co-located events, the Ask Me Anything Guests, session chairs, co-located event organisers, and to everyone who attended!

## References

ACM Presidential Task Force on What Conferences Can Do to Replace Face to Face Meetings. (2020). *Virtual Conferences, A Guide to Best Practices.* Retrieved from https://www.acm.org/virtual-conferences

Bonifati, A., Guerrini, G., Lutz, C., Martens, W., Mazilu, L., Paton, N., . . . Zhou, Y. (2020). *Holding a Conference Online and Live due to COVID-19.* Retrieved from https://arxiv.org/abs/2004.07668v3

## A.    Appendix: responses to free-text questions

### A.1.    Responses to "Which meeting mode do you think would have been best for the PLDI Research Papers track?"

| | |
|---|---|
| PLDI 2020 *but* speakers get some visible feed of audience. Speaking to dead air is tough. | I cant decide between option [broadcast] and [asynchronous]. Maybe next time we should try [asynchronous], just to get the difference to [broadcast]? |
| A mix between broadcast and async: i.e., pre-recorded videos available far in advance (even 1+ week; this is to alleviate time-zone issues, too) coupled with live re-broadcast combined with live Q&A sessions. | Prerecorded talks but live Q&A with typed questions |
| | Meeting mode, but with a smaller live audience, make talks more interactive |
| Just like PLDI, but with an option to drop into an extended discussion with the authors (including audio / video?) either after the talk or after the session | pre-recorded video, but a live session for each paper Q&A as well as asynchronous Q&A messages. |
| Combination of asynchronous for longer presentations and broadcast for short presentations. | Stick with the PLDI 2020 approach *but make pre-recorded videos available earlier*. That is, be primarily synchronous, but allow people who cannot be awake during the main session |
| preview the talk so they can ask questions in advance. | PLDI mode but with pre-recorded videos available, if for some reason broadcasting the recording doesn't work or if the timezone is not compatible |
| Asynchronous videos available a week before the conference, then a one day meeting that is synchronous with live Q&A | |
| I like broadcast, but missed all the papers I wanted because of time zones & conflicts. Need both broadcast and live. | See the idea I wrote on the previous page. Failing that, probably broadcast mode, or asynchronous but with live Q&A at designated times. |
| I think videos should always be pre-recorded and available, regardless of the approach. Q&A seems to be better if it is done live. | I like things that are live -- not only QA but also the talk. So I'd prefer authors to give live talks. QA should also be live but it's okay there is no video (slack/slido works well for me, if not better). |
| Meeting mode + a text-based platform for Q&A of virtual attenders | |
| Broadcast via PeerTube (rather than YouTube) followed by live questions on Jitsi (rather than Zoom) | |
| Again: the presentations should be available at the beginning of the entire PLDI event (since they are recorded anyway). Only discussion gains something from being live. There should be no cap on number of participants - that is ridiculous. Asking questions in text worked fine. | Asynchronous mode as described above, but additionally with a live session for Q&A for each paper. Every 20min, a new live Q&A session would start, but if people want, they can continue a Q&A session for as long as they want, while the next Q&A sessions already are happening in a different room |
| broadcast mode but with a better app for questions, with up votes and the like | Asynchronous long talks, synchronous lightning talks, live Q&A. |
| The PLDI mode, but enhanced somehow for the audience to make their presence and their feelings more manifest. Eg, a bar with the name and photo of each participant. And the participant should have gestures, eg "clap", "laugh" as in Teams. And the discussion should be organised as a tree, so that participants can ask follow-up questions, or refine another participant's questions. | make all talks prerecorded, broadcast them on a central channel (like TV). All Q&A async online via a forum with text questions and responses (though encourage the author to promptly answer questions after their talk is broadcast). The session chair could read out interesting questions and answers after a broadcast if they wanted. The broadcast could replay a talk multiple times during a conferences for international people. |
| Asynchronous, but with opportunities to meet virtually for Q&A and networking. | Live sessions but made the talks available earlier |
| A number of attendees should be present in the Zoom meeting, with the others watching the live stream and asking questions in text format. | combination of 1&3 |

## A.2. Responses to "Which meeting mode do you think would have been best for the PLDI Ask Me Anything track?"

| | |
|---|---|
| PLDI 2020 approach *followed by* a social session in a tool like gather or clowdr | be "upvoted" by the community (akin to how AMAs are done on Reddit), which the guests then answer via broadcast. The difference here from plain Broadcast is opening up questions early, which also allows guests to come up with more thorough answers ahead of time. |
| Same as previous question (so, PLDI 2020 with some audience to look at) | |
| Broadcast mode, but with no interviewer. | |
| Hybrid of Asynch and Broadcast: use a text-based platform to pre-queue questions that can | Meeting mode, but with a smaller live audience |
| I like the broadcast mode, because it removes the risk of AMA attendess playing the "here is a statement of my opinions disguised as a question" trick, or taking too long to get to the point of their question. | Asking questions in text, and having them read out loud by a moderator, seems fine. \*\*\*As long as the moderator is reading questions from the audience\*\*\*; not having his own conversation with the guest. |
| 1st option, but the Q&A should be integrated with the broadcasting platform | broadcast mode but with a better app for questions, with up votes and the like |
| Did not attend | The PLDI mode, again, but with the enhancements I proposed for Research Papers. |
| Did not attend this track | asynchronous mode (pre-registered talks), with authors available in a large meeting session (e.g. zoom with 300 participants), so that one or more authors can have live Q&A |
| It was good as it was run | |
| The Simon P-J AMA began in PLDI2020 mode, and moved to a more informal chat on Crowdr after a while. This worked well, and might be worth repeating (a PLDI2020, more formal segment to start, moving to a more informal "meeting mode") | A number of attendees should be present in the Zoom meeting, with the others watching the live stream and asking questions in text format. |

## A.3. Responses to "Please share your thoughts on how well the Q&A process of Slack-based questions answered live by authors / guests worked."

| | |
|---|---|
| The AMAs were so great! | Q&A were restricted in each thread, but discussing on a thread is difficult. Discussing on the main channel (out of threads) would practically work for non-parallel sessions. I annually attend a local symposium in Japan, where people talk on IRC-based chat in parallel to the presentation. A single channel is working fine there. We might want to have more asynchronous discussions on Slack, not just for posting questions for each presentation, because we had timezone differences. |
| I found that one strength was that slack provided a natural place to continue a discussion with others after the end of the formal Q&A. | |
| The interactions went pretty well. However, the slack UI is inconvenient for live Q&A. I would prefer a platform with permanent links, multi-tab browsing, powerful searching and low resource consumption. Presumably some Hacker News/Reddit-like/forum-style UI but with live comments/posts updating. | |
| It was pretty good. I find the fact that co-authors can answer live on Slack a benefit, which I've seen work in other, physical, conferences too. I also like having questions moderated and relayed by session chairs. I think that really helps avoid grandstanding by big names. | Q&A was typically too brief, but I appreciate some talks attracted less questions and so it depended heavily on the level of interest expressed. |
| | I liked the Q&A process of PLMW, where questions could be upvoted. In this way, the questions got answered that most people were interested in, and not the questions that came in first. |
| I thought this process worked incredibly well! I would highly recommend using this same message in future years. | I enjoyed the Q&A sessions. Text-based questions forced the "askers" to be very concise and the host was able to relay questions in a way that connected different topics well. |
| This seemed to work pretty decently. It was nice to have the opportunity for authors to answer more questions than time permitted in the actual session in the Slack. | I liked it better than live conferences. Far more effective medium for asking questions (no waiting in line, no worrying about how loud to speak, confusion over accents, etc.) And, easy to spawn followup discussion. |
| I thought we could have more freedom on Slack. For example in research paper sessions, | |
| Barring one, all my questions were asked. I especially liked the way some of the session chairs of AMA sessions grouped the questions and paraphrased them (sometimes) to maintain the flow of the interview. | I think it worked at least as well as live. I'm not sure whether meeting mode and occasional back and forth might have made things more interesting, but it was fine as is. The improved overflow question handling and ability to defer answers is a significant advantage over live. |
| I think one-channel-per-paper from past conferences fits better the Slack interface, and I remember no complaints about that interface. Channels are easier to find, navigate, and subscribe to, and still allow posting questions in advance, while threads can only use a small right column and lack subthreads (which would have been sometimes sensible!). | I liked the Q&A process a lot. I didn't ask questions, but I found the interaction mode, where moderators fielded the questions, simple and efficient. The one issue I saw was that some moderators are better than others at this process. Some just went from the top to the bottom, which is not great because it advantages people who ask questions many hours before the talk even happens and takes neither continuity nor general interest in a question into account. Others tried to ask the most popular questions and perhaps group those with a similar theme, both of which are nice approaches. This is a bit harder with Slack though since it does not sort based on likes or something like that. On the other hand, having everything centralized in the same service and saved for posterity is a big win. Perhaps there is a way to integrate something like Sli.do and Slack, which seems to combine the best of both worlds? |
| They worked very nicely especially given that the authors are monitoring their threads. | |
| Given the learning hurdle that the use of separate question threads had I think it went really well. I think having separate rooms or channels for each Q&A would have worked less well | |
| I don't think Slack is the right tool for this kind of a process. The questions were suppose to be posted into threads, and Slack is terrible when it comes to thread-based discussion. | |
| I appreciate the questions being repeated by the chairs (it's much smoother than waiting for the question-askers to check their mics, volume levels, synchronization, etc.) -- keep up the good work! I think that it may be fine to extend the Q&A though (as some of the sessions ran out of time). | Very well |
| | Slack is a mess. Many people and conferences like it, but it does not make any sense to me. |
| Worked really well. | The format was great. The asynchronicity of it allowed questions to continue indefinitely as well as to have several questions immediately at the end of the talk. The only thing that might be improved is to have the talk videos available sooner so that interested parties may be able to ask well-formed/thought out questions in advance (to be answered at the end of the talk, live) allowing follow up questions to their original questions. |
| I thought the Q&A process of Slack-based questions read by a moderator and answered live was excellent. I adored the "Ask Me Anything" sessions and was very happy that you let most of them run long. (45 minutes seemed about right for most of them.) | |
| Decently, but ideally one would have a dedicated solution for this where questions can be up/downvoted, discussion threads organized and moderated around questions with all paper authors able to thoroughly address questions asynchronously instead of improvising during a live talk. | A sli.do solution may have been better: gives people upvotes to make the interesting questions bubble up to the top. |
| | Fine |
| It went well. | Worked great. |
| It worked great! It could be improved by finding a way to facilitate more interactive Q&A outside the official slot (as suggested above) | It worked really, really well. In some cases, for example, clarifying questions or statements were provided in Slack, based on answers, and were picked up by the moderator. It yielded a |
| really effective back & forth.  In some cases, moderators simply read questions from the top, which was a little less interesting. Effective moderators grouped questions together to forms themes that directed the arc of the Q&A or AMA. Some clarity around voting for questions may have helped weaker moderators pick more interesting questions. | I think Slack works very well. (I was GC of ICFP 2017, which I think was the first SIGPLAN conference to commit to livestreaming talks. We used Slack there too.) |
| It worked out well. | Extremely well! One of the strong aspects of a virtual conference. |
| I think that it worked very well. In the papers it may have been better for the moderator to pick and choose questions instead of just asking them in the order that they came in, since that may just focus on questions from earlier in the presentation | I think it was the same quality as or even better than the Q&A process of a physical conference. |
| It is fantastic, the actual Q&A discussion can go on without stop after the designated Q&A session, which is roughly 3-5 minutes. | Worked well, it was nice that speakers continued answering remaining questions after the time scheduled for questions on video ran out. |
| Pretty good. In the AMA's it was very smooth with the interviewer picking up one question after another and if he knew about the topic as well, sometimes even improving on the question through formulation or explanation. Instead of asking a new question, one can also upvote another. It was good that QA could continue after the session. | I think most authors only answered about 2-3 questions from slack live, but most answered the remaining questions afterwards. If we do have a more spread out schedule, it might be possible to give authors more time for Q&A (say 10 minutes?) but this time is specially demarcated so their talks can't spill into it. |
| It worked very well, save for too little time to answer all the questions. | Prefer not to submit |
| | It worked pretty well. |
| Mostly very well, to the point that I think I would actually prefer asking questions using Slack at live conferences over queuing at a mic. | I did not quite understand how it worked. The instructions for all of the different online platforms were a bit confusing. |
| I quickly learned that I should put questions in the Slack as I encountered them, rather than trying to save them for the end of the talk.  This risks some redundancy in questions with the talk content, though. | I don't think that it worked at all well, because I don't find that slack is a very useful platform. |
| | I only asked questions at AMAs.  That was great!  The moderators did an excellent job, as did the AMA-ers themselves of course. |
| From my experience, it was superior to QA on a physical conference. There wasn't any problem with not being able to hear questions properly, it was easier to ask questions (in particular for introvert people), it was possible to ask and answer more questions. I don't think that the Slack-interface was ideal through (in particular with respect of threading) and I think that it is a shame that QAs are not persistently available. | I think it worked fairly well. I like the fact that there is a host moderating / asking questions and interacting with the speaker.  I think having a large meeting session with some capped virtual attendees can work as well, but will still need some host moderating so that not everyone speak at once.  The only main advantage of the virtual attendees with audio is that once a question is asked, there can be more smooth back and forth between the speaker and the person asking the question clarifying.  However, depending on the size of the meeting, there may be network bandwidth issue and fairness issue (who to let in and should it always be first come first serve?) |
| It worked fairly well in my tutorial where a couple my students helped to moderate the questions, and directly answered some questions directly. Something like that can also be done by author teams. | I really liked it as one can ask questions in the relevant channel and/or relevant paper. |
| | I thought it worked well.  I liked that people used thumbs-up to elevate common questions. |
| I sort of like the idea of people asking questions on video, if possible, but that may be tough to sort out. | it also happens in a physical conference, it's just a little less noticeable because it doesn't leave a written trace. (I was a big fan of Slack-based questions used in physical conferences in the last few years.) That said, we should re-consider the use of Slack. I think open-source tools like Zulip are now as competitive as Slack. |
| no difference from physical conference | |
| Need more time for questions! I only had time to answer one question. I also didn't like the thread thingy. People should have been allowed to ask questions on: youtube live chat, slack by flagging the @author or even twitter by using a special hashtag. The latter in my opinion is a missed opportunity. | It was very confusing to see different tracks using different ways to handle Q&A (external service, pinned message threads in slack, real time post in slack, etc) |
| I was a co-author. The Q&A process worked well. It is possible to get overwhelmed since there is no physical world-based rate limiting. I believe we spent about 2 hours handling questions, post-talk. But we were very pleased with them. | When the chair selected and asked the questions it was great. When authors/guests read questions themselves it was a bit awkward. |
| | Pretty well. I think in a physical conference, many times questions time is over and then the question is never made. Here the questions can be made even after the session continued or it was even over. I think this was a huge plus |
| It was okay | |
| It worked really well I think! In the PLMW sli.do was used, which also allowed Polls for the viewers and upvoting of questions, which worked really well too. I think a little bit more sophistication for the Q&A process could help, but it definitely worked well in my experience! | I loved the fact that, if not all questions could be answered during the live session, the authors could still engage virtually and answer all questions anyway. |
| I'm not sure if the host is needed to go through the questions. Paper authors should be able to read questions in Slack even faster than the host. Maybe even during the video since they are not actually presenting anything. | It worked very well! Thanks to the slack threads, many more questions than physically could be answered by authors in a way that is visible to everyone. |
| | I think it was excellent. In fact better than a conference because all questions were at one place and the authors could respond to them in detail. |
| N/A | |
| As I already mentioned above, I think it worked pretty well and I hope it stays also in the future. | Overall good |
| It was OK. But allowing questions to "bubble up" when upvoted, and removing questions that have been answered (as done by Sli.do) would have helped. | Some moderators asked the questions in linear order until time was up. There were some cases when it might have been better for the moderator to choose questions further down in the list. Overall, that part was better than at a physical conference, because the authors typically answered the remaining questions via responding in the Slack channel. Thus, more questions were posed, and more were answered overall. If one had a hybrid conference, that aspect would very likely be lost (assuming that speakers were physically at the conference, there might be distractions to getting answers to all questions posed on Slack). It would have been more convenient if Slack would have supplied Name + Affiliation, rather than just the Name. It was a minor |
| Very well. The thumbs up / voting system that naturally formed was also useful. | |
| I feel like this process did a good job! | |
| I like it a lot, much better than people getting up to ask questions. Questions can be a lot more crisply formulated, over an extended period of time. Chairs/speakers can choose good questions rather than wasting space on prominent audience members asking silly questions. Discussion can (and did) continue after the session. It's inevitable that some speakers don't get a good Q&A afterwards, and | |
| nuisance to have to remember to always provide one's affiliation. | I think Q&A was one of the highlights of PLDI. I think I attended those sessions more than anything else. One improvement could be to take questions on an asynchronous platform such as a Discourse forum so that the guest's responses to those questions are later available. Think of Reddit AMA, for example. That discussion holds far too much value for posterity to be lost in the walled garden of Slack. |
| One thread per presentation is not desirable. It is difficult to find the questions, or add to a given question, or for the speaker to answer offline - Questions should be handled in the slack channel, allowing threads for each question | |
| It worked quite well. One suggestion: it would be nice if questions could be upvoted (similar to Slido) such that the best questions got asked in person. I found often interesting questions with many likes got skipped over in favor of less interesting questions simply due to the order in which they were asked. | worked well |
| | This worked great! I do not think I have seen anything like this and I think the organizers figure this one out really well. As I wrote before, if researchr program and slack threads were one thing, that might be a bit simpler for everyone. |
| It's better: q&a is recorded; and the authors have time to prepare for answers. | |
| It was great. Pretty smooth. | The Q&A format on Slack worked really well for the AMAs. |
| I think it was okay, but one impedance mismatch is that it's hard to map Slack channel questions to the video, time-stamp wise. | too much was going on for it to be meaningful for me. I mostly ended up chatting privately with people |
| PLDI's live Q&A session where the author speaks is critically important, and contributed a lot. Text-only Q&A (in the style of ASPLOS) is just not an acceptable substitute. The one downside is that the full videos were not posted in advance, and the audience was not encouraged to ask questions in advance. It would be better if PLDI encouraged some people to ask questions in advance (like ISCA). | Great, with deep chat discussion |
| | It worked great |
| | I loved it! |
| | I thought it worked really well. It also contributed to the feeling participating in a communal event. |
| It worked amazingly. I think it's one the most-improved features in comparison to live events. | It worked okay, but there's room for improvement, e.g. one thread per presentation was difficult to navigate. As an alternative, Slido worked better for seeing which questions had most votes or were already answered, but doesn't offer a good story for authors later answering additional questions in chat. |
| Right questions were askes, with good reponses. | |
| Right questions were asked, with good responses | |
| Slack Q&A + live answer session was excellent. | It was surprisingly good. For AMA slido-like voting might have been more appropriate, but if sessions are allowed to run long, Slack is ok. And people organized to like questions, which was nice. Regular talks worked just great, and authors followed up on most questions |
| Great | |
| My Q didn't get answered, but I totally loved the AMA interviewees, and their answers. | |
| Slack worked reasonably well and can also allow typing in questions while the talk is going on. | It worked fine, although the interaction feels weak |
| Seemed fine. Would be nice if the platform allowed threads-within-threads, so questions could be grouped with answers. | my main issue was with the slack interface and timing of pinned threads. I mentioned my issues in the #lessons-learned channel. Basically slack is not ideal for this, and would have been better to have a channel per video. Zulip |
| It didn't work very well. | |
| provides the ability to associate messages with distinct topics that might help with the UX. A reddit like interface too would be interesting to consider, however having all the textual communication via a single app would be good. | I think the Q&A via Slack worked well. The written format enabled clearly enunciated questions. |
| It was very nice to have written questions visible by all and also reponses from authors. I did not found how to same them during the conference. It's a pitty. | I could not participate because I found no way to use Slack without running proprietary (Javascript) code on my machine |
| | Very good. The use of a moderator to select questions improved quality. |
| Seemed fine. | Reasonably. The slack threads could have been published beforehand |
| There was nowhere near enough time at the end of talks for questions. I had to wait for the presenter to respond on slack, which was a much worse experience | Very well. The asynchronous, possible-to-be-ignored nature of slack questions dramatically lowered inhibitions to asking question, and almost every paper spawned significant and interesting conversations in the corresponding slack channels, which it was nice to be able to read. |
| I think it was a huge improvement over Q&A at physical conferences. There was more time to formulate questions, other people could discuss and share insights, and authors/guests could interact on Slack after the live session. Having per-question threads would have been nice, instead of a "chat room" and flat experience. I also liked Slido's voting/sorting mechanism. | Overall ok. There could be more time for questions. |
| | It worked very well, especially after student volunteers started posting the link to the current talk as the talk started (so I didn't need to scroll back to find it) |
| It seemed to work well. The panel discussion I watched had a live voting system that worked well too. | The Slack approach was quite nice. I don't think this should replace ordinary Q&A at an in-person meeting, but for a virtual meeting, I think it was great. |
| It worked well, arguably even better than 'real' conference Q&As!, but presenters needed prompting on occasion to repeat the question | It worked very well. An improvement might be that upvoting a question reorders the thread to help the session chair select the most popular questions. |
| I think the questions in threads approach was a bit confusing. | |
| Simply fantastic in all respects; quality of Q and A was very high, I like the fact that Qs can start during the talk and continue till well past the talk, ideally we should preserve the text too. Q&A was fantastic, I would not fiddle with it. It would be nice to "preserve" the text though -- it was very high quality IMO. | Worked well enough. But the Q/A worked better in sessions with fewer people.In big sessions, authors were answering long after the audience attention went elsewhere. |
| | I thought it was quite efficient |
| | as mentioned, I think Slack is not great for this. I prefer apps where you can vote questions |
| Did not use Slack | Pretty well; that part came relatively close to a real conference (module time zones). |
| Pretty well. Questions from a greater cross-section of the community than usual, I think. | Slack-based Q&A was great. Way better than standard Q&A |
| Very well. Many questions got deeper offline discussion afterwards. | Better than in a physical conference, since the authors had some time to *read* the questions ahead of their answer. There are much less miscommunication issues. |
| I didn't like it. I prefer to ask question in the same platform I'm watching the live video. Posting question on YouTube as I watch would have been way better. | |
| | I can't log into Slack at all |
| The authors gave their best efforts. | Overall, much better than in live conferences, because the questions, being written, were |
| It's great. | |
| expressed better. But sometimes the chairs did not apply judgment when picking questions, and just asked them in sequence. | One of the important part thats missing from virtual conferences is the opportunity to talk with potential supervisors for future opportunities, if that can be introduced somehow, it will be very useful. |
| It happened 1 or 2 times that there's no question at slack at the moment the presentation ended because youtube live streaming have some time gap with zoom. this isn't a big deal since they could still answer those questions on slack. | very well. but I would use a channel per session instead of a thread, just to allow for more complex interactions |
| | Q&A + Slack was great, and could even be used for physical conference to keep track of questions and answers that the author did not have time to give at the end of talk. |
| For the talks, I thought it worked great -- my only complaint was that I wished the question sessions were longer! The AmA sessions were very good because the interviewer could stand in for the audience. Really requires that both interviewer and guest have good mics and fast internet connections though! | It seemed to work very well. |
| | Very well. I do think the quality of the AMAs depended a lot on the moderator, best ones were the ones where the questions were weaved into a narrative (e.g., Cook's, Lattner's). |
| I thought it worked very well. I was a session chair and was very pleased with the quality and depth of questions. It allowed me to batch multiple questions together, encouraged a broader range of questioners, and gave the authors a chance to answer all questions, even if only on Slack later. If anything, given the amount of questions, I'd argue to make the talks a couple minutes shorter and provide *more* time for questions. | I liked it a lot more than normal QA. Adding a mechanism to upvote a question might be a good addition. |
| | I read some Q&A on slack and they are helpful to me |
| | It worked really well. I liked the slack-based management of questions: they remains in the thread and could get answered later fostering discussions. Moreover questions more interesting gets more feedback and could be preferred to those less interesting for the community, when there is short time. Cool. |
| It seemed to work fairly well, although it wasn't transparent what would make a session chair choose to read out one question from Slack vs. another. This is of course similar in a physical conference, but could be biased in different ways: the question is known beforehand, and some names are hard to pronounce. I liked how even if your question wasn't asked live, the authors would typically have time to answer your question later on Slack, which is easier on the part of the questioner than at a physical conference. | It worked reasonably well, but the Slack interface is not optimized for such as format. Perhaps one channel per talk would have been a better organization. |
| | The idea of bundling the questions in a threadbased system like Slack is great, but the problem seem to be that Slack threads are linear, with the Reponses from the author it often became quite messy. Something with nested reply (reddit style) would make more sense to me. |
| Worked wonderful, big fan. Follow up questions could use some work maybe | |
| Generally good; would be better if more questions submitted ahead of time to facilitate a better interview. Maybe cut off 1 hr before, then allow further questions and points during? | It was okay, but the Q&A interface within a Slack thread was somewhat inconvenient. The linear structure stifled conversations. Ideally, one would have some specialized tool. |
| It worked surprisingly well. | It did work quite well, and it definitely saved the time that would be required for getting asking people properly unmuted. |
| Q&A felt often quite short, but it was OK when authors followed up later. | |
| Often better than f2f | this mostly worked - though the opportunity for followup and post-talk hallway track discussion |
| needs very serious attention.  And Q&A time was often cut short | It worked well. More interactions between the authors and the audiences would be better. |
| The Q&A worked really well, especially since it allowed extra questions to be asked even if the time runs out, and the answers to the questions are available publicly for anyone interested to see as well. | Several research managed to answer many questions quite effectively which was amazing. |
| | I think this is a great way to moderate questions and hopefully similar strategy can be used in the future. |
| Some questions seemed to missed. The fact that authors/presenters can go back to the slack thread is a good point. | should put the authors after the talks in gather to answer further questions and mimic physical conferences |
| Slack is very linear, and scrolling backwards at least in the Linux app is not easy.  So it was hard to find the links for the paper discussion channels, since they were all entered at once before the conference and other announcements came after them.  Switching back and forth between the program to find out what is going on and the slack to find the youtube link was also inconvenient. | The live questions went well, though I cannot comment on the follow-up Q&A on Slack.  This would be very important (to take the place of the after-track conversations at a physical conference). |
| | It was perfect! |
| | kind of ok, but we should have 1 channel per paper rather than 1 thread |
| Good | I really liked it.  I especially appreciated getting to read and consider the questions while my talk was playing. |
| This worked reasonably well. | |
| In some cases, it was harder for the speaker to understand what the question was about. It would have been better if there could have been a direct communication, such that the person who asked the question could clarify what they meant. | It worked well, though Slack's interface is inconvenient for this: would be good to have permalinks from outside, without need for scrolling. It was very nice to get to see answers to _everyone's questions. |
| | I think it worked fine. However, at times there was not much time for answering many questions. |
| I think it worked reasonably well considering the circumstances | it worked very well I think; having a single channel with a thread for each paper worked perfectly with slack |
| Reasonably well. | |
| Given the limitations of Slack I think it would be better to have one post (with attendant thread) per question rather than one per paper. That probably requires one slack channel per session to avoid getting out of hand; but maybe that's ok. It's important to be able to easily find questions from earlier talks for followup, and slack isn't good for this unless you post "following" in every one you want to check back on, which should certainly not be encouraged. Also, given that there's been quite a bit of subsequent discussion for some of the papers, I think it works better if the authors also post their answers afterwards -- some did (AFAICR) and that made things easier when checking back later. | Good |
| | QA worked perfectly |
| | Worked fairly well. One nicer feature would have been to enable separate private channels for each paper that anyone can request to join. Essentially it would enable multiple threads for different topics. I found answering questions to be slightly hectic. |
| Love it! I think Slack definitely works. I actually find it better than the physical QA. I also like Slido. | slido is better, but the Q&A in slack is definitely better than what zoom offers.  The key better features in slido are that questions can be moderated, questions can be sorted by how upvoted they are, and questions indicated as answered are removed from the "live" list. |
| I think answering questions live was not very helpful, but the slack was good and allowed for many questions and good discussions. | back to, if they so desire and have time. I also think an archive of all the questions will be quite valuable in the future. |
| It was helpful | |
| I very much liked being able to queue questions, which the authors can then come | |

## A.4. Responses to "Do you have suggestions for ways to provide more social interactions, networking or mentoring opportunities at a virtual conference?"

| | |
|---|---|
| More breaks in the schedule with an explicit call to meet up and chat. | mentoring conversations that I think would not have happened in London).  Final 2ct: I think virtual conferences have their place and we should continue trying to innovate. But I still would like to see at least 1 larger in-person community gathering per year. |
| See above re clowdr.  Offer routine group video meetings for the audience following a session, as an automatic follow-on from the presentation platform. Possibly even with automated subgroup partitioning (although one might want some opportunity to feed in private preferences: if there's a space with X, I would like to talk with them; please don't drop me in a group with Y). | |
| | Start asynchronous communications (e.g. Slack) enough earlier (e.g. ~2 wk) than the virtual conference. |
| More meetings of the Zoom coffee hour style, where conversations are small (roughly 5 people), time-limited, and randomized. In my experience, these events were orders of magnitude better at encouraging participation and interaction than any of the other platforms despite significant effort to take advantage of the other platforms, and provided many of the most valuable experiences from the conference. The other incredibly value tool was the #mentoring channel: having mentors introduce themselves and explicitly state that they are available for mentoring made them much more approachable, and I had multiple incredibly rewarding interactions with researchers I've looked up to for many years because of this channel | Schedule time for social interaction. The PLDI programme was too full, so it wasn't clear when people would likely be on gather. |
| | The list of attendees is not known. I mostly tried to look up (in Slack DM) the names of the people I thought are attending. This means I could be missing a ton of other people. |
| | During IEEE S&P there was a slack bot called Donuts that was starting a private chat with a random member of the audience twice a day. This helped (and strongly invited) to connect with people as there was no need for initiating the chat (as was necessary with gather) |
| | It would be nice to know the topic of discussion before joining a group. |
| | Consider non-video-centric social interaction (including text chat, audio-only). |
| The unofficially organized Zoom virtual coffee hours that Rachit and Irene put together seemed to work the best of the available options. Gather seems like a really nice idea, but would definitely need a lot of work on reliability before it could really be useful. | More scheduled short breaks where video interactions can happen. Potentially, conference official "coffee breaks" where people are semi-randomly or by topic assigned to rooms of 5-10. |
| I think PLDI did tremendously well with trying to establish social interactions. However, I think if you are not already in the "in-group" (I'm more in the SE community, ICSE/FSE/OOPSLA), it's very hard to initiate the same levels of interaction. The mentoring channel was fantastic and I think that worked probably even better than it would have in person (I had 5 | I think I need to become accustomed to these kinds of social interactions.  In person, I can remain on the periphery until some strong personality pulls me in, or until I feel comfortable speaking up.  It's more difficult for me to speak up in these virtual spaces, or even to reach out to one or two people I don't know, when everyone could be watching. |
| | Zoom Video Chat Rooms functionality for every participant |
| no there was good effort in the slack channels | Have a designated "social time" that specifies gathering in one of the gathering spaces (there were three if you include Slack). |
| I think PLDI'20 did an amazing job with social interactions, all things considered! One other possibility would be randomized breakout rooms within Zoom, to really mix things up and encourage people to meet new people. This could replace the somewhat limited randomness of sitting next to someone new at lunch at a physical conference. | I suggest sticking to exactly 1 social platform that supports both text and video. I am hopeful that we will have better options available for the next conference! |
| Not specifically. I think PLDI did an amazing job and I would not mind having every future conference be virtual if they worked similarly. | Clearly specify times when people should gather and have a single platform for interactions and make the user experience smoother (minimize possible technical issues, improve user interface, etc.). See my comments to Gather for more specific suggestions. |
| The organizers did a truly wonderful job. This is a first of a kind, nearly all possible options to create a great online PLDI experience have been offered. Of course, this is only a first of a kind. But I don't have concrete suggestions at this point. | Make talks live and interactive. The content of the research is what brings us together and provides a starting point for a conversation, also for new people in the field. The online environment provides new opportunities for having conversations around paper presentations. |
| Keep the mentoring and recruiting channels! Have more topic-focused chat rooms baked into the platform, for instance, right after a session. Have dedicated slots for socializing. Let's brainstorm more creatively to accommodate different time zones. | If this isn't entirely hopeless, maybe BoF sessions (where people sign up in advance and then show up in a virtual room to discuss a specific topic they all care about) would be the best idea. |
| I think this PLDI is the greatest example of how should we achieve this goal. | I initiated a clowdr room right after the session that had my paper in it, and invited the other authors from that session along with anyone interested. This worked well, and we had interests in common. I think it would be good to have designated post-session gather/clowdr times. |
| Mentoring sounds like a formal and difficult process. I guess thats not what they meant, the mentioning of job offers discouraged me from talking to them, as I am not looking for a job, currently. I would not know about what, how or why I should talk to them. I would appreciate the ability to ask people what they are working on. I really missed the ability to walk around a poster room with people standing nearby. Although the interesting posters are often occupied, the last time real life time I have been in a poster session, there has always been someone free and empty, that you can get into an in-depth talk on what, why and how they are working. | Maybe a signup form for people where people are matched into random groups so they can chat with each other |
| | Prefer not to submit |
| | Accessibility is important. Not having meals together was a major bummer. That is usually when I meet most people at conferences. I think some kind of shared online activity that is not a meal would help with that, like an online game. |
| Pick one platform and stick with it. Was I supposed to be on Zoom? Slack? Cloudr? Gather? Youtube? It was too technologically fragmented. | A help or overview of the different features would have been great--Gather, Clowdr, Slack, Zoom, and Youtube were all in play, but I wasn't sure where to go for what or when. However, I wasn't able to take time off to virtually attend PLDI, so maybe this was more obvious to people with more time to devote to it. But I really appreciated being able to attend |
| Arrange for small groups of people to meet on Gather. At POPL 2020, there were mentoring breakfasts with about 5-6 people per table. Similar events can be held over Gather, we just need some rules on how to make them work. | |
| remotely and casually, as I wouldn't have been able to attend at all otherwise. | Every time.  Also, now days a lot of these conferences are showing up on Hacker news.  So multiple pre-conference livestreams would garner more interest while nudging authors to get their talks/demos/sessions ready early. |
| It is necessary that every tool given is accessible by Free and open source software. That's the only way to make independent and sovereign research. | |
| | More casual interaction |
| I think a Clowdr/Slack integration might have been helpful.  On Slack, I got notifications when somebody sent me a message.   Having a #lobby channel might have done the trick. | No, but if you gather good suggestions please make them public |
| | Esp. for junior folks, it might be interesting to have a chatroulette-style random meeting with someone else attending the conference. |
| Have less options and more clear instructions so that it's easy to find the information. | |
| Move to a more asynchronous model perhaps and promote further interaction in the days before and after the conference. | More is not better. Slack felt a bit overkill for announcements and questions in threads in the "official" channels. There didn't seem to be a lot organic conversation in Slack. The /video and Gather thing seemed a bit overlapping and, being video+audio, very hard to be in both. |
| The mentoring 1:1 meetings worked really well. | |
| As someone who's had to extensively work remotely with frequent travel (literally travelling every 5 days for over half a year), I found prep work to be the most effective way to make meetings more impactful.   Having said that: the social interactions should be held at least once BEFORE the conference.   Live stream the rundown of papers to be presented or demoed, and then have available presenters answer questions from a chat. Basically building hype via marketing.  The easiest way to do this would be in small multiple sessions, like a company demo day. The idea is you only get 5 minutes to present your work (so keep it high level), and 5 minutes for Q/A (expect to explain scope of work). The result will then be people scheduling sessions with intent. If presenters aren't able to make the conference for what ever reason, they still got to put their work out in to the wild.   If the conference organizers wants to have more new blood show their work, use these pre-conference livestreams to showcase grad student work from the same lab/group. (again 5 minutes, but since the work might be early / preliminary and they might not have submitted in the actual conference: no Q/A) ---  Almost all the papers I've read of my own accord (not because I was expected to) caught my eye via youtube demo videos. I suspect this is the main reason I've disproportionately read SIGGraph papers. Just because they do a sleek preview of the conference papers by concatenating a bunch of video demos. Gets me to search the authors up. | I think there were people organizing spontaneous "coffee break" or "join me for a drink" ... I did not join any but I think this could be a nice way for people to interact |
| | Using open standards. Slack is not open-source and thus lacks a lot of inter-operability features. Youtube is neither open-source not free and provides a very poor platform to browse across videos. These platforms are very poor compared to, for instance, all the activity-pub platforms (Peertube, Plemora, etc.). In the future, please use open-source, free, and federated platforms. |
| | Some way that makes meeting new people natural, maybe blocking out time for a virtual social gathering or something like that. |
| | It will be great to have a dedicated virtual space for video chatting with mentors! |
| | One idea I had was to have some sort of speed-dating online for new students/researchers. I found it a little intimidating to use the software for social interaction given that I am a young researcher who does not know many people. It would be nice for social interactions to be quickly formed in this way (preferably near the beginning of the conference) |
| | I would feel comfortable with fixed slots in time in which mentors are available to chat. Similar to AMA but not facilitated/organized. |
| | There should be more organization in virtual space.  First, we need thematic video rooms (I guess, they can be setup on Clowdr). E.g. rooms attached to sessions and meant to be used to |

discuss presentation from different sessions. So like-minded people with close interests are more likely to find each other. Also, thematic rooms suggest more research-oriented conversations than e.g. in Gather.  Second, we need carousel-like timed breakout rooms for socializing especially getting to know new people (as opposed to chatting about technical stuff relating to the presentations; although this is not banned, just the emphasis is different). I participated in one in a series of such events on Zoom during this PLDI. It worked much better than Gather. But Zoom doesn't support randomization natively. I heard PLMW used more well-suited tool called Disco (https://disco.programming.systems/) but I haven't tried it.  It'd be nice to have more time to socialize in between sessions. 20 mins for a break is not enough.  A virtual conference should be more time zone neutral. One idea: create two streaming+Live Q&A sessions per one actual conference session in two vastly different time zones. Another idea: radically smaller time span for streaming+Q&A per day, aiming left-over time for socializing and sleep.

Somehow try to simulate the bumping into people with similar interests effect of being at physical talks.

don't hold a virtual conference

Something similar to the breakfast mentoring sessions organized at POPL 2020, but done virtually in small groups on Zoom and pre-scheduled. It may also get over the issues of various timezones as people sign up based on their availability and mentor's availability.  I also heard people really liked using Zoom breakout rooms based social interaction, so perhaps use a well tested platform for smaller social interactions instead of alpha-level software.

The program of the conference in researchr was always my starting point for interacting with the conference. I really liked that researchr can now show program in my time zone (though it did not save this permanently, maybe due to my cookie policy), and I also liked that it indicated the current session.  So, my preference would be that, once I authenticate as conference participants, as much as possible of interaction goes through researchr and especially slack threads associated with the paper. That way I could see who is talking now and I could could see discussions about just given and recent talks unfolding.  Adding slack-like functionality (like Mattermost) would be awesome and I would support encourage such developments.  I may have missed this, but having links from program to portal.acm.org versions of PDFs of papers would have been great.  Then, I would suggest having an opt-in way of showing who is browsing/clicking on which paper abstracts with an option to contact them directly.

more of the speed mentoring stuff or personal interactions. Too many people to make social interactions online useful. I felt burnt out so just disengaged after the first few days.

Having clear time slots set aside for social interactions!

Don't make a conf that should've been in London run on US time.

The #mentoring channel seemed very useful, keep it!

More dedicated time for social interactions. I was watching something the whole week and almost did not have time for gathering. Breakout rooms in Zoom with random shuffling worked great -- I would like to have something like that in the future. There should be more opportunities for social interactions for people in different time zones. Maybe 1 hour sessions 3-4 times in 24 hours.

I think providing a way for people to mentor or be mentored is a very good idea. I had limited time this week to participate on mentoring sessions.

No, I was happy with the mentoring but sadly did not get round to organising meetings with possible mentors.This was a timing issue on my end and nothing to do with the format.

Conference should explicitly schedule time for breaks/social interaction. Often, it wasn't clear when people would be in Gather or not. There were some planned Zoom calls; I liked the shuffle into random breakout rooms. Would also be nice to have scheduled Zoom rooms based on topic/session/interests.

Nothing comes to mind - the services provided seemed quite comprehensive

The Ask Me Anything sessions were great. ISMM's 1-on-1 sessions were also great.

- the #mentoring was great; but as Loris D'Antoni suggested we might decouple them from PLDI and just have "virtual mentoring days" scattered through the year.
- I think in-person conference attending should be kept in place, the virtual experience can exist in parallel with the in-person attendance.
- I love to watch live videos of the presentation in YouTube. However, I'd enjoy more if we use YouTube to post question and see other people interactions. I personally didn't use slack and didn't like the idea to jump in in a talk and have to go and open slack as well just to be a part of the conversation.
- Yes I have. Try to do more conferences to enlarge our knowledge.
- I think organizing one mentor : small group of mentees would be something to consider in the future.
- Make sure they do not require running proprietary code (incl Javascript) on the client machine
- Organize theme based birds of a feather
- Physical meeting/gathering
- Have more time dedicated in the conference schedule for socialising; this was OK on the last PLDI day, with the organisers making announcements on Slack
- I really love the AMA sessions! I'd love to attend more events like that
- Provide dedicated video chats with the authors after the talks for informal interaction. Have clear time slots in the schedule where people are encouraged to hang out together. It would be nice to accommodate the different time zones a bit better.
- I prefer the style of Zulip with threads easily visible for this. someone looking for mentoring had to scroll a lot to search for mentors.
- reserve some timeslots during the conference for gather-ing. The AMA were great, but the schedule of the conference was too dense.
- Rehearsal is suggested to reduce technical-related problem during presentation.
- Having a list of mentors' names available in advance, so I can plan better in advance!
- perhaps people could post (anonymously) what their concerns are, so that people who can help, can volunteer. Eg I did not offer mentorship, because I thought there was not much I could help with.
- Even I didn't mentor or be mentored with that channel, I would like to say this is one of the best parts of this virtual PLDI.
- dedicated time for everyone to go to Gather. Rounds of speed conversations where you are paired with 2-3 other people for 10 minutes at a time.  PLMW small groups was great, but did require more up front organization.  I'd like to see both those types of opportunities, and also more organic ones that can be organized on the fly.
- Make open to join interactions more visible. The barrier to approach new people online is much lower if you see people already talking. Not even for helping people with possible shyness, but also because logistically setting up conversations is quite annoying
- We should have some such universal Slack workspace for PL. Though the option of sending mails and scheduling meetings is always there, I found that people are much more interested to join such kind of explicitly visible opportunities to converse.
- Slack all the way.
- None
- There could be more break time, poster presentation time, etc.
- My research area is actually falls on PLDI areas, I mainly attended as it was free. I liked the "Programming Language Mentoring Workshop" a lot. This was the most informative and helpful session for me.
- Maybe have less ways to interact. I was unsure whether to hang out in clowdr lobby or gather or use slack. (I mostly used slack to find people)
- More events in the Gather space, and better integration with the slack channel to make easier for participants to enter gather space easier.
- Some number of (randomly assigned (?)) persons should be in the Zoom call, in addition to the session chair and presenters. I would recommend a discussion system that is ongoing even outside of the conference (the slack is a good start).

| | |
|---|---|
| Someone suggested a round robin of video chats just to introduce new people to each other, I thought that could be a good idea | the conference sessions and have little time for social. |
| More chats | Something like /gather for random 3, 4 people chats, but with **text** chats. |
| a more structured approach is needed. E.g. specific sessions/timeslots etc | Many students asked the same question. Having combined sessions would be better. E.g. a virtual round table (like gather), where a mentor announces that (s)he would be available. |
| I really liked the 1-on-1 meeting sessions that ISMM did, where they assigned people randomly for 1-on-1 meetings. I had a chance to talk to a lot of people who I may not have otherwise talked to. ISMM only matched up students with mentors, but I think it would have been interesting to talk to a few students as well. I actually would love to see something similar in a physical conference as well. The only improvement would be to make it clear that the assignments are random, because I had a few people who thought I had specifically picked them. | CAV used to do pre-conference pairing of students who are attending first time with other students or mentors |
| | It would be nice if there is someway to really encourage everyone to use gather at the same time. It's much easier to initiate an interaction if there are a lot of people there. |
| | Creating focused "virtual coffee" meeting to meet people I didn't already know was super useful. |
| | I'd like to approach a mentor but I got too shy for this :( |
| Yes: we should not have virtual conferences. For the duration of COVID-19, we should switch to a journal format and try to find online native social interaction mechanisms (a SIGPLAN forum or chat?) instead. The virtual conference experience was horrible, and much worse than simply blocking off time to read the papers. We should never do it again. | Better UIs, fewer registration hoops. For instance, in order to get into the https://gather.town/ space, you need to jump through ~7 UIs:  regmaster.com -> slackinvite -> slack -> type magic `/gather` command -> get gather.town link -> register there -> login there, and those who forgot the register on regmaster.com in advance (at least 2 people I know) didn't manage to access it |
| Last year I met a bunch of people at (a) the conference lunch, because you tend to end up sitting with at least some random people; and (b) at the pool hall night. Gather gives a little bit of (a) but it's not the same -- someone proposed setting up random discussion rooms and I think that's a good idea. It would also be good to have some kind of explicit social hour like the pool hall night -- not sure what but basically something fun people can do while talking and mixing, that doesn't take that much attention but enough to serve sometimes as a starting point for conversation. Few computer games are suitable, unfortunately. Also whatever it is would want to be stuffed into Gather and/or Clowdr, which makes it all a pretty tall order... sorry I have no specific ideas. | As much as possible, have a _single_ modality (as opposed to four: Slack, YouTube, Gather, Clowdr). Make sure there are "locations" that sort by interest/topic, and make it accessible from researchr. |
| | I think it is harder to rely on accidental encounters. Therefore efforts to allow people to discover others who share interests before the conference would be advisable. Efforts to match people based on predefined interests would make a lot of sense. I also found it really hard to interact with authors of papers |
| | Have clearly scheduled socialization breaks; probably half of the program time |
| These activities should be available not only during the conference, but also circa one week before and after the conference, making more people from different timezone to meet and have deeper communication. During the conference, I am more preferred to participant | Make the (initial) list of mentors available before the conference |
| | I was nervous about providing mentoring, because I did not want to get inundated.  Some limit on this would have been good. |
| | Still an open question. |

## A.5. Responses to "In what ways did PLDI 2020 fall short of a physical conference?"

| | |
|---|---|
| It lacks some of the natural socialising mechanisms, such as lunches and dinners, that make even reticent attendees socialise. That said, I think the organisers of physical conferences often don't realise that there is a long-tail of participants who don't actually interact much during events. | I still prefer physical meetings for small groups. |
| | Lack of meaningful interactions. You can't really replicate going for a dinner or beer with a group of people, or having lunch with random people. |
| social mingling | Social interactions, although it is admittedly a genuinely hard problem. |
| Informal interactions with other members of the community proved nearly impossible. | - Harder to pay attention in the virtual setting. - Fewer instances of serendipitously meeting new people, in hallway interactions and at meals, going out for drinks, etc. - Occasional technology blips like video cutting out made conversations a little less smooth when they did happen. - I liked the AMA sessions, but missed having more visionary, big-picture type keynote talks. - Co-located events were not all as well-organized as the main PLDI. |
| No easy way to have serendipitous interactions in the hallway. I just wasn't interested in pushing an icon around on a screen like in a video game so I didn't. I just didn't spend much time at PLDI so I didn't get nearly as much out of it as if I had travelled to London. | |
| People not being in the same time zone. Many "social" things in the late evening for Europe, where the conference was expected to be located. | Social interactions |
| | The physical conference provides a way to stay on the periphery and still make new contacts. Further, the multiple tracks of the physical conference forced me to prioritize the talks I wanted to see in person (this PLDI was intense when I tried to go to all the talks, and I entered brain-dead state a little earlier). |
| More spontaneous social interaction, the ability of people I know introducing me to other people without me asking (the online setting doesn't really allow for that), and the last big one: ATTENTION. If I'm home, there are a hundred other things to do. I cannot dedicate all my attention to the conference and the interactions therein. | - Social interaction - Adjustment to local time |
| | not getting to meet the attendees... |
| I wish there was more interaction on Slack. There could have been more informal spaces to discuss talks via text and meet people rather than the video chat options being the main choice. I appreciated the PLMW channel for this. | human interaction is best in person |
| | It felt very hard to get involved. Being at home, I had many things distracting me. In person, I am committed to the conference. The fact that it was free and I paid nothing for travel meant that I had "no skin in the game". |
| Generally speaking it is easier to talk to people in a physical conference. | Very hard to carve out time for the conference when I am still sitting at home with other obligations. |
| Social interactions were hard, especially for outsiders (I haven't been active in the community in ~8 years). | |
| Virtual PLDI was much less immersive and much more distracting. | Many fewer random meetings with colleagues. |
| Difficult to meet people that you didn't already know. | Exploring a new interesting locale and its food and beverage scene with new friends is an advantage of physical conferences that I don't think would be replicable for virtual conferences. |
| Hallway interaction, dinners and social activities. | |
| Peoples times were out of sync, leading to less incidental meetings and discussions. | Gathering enough momentum to be able to block hours of online PLDI time for attendees. It |
| is very hard to spend anywhere near the same amount of time connected with fellow participants when sitting or standing at a computer desk anyway (or any connected environment, really). | we should give the impression that we encourage this through the way we structure our meetings, allowing attendants to impose this onto themselves. In addition to that, some talks happen simultaneously like a normal conference would make it obvious that not every talk needs to be watched. Finally, inside the 8h should an offline break time in the middle, so one get / make something to eat outside of the conference, stand up and walk around, e.g., a time where no talk / AMA / sponsor-event is happening and one can leave the computer. Finally, we need official designated times with no conflicting other events, where people who want to talk to someone, can stand next to their poster, such that other people who want to talk to someone, can look at the posters and talk to those people that have a poster whose content looks like something one might know something about. |
| The social aspect fell short of course, but all things considered it was a good first attempt. | |
| It felt like there is a bunch of high-profile people+cohorts that all know each others and enjoy connecting with their buddies, and newcomers/outsiders/introverts may have a hard time finding a way in or feeling to belong in the community/event. This is probably true for physical conferences too, but online makes it much more easy to lurk and to ignore who you don't know (they are just absorbed by the huge contact list and you end up searching for who you know) | |
| It's hard to reserve time for a virtual event. | |
| time zone | |
| Lack of long conversations over dinners with new and old colleagues. But this is quite obvious. Virtual PLDI 2020 was mostly amazing! | Watching recorded talk is too much of a strain |
| | It's harder to have "hallway" discussions, the discussion in Slack is slower but often harder to keep up with due to much higher number of participants. |
| Listening to talks on youtube is suboptimal. | |
| Impromptu Interactions were limited. It was difficult to meet senior folks. People were primarily chatting with their cliques. | Informal interactions with other people -- online mechanisms seem a bit forced Ability to dedicate time for the conference while working from home Inconvenient times Uncomfortable to attend a conference for an extended time in front of a screen |
| No outside-of-conference kind of interactions: Dinner, running groups, etc | |
| perhaps the small talk during breaks was the major things I missed, otherwise, I believe it worked even better than a physical conference! | None that I think of except maybe I did not get the chance to meet the speakers. |
| I didn't get to talk to people physically. | Obviously interactions. Gather had a very small crowd. Also, for me a big part of interactions at conferences was about lunches/dinners, coffee breaks, drinks. All of these were lost. |
| See also other free-form text-fields... Regarding the scheduling below. I want to stress that a conference should similar to a normal working day be limited to an 8h span per day max, counting breaks, so actually conference time more like 7h, as to help work-life balance. 8h work per day and then END. I don't care whether that is at night or at day for me, because I can just sleep at day and wake at night, which would normally occur because of travel jet lag anyway, online time should be limited to 8h max. If someone would hypothetically require you to work longer than 8h a day for a whole week that would be unethical, and although we did not require this, | |
| | Colleagues and family did not understand that I was at a conference and thus do not reduce the expectations during this period. It makes it difficult to have time to read the articles and follow the talks. Social interaction and technical discussions are also much more difficult. Accents are also much more difficult to understand remotely than in person. |
| | I found the virtual social interactions somewhat frustrating, and definitely nowhere near what I'd get at a normal PLDI. |

1) Tools for live interactions were not mature and because of my lack of experience with such virtual events, I wasn't able to make an optimal use of it. I expect my experience would improve over a time.  2) Timezones.  3) Screen time. This was a big major problem for me. On a physical conference, one can walk a bit between talks, grab a coffee or a tea and meet people when doing this. On a virtual conference, one has to stay in front of a computer to be able to interact with others. I found this exhausting in a longer term. And it is even more noticeable for lunch breaks.  4) The lack of social events. Social events (dinners, tours over a conference place) are good opportunities to meet people and again, it can be quite exhausting to have these conversations in front of a computer.  5) I believe that a virtual communication will always be (more or less) inferior to a physical one. One the other side, no communication is inferior to a virtual communication and a virtual conference makes it possible for more people to communicate.  And of course being in a bad timezone amplifies 3), 4), and 5).  In my opinion, many of this can be solved and a virtual conference experience can be close to a physical conference (or even better) in many aspects. On the other side, I don't think it can match for a physical conference in all the aspects so I think that alternating a physical and a virtual format is the best solution.

I did not get to meet people.   The prerecorded presentations of the main program were not engaging. (But often I also have a hard time following talks at physical conferences.)

I miss seeing people in person. And I miss being forced to leave my home environment, with all its distractions.

Near-zero social interactions. Near-zero synchronization with others. Maybe it's just me, but I had a very limited attention span for papers. Once I got the idea, I knew that I could watch the talk whenever, so I just tuned out. In the end, I felt that compared to a physical conference:  - I learned much less about what everyone else is doing.  - I met near-zero new people  - I left the conference with no extra inspiration (whereas beginning-of-summer conferences are usually a big source of inspiration for me)  - I learned nearly nothing on the technical front.  - I didn't meet "my tribe", reconnect with old friends, or otherwise enjoy myself.

I was not able to adjust my working hours to that of the conference time zone, which I expected to be the timezone of England.

spontaneity of social interactions

I did not have my full time allocated to PLDI, so did not interact as much as I would have otherwise. Also, my attention span was divided between PLDI and other activities. But without virtual PLDI I would not have seem any of it.

I didn't meet as many people as I would have: the schedule for each day was very long, and I had a lot of screen fatigue and didn't have the energy after the sessions to spend more time on gather/clowdr. If it's anyway virtual, I think the sessions could be spread out over more days, and grouped around fields. Then on days when I'm not interested in the talks I can make more use of the interactions. Lastly, I have to say I did miss the fun of evening social events (dinners/drinks) that would have happened in a physical conference. :)

Felt difficult to meet people

The immediacy of "being there" vs. just another thing I can tune in.

It manly lacked informal discussions that we were possible in person.

It was still a bit difficult to interact with people. There is no question about it. It was my first PLDI and I also got to present my 1st paper. If it was physical, my advisor would introduce me to a bunch of people there that they already know. But that was not easy in a virtual setting.

hard to meet new people! hard to focus on talks!

I think there is no way to replicate the social experience of an in-person conference online. PLDI 2020 came closer than I expected, but it is just not possible. Moving conferences to be fully virtual would be taking away one of the key benefits to younger students and replacing it with something strongly inferior. Maybe once we have perfect virtual reality, but until then, it only makes sense during a pandemic.

Not forcing me to relocate away from my normal duties meant that I did not care out the time to actually attend the sessions.

| | |
|---|---|
| I wasn't able to meet new people/renew connections with acquaintances as strongly as in person. | lack of attention listening alone at home. Lack of communication with other people. |
| Coming on the heels of a difficult spring, spending many more hours on Youtube is hard. | Lack of physical presence led to zero random interactions. Hallway track was nonexistent |
| Because I was physically in my home during the conference, it was difficult to shut off all the usual "at-home" interruptions.  I like being away from home for a conference because I can devote 100% of my attention to the conference.  I could never devote sufficient concentration to the talks or to the socialization. | It's not the same thing. We should have optimized further for a virtual medium, e.g. more time for questions, more use of slack. |
| | Unclear social components, especially for folks who are visiting this community.   Inability to see someplace new - different pixels on a monitor do not constitute someplace new. Also, virtual is a pathetic reward for students who worked hard on papers, etc. |
| Honestly, I think the committee did an excellent job making this virtual conference happen and I applaud them for it.  I still miss some things about the physical conference --- I still would have had more conversations with people / hangout with people more if the conference were physical (just because I still feel more comfortable chatting with people in person than over video, especially with people that I am not familiar with).  However, I also see benefits in having virtual conferences (lower carbon footprint, lower cost, and the recording may allow wider audience to the recorded sessions). | Hard to interact with other attendees in different time zones. |
| | Hard to attend all talks due to timezone differences, little less possibility for "hallway meetings" (but gather & slack helped A LOT!), but I think this can be 'learned' over time! |
| | I felt that I missed out on the social factor. I think the conference were probably split between too many medias. |
| | Talking to other attendees. But I think something as simple as forming groups with different backgrounds to rate how cross-functional a talk was would remedy the situation. (i.e. you would have to discuss and explain how things are in your field for the others to understand) |
| The 12 hours were more biases towards US attendance.  It would be good to be more inclusive of all timezones.  This can be done by having 3 different regions where papers get presented during the normal time in those timezones.  There have also been suggestions of having a second Q&A session at a different time so that other people can ask questions. | Casual interaction and easy signposting |
| | Nothing, except the lack of pressure for introverts to get out of their comfort zone and interact socially. |
| I only tuned in for a few hours.  Had I attended the conference, I would have spent more time. | - I missed socializing, though /gather helped.  - I missed, more, chatting about research ideas with experts in an uninterruptible fashion. Not comfortable enough yet to do that; don't know the virtual social cues. - Hard to get focused on the event, when not away from my normal situation. |
| See this Tweet (https://twitter.com/hgoldstein95/status/1274355407138557952?s=20) and the related threads. On a high level, I think virtual conferences make it harder for junior researchers to form their network/community. They are also *much* less fun, which while I think counts for something. In general, I think doing away with physical conferences would do a huge disservice to anyone who doesn't already have an established network in our community. | Not everyone is forced to be in the conference space / gather. So the number of people you accidentally bump into is a smaller fraction of the people who would be attending the talks and other events. Also not much of a chance to hang out with conference attendees for social events outside of the conference. |
| Same timezone. | 1) It was hard to pay the conference much attention. It just becomes another part of the continuous data stream of Slack, Twitter, |
| YouTube etc  2) It is pretty much impossible for young researchers who do not already have a network to make new connections in this setting. I don't feel I'm part of a community of researchers if I'm passively hanging out in a Slack channel and occasionally watch a YouTube video. | harder to devote time purely to the conference given the distractions of home. |
| | Screens being involved made participating for an extended period of time tiresome |
| Networking ("hallway track", and evening social activities). It was also more difficult to disengage with local activities, since attending PLDI didn't really take me away from my day-to-day job. | No means to escape any ongoing obligations. Between children, research, meetings and teaching there was very little time for the conference. Social participation was nearly impossible. |
| No chance for social interaction with the authors/speakers you do not know right after the talks. | It fun, just technical stuff around interactions have to be polished. |
| | - difficult to travel - difficult to waste the time for travel - difficult to approach people with flexibility |
| The social aspect did not work at all for me. This may well be my own fault. I was very much less fully engaged than I would have been at a real conference. It is very easy to just do stuff on the side and forget about the conference. | Absence of the workshops |
| | absence of the workshops |
| Time zone. I completely skip the last sessions | A physical conference forces you to make being at the conference your primary activity.  With a virtual conference it was too easy to let the conference become secondary to all the normal demands/meetings of my "real job" and just watch the 15 or so talks that I was interested in seeing and interact with a few close friends. |
| Time zone. I wonder who thought that using Pacific Time would be a good idea for London. I only see that as another attempt to centralise Computer Science around the USA, despite all the criticisms that has been expressed in the previous years against this centralisation. | Social interaction (going for dinner, having drinks, etc.) Experience giving a live presentation. Distraction when listening to video talks. |
| It was hard to meet people "by chance". | face to face meeting |
| Randomly (or not) running into people; being in the same timezone as the conference (many events were inaccessible due to their hour); being able to fully focus on the conference | In every way. Virtual is okay for established people who already know each other. My students attended virtually (all of them, since it was free) and they were totally disoriented and they said they didn't meet anyone new. I didn't talk to anyone that I didn't know already either. |
| Only thing that was really missing was city exploration and getting meals with friends. I did not find online PLDI to be bad in any other way to be honest. Organizers did a fantastic job of making it excellent. | I was exhausted with the number of talks I was interested in watching. Plus given it was virtual, I couldn't really take time off work so it was intermingled throughout the days.  It was easier in the physical conference to just join a group going for dinner for example. It takes more initiative to join a social group in the virtual space. |
| Physical interaction; time zone issue; Hard to fully concentrate due to your regular day. | |
| It was challenging to meet people.  I had only a couple of exchanges of messages, whereas at a physical conference I probably interact with 40 people over the three days. | |
| It's hard to match the impact of a physical presence on the environment. | There's an inherent tradeoff between it being easier and cheaper to attend, vs. having the discipline to fully commit to the conference for 3+ days. |
| 1) Social interactions were really lacking as compared to a physical conference, especially for myself as a younger researcher. 2) It was | I was still in my apartment with distractions around and thus participated less than I would |
| have otherwise. But I can imagine on another occasion isolating myself better. | what was going on when and how I get there. The Slack pinned messages and announcement channel did not compensate for this. Unfortunately, attending PLDI 2020 felt more like having an additional social network for a week than attending a conference with a clear timetable and events happening. |
| I think it would have been easier to network at a physical conference | |
| It was very hard to initiate actual stimulating conversations on various ideas. Even on gather, the few times something started one would quickly get interrupted.  It is really hard to be motivated to watch a video talk. | |
| | It is difficult to work and follow the conference at the same time |
| It was too overwhelming and too many people online virtually for me to make meaningful interactions with people | Timezone differences. Some of the sessions happen in weird time for some people |
| | Talking to people |
| social interaction, catching up with friends | Less interaction with colleagues. Harder to find time to attend talks (particularly due to timezone differences). No change of scenery. |
| The social aspect is much worse comparing to a physical conference. | |
| Feeding question to authors through a moderated Slack channel worked remarkably well under the circumstances. The one thing missing, I think, compared to a questioner standing at a microphone was the inability for a questioner to ask an immediate follow-up question.  So some interactivity was lost. | I got into fewer random hallway conversations (but at least some). |
| | Social interactions, and too much screen time per day. |
| | social interaction as well as the primary attendance interface: I spend nearly 15 minutes switching between Slack and the website just to find the youtube link. And I'm not the only one I knew to have this problem. |
| People did not walk up and sit next to me. This is how I normally interact at conferences. | |
| People would quickly enter Gather then leave--I certainly did. I would look at the list of people to see if there was anyone I wanted to talk to, then leave if none were there. | Even though the organizers did a fantastic job, it is pretty difficult having many social interactions in a virtual conference since it is very easy for people to just tune out right after the talks. |
| Social interaction | |
| Unfortunately, I was very much disappointed by my experience attending virtual PLDI 2020. I found Slack a completely inadequate platform for running the entire conference. Focusing on Slack resulted in a lack of clear information about the timetable (what is happening now, next?) and where to find the video streams for the talks. The entire experience felt very unstructured. The PLDI website which had a much better way to show the time table was completely unused - e.g. there were no links to where to see the streams on the website. I found the Slack interface for threads is terrible to use. The mixture between some threaded and some unthreaded conversations just adds to the confusion. That Slack allows to write the /video and /gather commands in any channel and they appear as messages that only you can see is a terrible confusing interface. Overall I was missing a webpage that clearly explained | Personal interactions are too weak, and I can't focus my attention during talks |
| | For me the main issue was the timing, as having children meant that the majority of the evening sessions were not ideal and alternate arrangements were not available. |
| | Social interaction and my full availability to concentrate on this while being at home. |
| | Social interaction, face-to-face, for obvious reasons. |
| | Gather is not well advertised |
| | Harder to interact with other people. |
| | Not traveling did mean a time zone difference, but being able to rewatch the past YouTube streams helped to mitigate that. |
| | Social interaction, keynotes |
| I interacted with way fewer people meaningfully and only met a single new person that I didn't know before. | I refuse to run proprietary software on my machine, so I could not participate in discussions (because of the use of Slack) |
| In my limited experience of attending physical conferences, interactions with other research groups at large specially at dinners is something that virtual conference cannot replicate. | sometimes the social interaction can be good at a physical conference |
| | Social opportunities are poorer |
| Difficult to reconcile time zone of talks with my own time zone | It is much harder to have informal chats, dinners with people, etc. My experience of conferences is that they feature an awful lot of peer-to-peer learning, as well as being something of an "immersive research retreat" for those travelling. PLDI online struggled with both of those things. |
| Gather < Hallway | |
| Random interactions, sequestration -- A virtual conference overlaps the rest of my life, and has to compete with everything else.  When I attend a physical conference in person, I'm committed. | |
| | Lack of engagement since it is virtual |
| I think the main reason why I had difficulty socialising was that all the socialising happened when I was asleep due to time zones. Also, like every conference I have attended so far, I was surprised at the low quality of the talks. | being able to focus on the conference, on the other hand, it is not my main community |
| | People not being in the same timezone fundamentally limits interaction opportunities (which, honestly, are the main point of the conference; i.e. talks are mostly conversation-starters, not the main attraction) |
| the "hallway track"; part of which is "synchronized breaks" and bumping into people, but I'm *very* optimistic this can be improved in future iterations. | wifi sometimes not stable |
| | The primary benefit of in-person meetings is serendipity. That kind of fun, accidental discovery usually happens in hallways and over dinner.  While I completely understand and support the organizing committee's decision to cancel the in-person meeting, I was very disappointed that we had to go virtual. |
| Interaction, inspiration, discipline to attend most of the talks, research-inspired international travelling | |
| Meeting people | |
| Social interaction (but if many more people had used Gather or similar, would have been closer) | |
| Interaction | hallway conversations, hard to justify schedule changes at home or work for a virtual conference |
| To meet people in person and talk about your work to make a real connection. It's not the same in a virtual chat, they serve different purposes in my opinion. So I'll have the conference physically and keep the virtual chat going. | Not being in the same time zone as the event is a disadvantage. Going out together for dinner and drinks in another city was missing. By not traveling and taking time off for the conference it was hard to balance the normal work week with attending the conference effectively. |
| | physical appearance of a speaker was missing in virtual conference |
| No comments | |
| I think it was harder to informally interact. | There is a strange difference between the questions that are asked live, and those answered on slack. The former feels more satisfying, for some reason. |
| I did not have chats with other participants. This tends to happen naturally by 'bumping' into other people. This may be because I felt an inhibition to use gather/clowdr, whereas it is hard to not meet people at a physical conference. | Spontaneous interactions with new people was almost impossible |
| | being at home is impossible to engage full time with the conference |
| Face to face interaction and socializing, opportunity to travel | |
| Not being fully emerged into the conference; Fewer social interactions; Time zones limit who you can realistically meet (which may have a strong negative effect on research in the long-term) | Social interaction |
| Interaction with other participants is still not as good as in physical conferences (partially due to not everyone using gather/slack) | The largest problem was the time zone. I couldn't attend half of the events that I wanted to, because they were throughout my night. And this cannot be avoided for people from certain countries if they are attending from their own time zones (though the OC should perhaps consider changing the times a bit in future virtual conferences based on analyzing the stats about number of participants from different countries). |
| Serendipitous interaction with a broad swath of the community. | |
| The timezone difference made things very difficult -- existing in 2 timezones 5 hours apart at the same time just doesn't work that well. | |
| A physical conference is a "festival" of the discipline, so has relaxing moments and aspects of leisure as well as work. Lack of physical presence makes attending virtually -- like Zoom meetings etc. -- inherently fatiguing, at least to my brain | As an introvert who hesitates in approaching people, I did not take part in any person-to-person conversation. While in any physical conference, being surrounded by people takes away some of hesitation and I usually talk to some people. But this is something I should work on |
| Somewhat lack of immediate interactions. | Not at all |
| Don't know; this was my first conference. | normally, I really enjoy chatting at poster sessions and conference dinners/social events. Other than that, attending talks via Zoom is absolutely OK, and sometimes even better. |
| Self explanatory | |
| social interaction, and too many distractions from home | Participants are in different time zones. |
| Gather / Clowdr isn't as good as physical walking around. | It was too hard to separate being at conference from normal life/work |
| time zone was one issue with virtual PLDI | I did not talk to anyone. I was not forced to do that. I am very uncomfortable at normal conferences and always need a week to recover afterwards, but, because my institution spent money on it, I do my best to make it count. With purely virtual setting, I can hide behind my screen and avoid talking to people. I can also make excuses that I have to get work done instead of participating. Not great for perhaps "socially awkward" people like me. |
| It was difficult to structure my own time without the strong signals of "session is starting now... now it's lunchtime... etc."  And it was MUCH more difficult to find people. | |
| From my perspective it was better (time-constrained industry person - I probably wouldn't have come across/found the time to read the papers, but the fact the conference was happening at a specific time meant I took the time to watch the 20 min presentations, and this enabled me to quickly find the subset of relevant papers) but I know I am not your core audience! | Gather is cool but will never match actual hallway meetings. |
| | I find it hard to focus on a prerecorded talk, unless it is very close to my area. Also You don't get to travel |
| No physical distance from normal day-to-day work. No physical meetings with others (which is very humanizing and cannot be replaced by current technology). There is not much opportunity to meet people who are normally in very different timezones. | Social interactions. |
| | I found interaction harder to get started. This may be because it is new. |
| | Of course, missing hallway chats is a pity. |
| Social interaction. Poor fit with local time zone. | everything is on in the middle of the night |

Social events live. A community needs to share real live, occasionally.

Despite the great efforts to virtualize the hallway track to me it is not the same and did introduce new hurdles to speak to people. Also attending the conference from home did not help in my discipline to really attend the sessions.

"Coffee breaks" should be synchronised, so more people would be on Gather at the same time.

Lack of networking, but this was well compensated by (1) the comfort of not having to travel and (2) the slack channels, which should imho remain active in perpetuity after the conference in some form or another. I really hope you will save the contents and make it available to the community.

I felt I did not connect to as many people as during a physical conference. The schedule was also clearly arranged to accommodate American / European time zones. In Australia it was a struggle to attend the sessions live.

Not enough social interaction

No "chance" interactions. Significantly fewer opportunities to interact with junior researchers, PhDs.

Not much interaction with other people

It could not force people into the same time zone, making social interactions between groups of people from widely different time zones much more difficult.

The Gather space did not fully fill the void of an actual hallway that you can talk to people in. I generally found that the gather space was mostly empty (or filled with people with video & audio disabled). I also missed not having a dedicated time for posters: I spent a few hours around my poster to see if anyone would come over to look at it but only had 1 person visit it.

Almost no interaction with other participants other than speakers.  The conference has to interleave with other obligations, which is stressful.  It is harder to concentrate on a video talk than a live one.

Timezones

PLDI 2020 failed at the fundamental purpose of a conference: it was super hard to find anyone to talk to.

I think there is a risk that people do not pay so much attention and work on their own projects during a virtual conference, as they know that everything will be recorded and they can watch the talks when they have time. Physical conferences don't offer this possibility, so they require us to be more focused on the talks and on social interactions during the conference.

In-person interaction, odd timing

Everything that needs engagement. The talks I watched were quite good, but I watched a lot fewer because of the missing "well, I'm already here"-effect. The social aspect also fell very short. Partly that's because of my time zone (Central Europe), so I usually was in bed for any virtual post-conference activities. Partly also because I felt quite uncomfortable approaching people on Slack - Gather might have worked better, if it had been used more.

Lack of interactions, networking. The tools made available were reasonable, but still not ideal and anyway interacting on a computer is rather annoying. We are also probably tired after being confined to this way of interacting for so many months.

Much harder to engage speakers after their talks. The social tools are not (yet?) an adequate substitute, even though they do have certain pluses, and note that Friday was much better than Wednesday which was also much better than Monday, as more people figured out how to use the tools and they started to gain critical mass. Also it's much easier to accidentally miss talks while in the hallway, because there isn't the effect of everyone getting up and going in. And then, unrelatedly, I've found it's much harder to concentrate on dull talks by livestream, even though I carry most of the same distractions with me in person. Not sure why this is. (And some talks will inevitably be dull for some people, that part can't be helped).   Also I found the constant alternation between talks and AMA draining -- maybe the answer is to skip more of them but that doesn't feel like the _right_ answer. I think there should be more time periods that are just breaks or specifically tagged for socializing -- there were a few but not enough. Even inserting five minutes at a time here and there

| | |
|---|---|
| to get up and make coffee and hit the restroom would be useful. | I think that the hallway chatting was much worse. I think that pushing events into gather or clowdr to force people to be present for those kinds of events might have made people stick around. e.g. people could have asked questions for the ama in some area of gather, then after the talks they would have already been present there and likely continued chatting. Maybe also having something like voice chat rooms like on discord could be a way to improve this |
| I think it's pretty cool actually! I wasn't as focused as in a physical conference though, because if I attended a physical conference, somehow I am "forced" to attend talks or talk to people but if it's virtual, I may alternatively choose to work on stuff or do something else. This may be due to the fact I personally prefer things that are more live -- since PLDI 2020 only offers recorded talks, I found that less attractive (but I enjoyed live QA, PLMW, and AMA very much!!) | |
| | Time zone. I had to sleep when most the conference was going on. I find it strange that the conference was tied to a that specific time zone, when events could have been staggered throughout the 24 cycle. Accidental encounters were gone. I also found it very hard to discover work outside of my immediate interests |
| It is not very friendly for people from some timezones (e.g. in UTC+8, we have to stay up very late if we want to participate the afternoon sessions) | |
| | Social /gather and /video is not the same as in a physical conference |
| ability to socialize with friends. But may be we should have other venues (such as workshops) for this. | no convenient excuse of eating for socialization |
| | Socialization |
| For me it was better than the physical experience. Getting acclimatized to the atmosphere and food habits of a different country takes some effort for me. | The social interaction was really lacking. |
| | Social interaction |
| I know the organizer tries really hard to improve this, but the virtual PLDI still lacks the atmosphere of social interaction. And since we are all at home, it is somewhat difficult to focus on listening to talks and sometimes distracted from other business. | Social and networking |
| | I would prefer live talks to pre-recorded talks (although I understand that the latter are safer) |
| some of the talks are not clear, perhaps the authors should upload better videos. | Not everyone was in the same timezone. Also, it would have been better if there was a video conference with chat setup for people who wanted to continue talking about the last paper being presented. |
| Social interaction, inherent "push" to engage with talks and ask questions | |
| Fewer hallway track conversations and meeting people I don't already know. | Hallway track leaves much to be desired- spontaneous meetings are nearly impossible. Dinner/tourism is a big part of PLDI that can't be replicated |
| I was distracted by the rest of my life, since I wasn't at the venue | |
| | In physical conference, you are more like to interact with people. |
| fewer spontaneous interactions, inability to do necessities such as walking, eating, waiting, etc together with other attendees | I think part of the value of a physical conference is blocking out time to attend an event; it's hard to do that virtually. |
| In virtually every way | |
| Very few accidental interactions. I can't get drinks with my friends. It's very easy to skip talks. I don't normally have my partner and child with me. It's harder to beg off other work obligations when you're not physically away. | social aspect |
| | Social interactions |
| | There was a lot of content at the conference itself, and as I had to actively initiate social interaction (e.g. by actively deciding to go to |
| Informal personal communication with attendees during lunches, dinners, and coffee breaks. | |

Gather or a video room), I ended up forgetting that in favor of following the presentations.

## A.6. Responses to "In what ways did PLDI 2020 exceed your expectations?"

| |
|---|
| I think combining a core of proven technology (Youtube, Slack, and to some extent Zoom) with younger experimental services like gather and clowdr was bold and effective - certainly much better than another virtual conference I attended.  The organisers clearly put a lot of work and passion into pulling off the conference in difficult circumstances which I think was key to making it work this well. |
| own-timed viewing of talks |
| The Q&As are more accessible. The per-presentation offline discussions continued easily without time limit. |
| The Zoom coffee hour, mentoring channel, and AMAs really exceeded my expectations. The most significant value of attending PLDI seemed to be the encouraged interaction with other researchers in the community, since the majority of papers and talks now are published after the conference anyway (not to say these aren't valuable, but the social interaction is hard to obtain any other way), and all three of these provided a fantastic and welcoming platform to meet and learn from both other students and senior members of the community. |
| The paper presentations were very polished, and the Q&A format worked well. The AMA sessions were also quite nice. |
| Mentoring worked extremely well - keep doing that! |
| It was extremely well-organized! |
| I think the team did very well putting things online on short notice. |
| I got more chances to talk to professors and other senior researchers. In a physical conference, students usually spend most of their social time with other students.  The AKA sessions are great. |
| The slack channels for questions were far better than my experience of live conference questions.  Also, the control over my schedule for watching videos was really nice. I think gather was really promising, just not quite there yet. |
| Very accessible. Nice with the slack channels for questions. Easy to just take part in the pieces I was interested in, without having to spend a lot of time travelling. Great to be able to listen to talks at higher speed to get an overview, and to skip things I was not so interested in. |
| Loved the AMA sessions! |
| The talks are extremely clear. Asking a question is now very easy as well as following it up in the slack thread is very very helpful/productive. |
| The talks generally went very smoothly. It was very easy to attend the sessions or catch up when one was late. Both session questions and Q&A's worked really well. |
| The possibility of re-attending talks allowed me to attend most of the seminars I was interested in.  I also hope they will be kept online in the future, so it will be possible to re-watch them. |
| I enjoyed the Zoom meetings run by Rachit. |
| I liked the fact that when I felt like it, I could have tuned in and watched a talk from my couch. |
| Interaction with the authors, academic/professional interaction, *excellent* organizationally-wise (thanks to the efforts of organizers and student volunteers it was really easy to find out "what's going on right now", including active sessions, Q&As, events). |
| - The Slack seemed more bubbling and engaged than the virtual ASPLOS Slack. - AMA sessions were great and super interesting to watch as a junior researcher. - Moderated Q & A seemed to improve question quality, fewer "it's more of a comment than a question" type responses. - Mostly single-track made it easier to watch all the talks I wanted to see live. - Video abstracts were awesome, great to be able to get high-level overviews of lots of talks before seeing the whole thing. |
| The Slack worked really well. I like that you can go back and look at questions that people |
| asked. I think it works better to ask questions in text than orally. |
| It was also really nice not to have to decide which talks to have to miss, it was *amazing* to be able to catch up on talks I had missed due to outside issues, and I loved being able to catch up on a talk on slightly faster speed when I was having difficulty following the normal speed of the talk.  I also adored being able to scan backwards immediately when I missed a key point and then to catch up again to the live version on 2x speed. |
| Slack interaction was easier and more social than expected |
| - High exposure to many listeners for my paper
- I expected things to not work well, but everything was very well organized. Chapeau (hats off) to the organizers... |
| In all. This was my best virtual conference experience so far. |
| AMA sessions were very good. |
| people making the effort to proceed as if it were in person |
| The papers presented were incredibly complex and advanced. |
| Fantastic Q&A on talks, really liked talking to students who approached me via #mentoring |
| Probably better attendance than a physical conference, and probably slightly more polished talks. Talk Q/A also worked well, maybe better than a physical conference, since overflow questions could easily be answered electronically. Easier to attend the subset of talks I'm really interested in. |
| In pretty much every way. I was very leery of how talks, questions, and social interaction would work and I think they were all stellar. |
| Quickly moving to a virtual format has surely been not easy, and the outcome could have been really unsatisfactory. It was, instead, excellent! The organizers did a majestic job, and I want to thank them all. |
| Truly amazing effort, I did not expect the means for interaction would be so effective and diverse. |
| Recorded talks were of higher quality in my opinion. Having the high-quality recordings available later is a big win as well. I think live-streaming them was effective as well. |
| Resources were immediately open and free to use. Streams were on Youtube which allowed easy and convenient rewind, watching anything you missed, or speed up. Volunteers/organizers coordinated everything well and were well prepared. |
| I thought the organisers really did everything in their power and beyond to make it work and try new things. Really impressive dedication. |
| There were lots of people interacting. |
| The availability for casual talks and gathering |
| I was surprised by how intense the whole week was and by the number of interactions I was able to have. |
| Discussions happening in Slack were interesting, but didn't distract from speakers' response to questions. It was fantastic to see speakers following up on Slack, and to see all questions asked even if there wasn't time to answer them.  ISMM running repeat sessions was exceptionally useful, since I could pick a schedule that worked for me.  LCTES chose to simply provide pre-recorded talks on YouTube, which is fantastically convenient for viewing on my own time. |
| The various media to make interaction better was just excellent. |
| I think the sessions ran smoothly. Asking questions on slack was easy. |
| I think that the pre-recorded paper presentations were of higher quality, since the presenters were able to edit, etc. |
| in every possible way to be honest! you folks did an amazing job! |
| The opportunities to reach out to senior professors/scientists. |
| It was super cool to be able to attend ask questions over slack and slido and have them answered, while not having to travel anywhere, less effort, less cost, less carbon footprint, etc. Also the Q/A were very smooth with selecting written questions. The ability to look at recordings within a 4h time-shift when you are fed up with viewing something and want a break, and to view stuff in double speed to catch up. Clowdr and gather were nice experiments and I like experiments. The one person that I tried speaking with on gather was ok. Seeing the lively feedback on #lessons- |

learned. Seeing the business-meeting to figure out what PLDI actually is, and where it will be moving. All very good.

I got a chance to listen to all the talks if so I wish - I no longer have to choose between 2 parallel talks  I like AMA sessions a lot

The lineup of speakers and events was above any prize!

The one-on-one meetups were great.

Variety of communication channels AMA sessions  Mentoring talks

The organization of events were good

presentations and Q&A sessions were strong, personally I preferred this style of presentation to being crammed into a room.

I was impressed by the effort put by the organizers. Gather was working better than expected even if I was not able to meet people that I have hopping to interact with.

- The ability to watch a presentation without distractions, rewind, pause, go back. - Significantly lower barrier to questions/responses.  I could just put the question on the slack channel.

The technical tracks went very smoothly, and I liked that the tracks were at non-overlapping times and that I could catch up on the livestream if I missed talks due to scheduling etc.

I wouldn't be able to attend PLDI if it wouldn't be virtual and there were many advantages of attending it virtually compared to only be able to read proceedings or watch recorded talks. In particular:  1) I found great to be able to ask questions during the talks virtually. From my experience, QA for talks were better than in the case of a physical conference. No problems with people speaking not loud enough, easier to ask questions for introvert people, possible to ask and answer questions before (in particular in the case of pre-recorded talks) / after the QA session. It is a bit shame that Slack QA are not accessible persistently.  2) It was great that all the talks were recorded and thus accessible for everyone (this is already the case for many physical conferences though).  3) It is great to have an opportunity to interact with people (gather). I think that there is a big room for improvement both in terms of tooling and people learning how to deal with online interactions.  4) It makes the conference accessible to much more people (including people with a handicap of some sort).  5) Eliminating travel makes the conference much more environment friendly. Indeed, large physical conferences as a way to spreading a knowledge and enabling (non-obvious) collaborations are probably not sustainable.  I think that 1) and 2) should be integrated to physical conferences. Since from my perspective, the biggest point of physical conferences is to have non-virtual contact with people, I don't think that trying to integrate 3) to physical conferences is a good idea.

Extremely well organized. I felt a bit confused at first but it surpassed my expectations

The AMA sessions were great.   My tutorial had a much larger audience (by far) than I would have expected in the physical conference, reaching many people who would not have attended otherwise. The live interaction on Zoom and Slack was excellent.

It's fantastic that it was free to register! It made it viable to attend only a fraction of it, which is not normally tenable.

The organizers did an amazing job to make this feel like a real conference. I think the shortcomings may be inherent to the format. I also liked the Q&A for papers. Slack was a good medium for that. And certainly I liked the convenience of not having to leave my home/office, tuning in whenever I wanted without disrupting my work flow, being able to watch asynchronously. And one can't beat the cost.  I found there were several ways in which a virtual conference is superior to a physical one:  - 1.25x speed for talk playback. Handy.  - Time-shifting talks, pausing them.  - Being able to let sessions run longer. E.g., AMAs.  - Slack for questions, so that many more can be answered. (Though I would *not* mix physical questions with slack, though--the slack channel should be strictly for asynchronous questions if there is a real, physical questions track.)  - No multi-track conflicts of interesting talks.

Free online attendance was great, and I managed to join a couple of the talks that I found interesting (in the early hours)

Live streaming out to YouTube live was an excellent feature, since they were recorded.

| | |
|---|---|
| This allowed me to get on with my day job, occasionally stepping away from PLDI talks, to watch them in the evening at my leisure. I did this for many talks. In that sense, I probably watched *more* talks than if I had been at PLDI physically, since urgent work matters sometimes gets in the way of focus and attention at conferences. | Many speakers were extremely inventive in taking advantage of the opportunity to pre-record their talks.  Examples: the Armada talk, with its costume changes, and the French guy who put the transcript of his talk in a sidebar, thus avoiding pronunciation difficulties.   Also, I was able to rewind to repeat something I missed, or time-shift the talks, etc. |
| Ease of asking questions | The sessions were well organized.  I do like the concept of gather even though I didn't utilize it much.  I also like the fact that things were recorded, and I can watch them later if I happen to miss them in real time. |
| Very easy to make connections and chat with people! | |
| It was easy to see other people's reactions to presentation and then look at them at a later time. | Having access to all recordings of presentations, to be able to replay them as didn't get to watch many live (due to timezone issues).   The Slack channel for questions worked really well, I got to interact with people just as if at the conference in person. |
| The asynchronous question mechanism was amazing, I got a lot more discussion on my paper than in many regular conferences. It was also great to be able to rewind/pause the talks on youtube, and even to cook dinner while watching talks, etc. The slack workspace was very active, and the ask-me-anything sessions were particularly valuable. | |
| | The "Ask Me Anything" sessions were a great addition. |
| the format for presenting papers (recorded video with q&a afterwards) was great. | The talks were extremely high quality, which seems to be related to their pre-recorded nature. Also, the ability to tune in and out of talks, pause them, speed them up, etc, was quite a nice way to interact. Also, far more researchers were able to attend this conference than would have been able to attend physically, which is a wonderful thing. |
| Having youtube streams made it very easy to watch presentations in my own time, and to share interesting talks with colleagues. | |
| It was available to without any cost and we could attend/watch the videos even after we couldn't attend them in real-time (due to time zone differences). | If it was not virtual this year, I would not haven joined it. Since it is virtual and free,  I was able to join a tutorial about quantum computing I liked!  I do appreciate everything is recorded and available online for free! That can be recommended when physical conference is possible too. |
| It was still very well organized. Also it was awesome that a lot more people got to watch the presentations since it was free and on Youtube. | |
| i like how we immediately have access to talks via recordings, this gives much flexibility when it comes to finding talks one wants to attend | Free? I guess. |
| | I like the fact I can see all the videos online, rewind, ask questions later, etc! All of these things should be retained for a hybrid physical conference too. |
| It was free and easy to attend a little bit while working on other things. That was nice. I'd prefer all future conferences to have just as strong of an in-person component, but also offer a free virtual component for the sake of affordability for people with less funding and to respect the time constraints of people with conflicting deadlines or with children. | Slack questions seemed to work very well.   In some ways, for our paper, it was even better than the usual in-person Q&A.   A lot more work, but also more rewarding.  Presentation quality was generally pretty high. |
| | Low effort participation. Accessibility to a wide audience. |
| There seemed to be a lot of people engaged! | |
| The organizers were enthusiastic and did a great job given the circumstances. | AMAs were really awesome, video streams worked really well. |
| Readiness to answer all the questions even way after the talk | No technical difficulties at all, the talks were interesting and well presented (maybe even more than in a physical conference) |
| The production value of the talks were much better than previous years. I think this can be attributed to the talks being prerecorded. I think this would be a great idea in the future, even if the conference stops being virtual. | Being able to attend "lightly" without committing a full week+travel was nice |
| | Very organized and strong presence on the Slack, it was very easy to get information when needed. |
| Never had expectations, but The fact that I stopped watching live streams only to return to 7 hour long uploaded videos should say something. (I watched almost every talk/session. Really loved the the probabilistic programming portion. Good to see Neil Toronto's work wasn't a "shot in the dark" ) | On YouTube live, I can catch up the sessions though I missed them. |
| | The Q&A over slack was incredible. Infact this could be a feature worth keeping even if we go back to physical conferences. The AMAs were outstanding. |
| Speakers so willing to discuss and answer questions | I happened to listen to more talks than I usually would. |
| I really liked the use of Slack for asking questions and continuing the discussions after the talks. I hope this stays, even for non-virtual events. | I was impressed by the number of people who participated. If I had been better at using the tools provided, I believe that there was a great opportunity to interact with people who might not otherwise have attended a physical meeting. But I have to say that even for basic things like getting to the correct YouTube channel and thread of the Slack channel, it was a bit confusing until Day 3 of PLDI. |
| - I like the Youtube stream coupled with Slack questions (why not Sli.do?). Nice to be able to rewind the stream, or pause, when you get distracted. - I am glad for the single track over a longer day; being able to watch talks whenever helped deal with timezones/day-length. - AMA sessions were great. - Slack-based communication was more informative/useful than I expected it to be. | love the youtube -can start late, variable speed, and rewind if needed and end on time for the live Q&A |
| | Felt inclusive |
| Much more accessible. I wouldn't have been able to attend this year if it was not virtual. The YouTube streaming feature allowed me to attend talks whenever I wanted. Chatting with folks over slack was easy and I could find whoever I wanted to just by name rather than walk around a large building for hours. | 1) The slack channel worked really well for question-answer. 2) The AMA sessions were great! 3) Many more people watched the average research talk than I have seen at past in-person conferences. |
| | Unexpectedly smooth in operation. |
| The organisation was really good. I'm really impressed by how the organising committee(s) rose up to this challenge, despite really not having signed up for a challenge of this scale. | Not in any way. |
| | It was more engaging, especially Q&As than I'd expect from a virtual event. |
| It was organised very well | It was a good organization and mostly worldwide. |
| The AMA were fantastic. I think most of the papers got more questions (and thus more interaction between authors and attendees) than in most physical conferences I have attended | it was a good organization and mostly worldwide |
| | The slack Q&A sessions were excellent. Far more useful than the normal Q&A at a conference. We should definitely try to preserve this when we get back to mostly physical conferences. |
| The video talks were nice. Some researchers did make use of this new channel to express themselves. | |
| Super easy to approach people. You don't have to hunt down somebody in the hallway. | I was able to attend many session that I couldn't have attend otherwise. It was great. |
| Speakers answered to the questions much better than physical conferences. Even after the talk ended, they answered with Slack. | How smooth it went, and the preparation that facilitated that |
| I absolutely liked the AMA sessions. | The AMA track was awesome & a good use of virtual-ness |
| it didn't | AME |
| That I was able to participate in conversation in a more democratic manner in the medium of chat on Slack and taking my own time to do it. I find it much harder to interject if senior people are talking and I don't have anything worthwhile to add in the moment. | The talks and Slack for questions worked quite well. The gather was OK. |
| | The ability to watch presentations at a slightly later time or rewind sections was good compared to a physical conference, but there was still a lot of synchronous communication that you don't get when just reading papers or watching talks on YouTube. Good combination! |
| There was an actual community participating and working hard to mimic the conference experience. It really was impressive, far exceeding most people's expectations! I think some of that was from coronavirus -- many of us wanted a conference and were eager to contribute to making a virtual one succeed. And the conference leadership was stupendous. | Videos with talks were often better than real talks in rooms with small screens and/or bad aqustics. Q&A in Slack were awesome -- plenty of questions and detailed answers helped me learn a lot. Social interactions were not as bad as I expected. I wish I had more time to talk to mentors and other attendees. |
| The sessions worked very smoothly. It was easy to follow the actual talks that had essentially perfect video and audio (and even subtitles in some cases!) It was much much less stressful to have it virtually and much less time wasted on things having nothing to do with science (like getting visa, boarding airplane), let alone all the cost and environmental benefits. | super accessible from anywhere on the earth. Also the ability to rewind/replay the (live stream) video instantly really helps |
| | I felt many people open to discussion and conversations in the slack channels. |
| | I thought that there would be no social interaction at all, but the organizers did a great job with Gather and the mentoring sessions. |
| I really liked the Q/A on slack, the mentoring sessions, and gather (even though it had shortcomings). | The organizers tried really hard to bring back the hallway interactions |
| 1 on 1 mentoring sessions were actually better this way | The Slack channel allowed unfettered access to many well-known people in the community in a way that allowed both parties to communicate on their own terms and outwith physical constraints that would at a physical conference. The live streams and AMA were useful as well. |
| Watching talks online from home is nice. | |
| If there was not time for an author to answer all questions, the questions nevertheless had been captured in the Slack channel, and in some cases authors took the trouble to answer such questions on the Slack channel a bit later. | Virtual talks are much better than in person talks. I think that asking for a prepared video even for a normal event and letting only questions during the meeting is better. |
| Talks seemed a little more polished. | Excellent people all around. Super nice talks. |
| It was _really_ nice to pause/play @ 2x speed the talks. | during presentation, I can roll back a little bit to the point that I missed |
| Very few hiccups in streaming and the scheduling of all events. | Very well organized! Much more interaction over Slack (especially Q&A) than I expected. Mentoring/job adverts channels were useful |
| The mentoring channel and the AMA sessions were both excellent ideas that should be kept! | |
| and should be replicated for physical conferences. | Watch live videos on YouTube of the presentations was a great idea! We should always do that!! |
| I enjoyed getting to see a lot of research and presentations that I otherwise wouldn't have been able to justify traveling for. | I wish I could attend the physical conference that'll be better for me to gather more knowledge. |
| Slack questions, Ask Me Anything session | I was surprised by how clear to understand the talks were — as an undergrad I thought it would be opaque to me. I was surprised by how friendly everyone seemed on slack :). |
| Gather was actually fun and entertaining, for catching up with people you already know. | |
| I honestly did not think I would be able to attend so many parallel talks and still get a reasonable gist of what's going on. Replaying the talks to catch up on things was very helpful. On a different note, the virtual 1-1 interactions lessened my social anxiety of the researcher I wanted to talk to. The mentoring sessions by Alastair Donaldson and multiple other panelists including the session chaired by Nadia were very nicely planned and helped me find a new direction in tackling research and pressure. | 1:1 meetings, Ask Me Anything, Replaying so that my time scheduling became pretty flexible |
| | free, more inclusive |
| | I felt the 'conference atmosphere'. I like that a virtual conference is integrated more seamlessly with other aspects of life (the other duties of an academic, private life at home), whereas a physical conference means a 'rupture' (for the better or worse). |
| | Ease of interacting with authors via questions on Slack |
| Ability to watch any talk I wanted to at my own pace, albeit without interactivity | This was so much better than a physical conference. I was able to attend my talks of choice and use the remainder of my time for other purposes. I didn't have to waste a whole week in miserable airports, planes, hotels and conference venues. |
| Ask Me Any Thing | |
| Zoom, YouTube, and Slack interoperated better than I expected.  We'll get better at this with time. | |
| | Quality of the  AMAs |
| The Q&A during talks was amazing,  students found it easier to approach people, overall I was much more EXCITED during this PLDI than I have been at others in the past, probably because I was not EXHAUSTED from travel, could quickly pop out to hug my kids, could rewind and fast forward talks that I'd missed, could ping people on slack instead of running about a stuffy hotel looking for them, the food was better etc etc etc. In brief the elimination of the substantial travel cost and the organizers' METICULOUS attention to detail, putting together such an exciting program has made me EXTREMELY bullish about inevitable virtual conference future. | The slack channels for discussing papers were excellent, encouraged more discussion in a shared forum than often happens, and allowed real time "what did that notation mean" discussions. I felt much more empowered to add a question to the slack channel --- which is async, and unintrusive to people not interested in --- than I ever would have in a normal conference (and to my pleasant surprise, my questions do not appear to have been a waste of time for anyone.) |
| | it went very smoothly and Slack was very active! |
| Great online set up that we can go back to in order to relearn although I'm not %100 sure we will, let's hope the best! | The mentoring channel on Slack was amazing (I'm a first year grad student); I'd strongly encourage future conference organisers (even for physical conferences) to adopt something like that |
| Pre-recorded videos worked well | |
| Large, diverse participation. Ask Me Anything sessions. | |
| Everything recorded and easily accessible. | great flexibility |

Talks were better than usual, I think because authors were given a way to control the final product a little more precisely. In a way, this was a big improvement over typical talks.

REMS/DeepSpec had great content

Because it was free and virtual, I was able to attend without asking my advisor for funding etc. The mentoring channel was a great resource! I was able to have some unplanned conversations on gather and clowdr which I did not expect from a virtual conference.

AMA part is very interesting

Authors answer questions on slack, which is easier than the time constraints in physical meeting.

Having questions answered through Slack after talks was nice

the AMAs sessions were fantastic! and Gather is really fun

Gather at least partially mimicked the real hallway track (module time zones and nb of people who actually used it)

Gather/clowdr/slack far exceeded my expectations in terms of interactions with other researchers. PLDI was a far better experience than other virtual conferences I attended.

I loved the AMA sessions

The talks were better than I anticipated. The AMA sessions were universally awesome. Mediated Q&A was more useful than traditional in-person Q&A.

I was really impressed with the virtual "physical" meeting space!

Integration between Slack, Gather, researchr, YouTube and Clowdr was better than I expected. The YouTube live streams did achieve a greater sense of "presence" than I expected.

I can switch between sessions to listen to what I was interested in.

talking to mentors and AMA sessions have been very useful (I'm starting my PhD in CS soon, so I got many useful tips and advice; was able to talk to these professors whom otherwise, I would not have a chance to talk to).

online access

there was great mood of mutual support. quality of talks was superb. the discussions on slack simultaneously to the talks, and the fact that the questions were _written_ and thus better thought out, made the talks much more useful.

nice tutorials

Gather!!!

very well

Such a relief not to have to fly across the ocean! And I did have several random / unexpected interactions with people I had not met before.

Tools: Slack was great, Gather/CLOWDR show potential -- I'm eager to see how they develop in the coming months.

The youtube-streamed presentations were great because of the features: a) rewind when you missed what the speaker just said b) pause to go make some coffee c) watch later (timezone issues). The presentations on zoom disappointingly lacked these features!

It was very well organised, with many creative ideas for making the most of the conference being online. I experienced no technical problems at all in attending talks! There were various opportunities to meet other attendees during breaks. Attention to mentoring, AMAs, and the channel for SIGPLAN CARES were very nice. Despite the high influx of people (because online, and free), the conference had a pleasant, positive atmosphere.

Social interaction

I hadn't expected it to be so great in a virtual manner to be honest. Kudos!

Much better organized that I expected from a virtual conference

Watching talks from my office made it easier to take notes, look up references.

Being virtual, allow me to take part in PLDI, which is somewhat outside my main research area.

I was happy I could attend great talks even during these difficult times, so the PLDI organizers did an excellent job

I liked the lack of parallel tracks.

AMA were fantastic, Q&A through slack are much easier to follow than IRL (both as an author answering and as someone asking), they also let you ask/answer more questions if you're interested.

| | |
|---|---|
| It was very well organized and of high quality. It helps to have material positioned in a way that is easily visible and explorable. All the slack channels, Ask Me Anything, and talks were very well managed by the hosts. The conference website was also very impressive, that made it easy to follow. | interaction after research paper presentations: I received more questions than last year and had an easier chance to discuss them. |
| Quality of the workshops I attended, talk Q&As, AMAs. | The 20-minute session per paper were great! It was possible to understand the work well in each talk, and thus there were good discussions afterwards. The AMAs were also a bonus as compared to a live conference. |
| Being able to go back to talks that I missed and the questions on slack. Being able to catch up on talk at 1.5x speed. Being able to find people on slack to talk to. My own food, not crappy hotel food. The AMAs were great. I could watch talks from workshops that conflicted with workshops I went to. | Gather is really promising |
| | The videos were online sooner, but in all other ways it was worse. |
| I can repeat the talks anytime I want and discuss with the authors via slack anytime also. These two points are very helpful for people cannot afford to travel to PLDI. | The "Ask me anything" sessions were great. Also the mentoring channel. |
| | It was nice to be able to get in touch with paper authors themselves even after they finished |
| | Talk qualities, AMAs |
| Organization of session is perfect. Slack channels and threads for questions worked like a charm. Discussions remains in the chats and questions get answered offline in a much more solid way. This is even better than live conference. | I expected the social tools to be much worse (particularly gather, which seemed totally unpromising at first)... also you can't beat the registration fee :-) |
| | In a lot of ways! I didn't expect that it runs so smoothly! I believe many people worked very hard to make all this happen. I really appreciated that. I think many people can feel more free to answer questions than in a physical conference. Another good thing is that it may be easier to find certain people to talk to, because in a physical conference, you need to first find them but in a virtual conference, I can always find them in the slack channel. |
| The number of participants (=reach) was amazing. | |
| Lost of opportunities to connect with people, reduced communication barrier, very well structured Q&A sessions. | |
| The interaction (through slack, sli.do for plmw) was very good. Questions on slack | |
| The organisation was well prepared for the virtual aspect. They picked decent tools for the job and made sure that everything went great. There also seemed to be a great group mentality in the Slack workspace | The slack platform is a good way for people to interactive with the authors. |
| | YES |
| | Makes future interactions with fellow researchers easier. Saved on travel effort. Talks are available for viewing later. |
| The sessions and Q&A worked quite well. | The publicity of the work because there are a large amount of people registered, and there are a lot more questions on Slack compared to a normal QA session so it is nice. Also, I really like the "Ask me anything" session and it is really inspiring. |
| Questions were nice and the talks too. | |
| Overall, tech worked surprisingly well, without major breakdowns or delays in the program, that's awesome work by the organizers! | |
| It has been organized really well. Everything worked most of the time. | I can talk to many people. |
| Things ran smoothly for the most part. I also really liked being able to roll-back a livestream and watch something I may have missed. I also think that the Slack was good for getting more | Surprisingly accessible, especially Gather. The ability to watch talks in convenient times for my time zone was great! |
| | Better talks and answering questions. AMA was the clear winner of the conference. |
| Quality of presenters, quality of audio/video | None |
| Everyone's so nice =) | The quality of the presentations. |
| not really, unfortunately, but I realized that slack discussions for each paper *could* be pretty good, they could go on for all three days, and even result into spontaneously scheduled small video calls, which would have exceeded expectations and also what's possible in a physical conference, but I think much less happened there than what could have happened, probably because only one "thread" instead of one "channel" per paper was available, and because it was declared as being "only" for questions, and the threads were very hard to find | question-answers |
| | Virtual worked better than I expected |
| | Organizations went better than expected, nothing major went wrong at all. |
| | Already had high expectations that were met |
| | I saw people interacting on /gather. |
| | People interactions feel to have been enhanced by the virtual nature of the conference. Easier to talk to a "stranger" on slack or gather than approaching them in person. |
| It was extremely well run given the limitations of it being virtual. | The AMA sessions were thoroughly enjoyable. It was really nice not needing to travel and being able to partake in what I wanted to. |
| Strong organization on a short timeline. Talk quality was higher than usual (thanks, pre-recording!), as was Q&A. | The amount of interaction was better than I expected- though still very lacking. |
| All the talks were streaming and recorded. It enables us to check some talks back immediately as needed. It allows me to easily catch up talks (even in the presence of language barriers) and to have short breaks safely even within a session. | I really enjoyed the AMA. |
| | Relative lack of tech issues for presentations |
| | great interaction, everyone can know clearly about others' questions after a talk |
| | The infrastructure prepared was very well explained and didn't seem to lack anything in particular. |
| I think that the workshops were better than usual. The number of people that were streaming the workshops significantly increased. It is probably easier to get everyone who is interested in a topic on a chat channel rather than just those who managed to find time to go to the conference. In other words, I think that traditionally small events (<30 ppl) should move online and large ones (>100) are better in the real world | I think the entire virtual aspect was much better handled than I expected. Granted, I originally envisioned just zoom sessions and slack when I heard the term "virtual conference." |
| | Well organized |

## A.7. Additional factors specified in response to "If you would have been unlikely to attend, which factors would have influenced this?"

| | |
|---|---|
| had not planned to attend | Avoiding travel (even with an emissions-free option) |
| PLDI is not my main community. I would have definitely gone to ICSE, for instance | getting required corporate approval |
| VISA issue: have difficulty returning back to U.S. | Probably, that I did not submit a paper? Also I dislike traveling. |
| Insufficient interest | I don't enjoy being around other people |
| I didn't have a paper published at PLDI, so I probably wouldn't bother to go. | |
| In the end my physical attendance would have depended on whether travel cost + conference fee would have fit in my company's Training Budget, so I have no idea if my training request would have been accepted | Attending another conference |
| | Topics not sufficiently relevant to my research |
| | I'm an undergrad and wouldn't have felt qualified to go. |
| Too many overseas conferences, flying out of Australia is expense and time consuming.  I can only limit attendance to a few conferences each year. | No paper and networking is only reason for travel |
| | No justification for reimbursement. |
| Internship | Lots of other travel |
| Getting management buy-in | I did not know about PLDI |
| I'd rather spend my company education credit on a more practical (i.e. Pycon, Scala, Salesforce, MS, Google) conference or on Continuing studies or Coursera classes. Goes a lot further. | relevance |
| | As an interested observer from industry, only a small part of the conference was relevant to me. The virtual setup made it easy to attend the subset of interest. (My company would have been much less happy if I wanted to take the whole week to physically attend!) |
| Not an author, so no grant money to cover expenses. | |
| I don't really feel as belonging to the PLDI community | As an industry programmer, I only make time for one academic conference per year.  This year I would have attended ICFP or one of its colocated events. |
| incompatible with the day job | |
| had to make arrangements, now, none were necessary | Not from PLDI research areas |
| POPL deadline | University doesn't provide funding for travel unless a paper is accepted |
| I typically attend only if I have a paper at the conference. | Did not submit |
| PLDI was at the same time as my daughter's High School graduation.  Can't be in two places at once (except virtually) | I am involved in several research communities and can't attend all conferences of all communities every year |
| | Time and attention needed to make travel arrangements beforehand (during semester) |
| London is too expensive and not a good place to hold a conference | Time conflict for other business |
| logistics, conflicts | job responsibilities intrudes on conference travel |
| Ridiculously low quality reviews that I saw at PLDI this year | |
| | Getting approval for student travel and getting travel grants |
| Conflict with some of my original travel plan that was cancelled after the breakout | did not have anything to present at pldi this year |
| not many talks interesting to me | |
| WAY Too much travel this year. | Turnaround time |
| I am part-time student and hence getting the necessary funding/permissions is always a hassle. Further I would be treated as an industrial participant blowing up the cost. | Not being an author |
| | My interests usually overlap more with other SIGPLAN conferences |
| I would probably have agreed because I do *love* London, but I would have hated all the travel and cost and being away from my kids. | |

## A.8. Responses associated with "I prefer to self-describe" for the question "To which gender identity do you most identify?"

I find this question offensive and I'm ashamed of my research community to find it here. I am male, that is my sex, and it would be great for PLDI to record this legally protected characteristics. Gender is not my identity: gender is the collection of stereotypes and biases imposed on people for their sex; gender identity is its introjected self-imposition. People should not be asked to identify into their oppression. Please, listen to some of the groups of women, and especially lesbians, campaigning against the degrading impact of gender identity theory. And, if you want to address discrimination, ask about sex. https://sexandgenderintro.com/gender/ https://www.economist.com/open-future/2018/07/03/the-gender-identity-movement-undermines-lesbians https://womansplaceuk.org/ --- even Jesus and Mo know this https://www.jesusandmo.net/tag/genderism/

| undefined |
| Giraffe |
| I prefer not to describe. |
| Programmer |
| transitioning transfeminine |
| Today it's unimportant! |

## A.9. Responses associated with "Another Ethnic Group" for the question "What is your ethnic group?

| | | |
|---|---|---|
| Jewish | I prefer not to describe | All of the above + some turqouise skin |
| n/a | Mixed | Nepalese |
| idk | Jewish | Slavic |
| Nerd | decline to state | Arab-Berber |
| human? | African | Tatar |
| Taiwanese | Human | Turkish |
| Jewish | | |

## A.10. Responses associated with "Do you have a disability or special need that impacts your access to ACM conferences, special interest groups, publications, or digital resources?"

| | |
|---|---|
| I am from India i dont earn enough to travel to Europe. | I am neither able to access Youtube, nor register a Zoom account for free (although I have links that you sent to me) |
| colourblind | |
| I need to dive in too deeply, kind of leaves me crippled in fast paced environments | |

## A.11. Additional information provided by respondents who answered "Yes" to "Do you identify as being a member of an underrepresented group?"

| | | |
|---|---|---|
| LGBTQIA+ | woman | women in computing |
| Women in STEM | Chinese | Female |
| women | Me | Latin American |
| Black peoples | Women | LGBTQ |
| Nigeria | Asian | Transgender |
| lgbt | female | Woman of color |
| Female | π/\O | Women |
| Bisexual | LGBTQ+ | hispanic |
| nonbinary, ace | LGBTQ+ | women are urm in computing |
| female | Slavic/Dinaric | Gay |
| Female, Latino | woman | African-American |
| Queer and disabled | LGBT | Women |
| Hispanic | Women | |

### A.12. Suggestions for ways to mitigate the effects of time zones

| | |
|---|---|
| Prerecorded videos but with live discussion sessions led by session chairs and nominal video watching times signalled in the schedule to encourage watching the videos in advance while allowing attendees to rearrange their time. | three nearby timezones to target, 10 hours on, 16-18 hours off, 10 hours on, etc. |
| I select the "one time zone" option, but I think it's also reasonable for that zone to be uninhabited: mid-atlantic or mid-pacific, for example, give reasonable reachability from (parts of) two continents. | Hybrid conferences will likely be the norm for many years to come. Live presentations should be partitioned into remote and physical ones (e.g., alternating remote and physical sessions), with all questions being channelled through a Q&A channel. Moderators will decide which questions get answered live, others may be answered asynchronously. The Q&A channel may include A/V recordings of questions. Maybe experience will tell recorded material may use a full 20-30mn slot, while live presentations may be shorter. |
| Live activities, but more efforts spent on free asynchronous communications. E.g. everyone can create (public) channels and promote online discussions there. If the discussion gain attitude, attendees can create a room for video chatting. | |
| Have all slides available to review a day before. | I appreciate the paper presentation videos being available after the conference. |
| Maybe a combination of focused and asynchronous: the talks could be prerecorded and the Q&A asynchronous, but the social gatherings should be focused and not spread, otherwise people would not have a clear idea when is a good time to tune in. | Focus 8h, with parallel tracks, breaks, and official online poster session, plus an official offline break ("cook your dinner" or "family") in the middle ;;;;   Another alternative could be 8h per day, but instead of alternating yearly the timezone, every second day could use the other timezone. |
| It could be interesting to try to rotate the live activities within a conference.  E.g., picking | Spread, but replicated at staggered hours other than 12 hours (8 or 16); alternately, pre-commit attendees for talks, giving their time zone, |
| which may reduce the overhead for talks with lower time zone spread | would say that Q&A can be split into two streams 12h apart (but shortened twice) and workshops can be focussed on one time zone with resulting materials available at least until the conference ends |
| I think that it is hard to say without trying it out. Focussed seems to facilitate more live interactions, on the other side, spreading the even could make it less exhausting (screen time problem) provided that one would manage to not work during the pauses :) | |
| | Do a demo day live (5 minute presentation + 5 min Q/A). Then do the full sessions spread out over 24 hours. The times should be set by a survey after the live demo event. Basically only what gets the most interest + Timezone cohesion gets a replicate live full talk/session (assuming presenter is up for it). Everything else is uploaded as a video at the start/end of the conference. |
| make talks live with extended time for discussions, perhaps repeat talks in multiple time slots, do more in parallel | |
| Haven't thought about this much. Don't want to offer uneducated opinions. | |
| Spread as it was, but improve experience for asynchronous out-of-timezone watchers by providing individual videos after the live session, instead of having to search through hours of livestream that happened while you were sleeping to find the talks that you want to catch up on. | Separate chat rooms for each sessions, to be able to very quickly find the interesting discussion. |
| | If the conference cannot be held physically, then just cancel the whole thing and go back to ACM journals. |
| Focussed, but each day based in a different time zone | stop having conferences and go full journal. That really is the only option. Virtual conference doesn't work at all. |
| There is no good answer to this question, which is why I don't think that a virtual conference is a good idea. | This problem is no worse than with physical conferences vis a vis jetlag. Rotation is good, etc. but it's odd to think this problem is worse for virtual conferences. |
| Given that the presentations are taped, I don't think replication is that important. | Let speakers indicate their preferred zones and apply a scheduling algorithm. Make text discussions more visible and longer lasting. |
| Present the paper talks at allocated times in each timezone followed by questions. Then have question time replicated in other timezones. This would work well for papers with authors in different timezones. So could be an optional Q&A session. | Parallel tracks spread over 12 hours. Each paper has its own session that can go over in case of many questions. |
| Make the live portions short so that many time zones can be accommodated. Videos provided in advance. | Make presentations available at the start, but schedule live video Q&A sessions (two for each presentation, 12h apart) |
| Spread conferences over a longer period of time, keeping talks and events within a time frame that works reasonably well for everyone. | Spread the conference across more days, with less screen time per day. Make regular videos available a bit in advance, repeat Q&A twice. Schedule multiple social interactions across 24 hours to accommodate more people. Make things like AMA or keynotes run live times but in different times, so that everybody could watch something live, but maybe not all of it. |
| similar to the first option but the Q&A should be scheduled in a time for real communication, as well as asynchronous message Q&A. | |
| More asynchronous communication | |
| no live sessions for paper presentation is actually needed unless one wants to verify the author identity. Making prerecorded videos available beforehand allows for better questions. Live steaming of question answering limits quality of the answers. Regarding non-paper live activities like workshops or Q&A I | Recorded video released early, online Q&A twice, 12 hours apart |
| | https://icml.cc/Conferences/2020/VirtualICMLSchedulePlan |
| | focus on the time zones harboring the most attendees |
| I think it will be difficult to predict which options are good, without actually trying them out. | everything if I get a lot out of what I can participate in.  I appreciated the lack of parallel tracks.  Video conferences are stressful and parallel tracks add to the stress. |
| ISMM's strategy worked well: record the talks; repeat the live Q&A sessions. | Spreading things out makes conferences worse at their most important purpose, which is bringing people together to talk. We either have to commit to being present at the same time, or use asynchronous modes of communication. |
| Honestly I just want the US to be inconvenienced for once. | |
| All talks at beginning.  Extended interactive Q/a and tool  demos if possible. | |
| Make all recorded presentations available at the beginning of the entire event. Schedule 2 live discussions (e.g. Q&A) for each presentation, 12 hours apart (or something). | Maybe make the videos available at the beginning, but make the Q&A sessions live (and two each 12 hours apart)...?   Also, in that environment it might be interesting to set up chat rooms to watch talks or groups of talks with smaller groups of people and then maybe discuss afterwards. Might make some talks visibly more popular than others, which is probably bad, but I think it would improve the engagement with the talks, which is somewhat difficult with prerecorded streams. |
| Good work! | |
| I think that live activities should focus on AMAs, mentoring, networking, .. but not pre-recorded talks. | |
| ISCA 2020 global hours for keynotes | |
| But keep the slack conversion and video, and chance for asking question virtually. | |
| I don't understand what you mean with "replicated", but the youtube recording available after event is closed works quite well. I was able to watch later all the events I lost due to timezone issues. | Some sort of replication so that everyone can engage at some level |
| | Uniform spread may not be the way to go. Maybe allow authors to choose their slots? There is a risk that the distribution of author/attendee timezones differ... |
| It is good to space out as much as possible, because most people cannot spend 8 hours straight staring at a computer screen. | async Q&A with authors via some actual forum (not slack threads) |
| Virtual pre-recorded talk presentations in virtual spaces on a repeating schedule, so that people can gather, watch and discuss them. Probably combined with async Q&A. | I know it should be up to the participants to organise, but I think your lack of official events meant there was a lack of unofficial events outside US time. Having something outside of a narrow timeslot would give participant something to organise around |
| less than 8 hours per day | |
| Having the presentations available up-front would allow more flexibility in allowing people to watch them. But I don't think the entire conference should become fully async, some events should be scheduled, spread between timezones to allow people from multiple timezones to participate (e.g. have an event that is in the morning for US and afternoon for EU, so people from both regions can interact with each other). | Have live QnA and live videos but also post videos in advance |
| | Use sli.do or some other app that enables moderation and better sorts upvoted questions.  When doing the live Q&A twice, alternate which 12 hour time frame goes first and then use some of the most voted for questions from the first time frame in the 2nd Q&A instance. |
| I thought the schedule was perfect.  I can accept not being able to participate in | make pre-recorded talks available all the time, have live Q&A |

## A.13. Responses to "Finally, if there is anything else you would like to share regarding PLDI 2020 then please do so here"

| | |
|---|---|
| It was hard to find information about papers, and I think a separate channel per pub makes a lot of sense. It worked well at ICLR, I thought it worked well for the tutorials at PLDI, and keeping everything to a single thread doesn't lend itself well to branching conversations. | impressive amount of effort (including the student volunteers), for which I'm extremely grateful as an attendee! |
| Hugely impressive effort, great work by the team. | I am curious as to how many side-bar conversations people felt they engaged in via the virtual PLDI or afterwards as compared to the physical PLDI. |
| Re the survey question about time-zones ... The UTC+12 time-zone didn't include Auckland ... probably the largest city by far :) | Fantastic job by the organisers! Thank you very much! |
| Thank you very much for giving us the opportunity to attend PLDI 2020 virtually for free! | I appreciate the amazing amount of work the organizers put in to make a virtual PLDI a great experience! |
| Amazing work! | I just want to say that y'all did an absolutely amazing job! I was deeply surprised to learn that we are already at a point where fully virtual conferences can be this phenomenal. I now believe that basically all conferences should be virtual. Nice work! |
| Great first conference for me. :) As someone with fatigue issues, I was able to enjoy a lot more of the talks than my normal energy level would've allowed had it been in person. | |
| | Thank you!! |
| Thank you so much for making this available to the broader community and for doing such a fantastic job, under the circumstances. | Where are the videos?? I missed most of the conference and am trying to watch afterward. Please post them. Thanks. |
| Huge shoutout to Ally and co for organizing such a magnificent virtual conference! This was my first PL conference and I enjoyed it, especially the AMA and mentoring sessions. I think I am gonna frequently visit PL conferences from now on to meet more cool PL folks! Thanks | I want to thank the organisers for all the work they have put in to pivot from a physical to a virtual conference in such short notice. At the same time, it's quite sad that it has to come to a global pandemic for us to be experimenting with all of these solutions when they have been deeply needed for YEARS to tackle climate change. |
| Thank you to everyone involved! | |
| I completely enjoyed the full virtual conference. The fact that I can also watch the parallel tracks later made me more comfortable about replaying something before I ask a question. It also provided me access to some great mentoring sessions with seniors researchers like Appel that I may have hesitated to ping otherwise. | Given the circumstances you did a great job |
| | Overall, a surprisingly smooth and rewarding experience! |
| | YouTube streaming of Zoom meetings seems to have caused delays and a sense of unsynchronized video and audio. I guess that streaming directly to YouTube, not through Zoom, could have avoided that. |
| Thanks for the great conference! | Excellent job by the General Chair Ally, Dan, and the entire crew in pulling it together. |
| Big thanks for the effort in organizing everything. | Thanks folks for pulling this off! It was clear how much time and effort you've all put into organizing the conference once it went virtual (and probably even more work planning it when it was supposed to be physical). This was |
| Thank you for the *excellent* organization of the conference! This was a tour de force demonstrating how we can move to virtual conferencing and must have taken an | |
| by far one of my most enjoyable conference experiences ever (physical or not)! | physical communication is superior to a virtual communication. But a virtual communication is strongly superior to no communication and many (in fact most of the) people are excluded from communication when the conference is physical.  That being said, I suggest to alternate physical and virtual edition of PLDI. Assuming that this would happen to other major PL conferences and there would be some coordination, there would be enough physical and virtual PL conferences each year.  For sure, there are things from virtual conferences that could be integrated to physical conferences (streaming / recording of talks, making it possible to ask and answer questions virtually, making QA persistent and linked with the paper), but I think that it is better to alternate physical and virtual conferences than trying to make "hybrid" conferences (for example trying to incorporate online gathering to physical conferences).  I hope that there will emerge a good platform for virtual conferencing, ideally open and shared by other conferences.  Thanks again for organizing virtual PLDI! |
| I wish to thank all the organizing members and student volunteers for the great event, as well as anyone who participated. I enjoyed the event very very much. | |
| It was nice, looking forward to next PLDI! | |
| I will be so glad to join the 2021 edition of the great conference | |
| Thank you to PLDI organizers | |
| Congratulations on a fantastically organized event!  I think virtual conferences have a bright future.  One thing I noticed was significantly more attendance in some talks and presentations that in a physical version of the conference may not have been as well attended.  That is a positive sign for authors whose work may, otherwise, have gone largely unnoticed. | |
| Thanks a lot for organizing virtual PLDI!  This was my first virtual conference. As a Ph.D. student and a postdoc, I attended physical conferences. I now work in an industry and although I still work in on PL (static analysis), I am no longer able to attend physical conferences with few exceptions (because of my employer - cost, family - three small children, and remote geographical location). I think that both formats have their own (dis)advantages. I have to say that I benefit from the online format a lot in my current situation, but I think I would benefit from it even as a Ph.D. student / a postdoc as I would be able to attend major PL conferences more often (from big conferences, I attended only FM once and ECOOP once during my Ph.D. studies) and thus better connect with other PL researchers.  Ignoring impact on the environment, cost, family situation, and the fact that I would be never able to attend major PL conferences as often as I wanted if they wouldn't be virtual, I would personally currently still (strictly) prefer to attend a conference physically. This might change when virtual conference tooling would be more mature though. Also, knowing that virtual conferencing is possible, I have to say that I would be uncomfortable supporting a physical format knowing that it excludes many people and harms environment.  I strongly believe, that a | |
| | Thanks for organizing it! This was a good attempt. But there are still things we can learn. |
| | I think Ally, John, and crew did an awesome job. I like the idea of supporting antipodeans by replicating live sessions (eg Q&A) at times that work for them. But how would that work, if the authors of a paper are all in one timezone, 12 hours away? There are lots of EU-only papers; for them to have a Q&A that works for an NZ audience, all the Q&As will have to be crammed into narrow windows around 9am and 9pm. Similarly for US East Coast papers and SE Asian audiences, and vice versa. |
| | The timezone problem was the major hurdle for me. I expected a London timezone when I signed up (as PLDI was supposed to be in London), but understand how it might have been adapted to speakers/attendants.  I don't think the approach change particularly, but the timezone should rotate over the years. |
| | Brilliantly organised, thank you very much for putting so much energy into this. Long may virtual conferences continue. It very much levels of the playing field, democratising the conference experience for junior researchers, PhD students, and those normally unable to |
| attend e.g. because of health or child care issues. | conferences), I think you should hire someone to do live captioning and to caption talks. I have a lot of trouble understanding audio over video, and despite really enjoying the few talks I tried to watch, found it so tiring to try to follow the audio that I watched only a few. This matches my experiences at in-person conferences when captions are not present; I usually just ignore most of the talks and spend the conference talking to people and making connections. While I think in-person interaction is strictly better than online interaction, I did find it very nice to be able to see who was in a room before walking around. This meant that it was easy to avoid people in the community who had hurt me, which made me feel much safer. |
| Thank you for the great work! | |
| It was a great success to adapt the PLDI conference to circumstances as you did. Being virtual (and free), it is accessible to many more people than a physical conference. I would not have seen any of it if it were physical. There is absolute value in a physical conference, but it also makes it exclusive. So for future conferences a hybrid approach makes a lot of sense. | |
| Big thank you to all the work put in to make an amazing conference happen online with mostly no glitches during such a crazy time! | |
| I'm not an academic and a lot of the presented material wasn't accessible to me - I'd understand with preparation, but I couldn't follow some talks live. But that's fine! Because the pre-recorded videos were available, all I missed if I didn't follow was the Q&A, and I wouldn't have asked questions anyway. It worked out. | This attempt at a virtual conference has convinced me that the idea of a virtual conference does not work. For me, the point of a conference is to move my physically out of my normal environment so that I can spend 16+ hours a day interacting with my peers. This is just not going to happen at a virtual conference, when I'm inevitably going to be trying to juggle all of my normal responsibilities, and will be in the wrong time zone. |
| One thing that should be greatly improved is the quality of audio for talks. Too many talks had poor audio quality, which can be a pain for the attendees. I guess one way to solve this problem would be that the people who give the talks could hear their own audio, which they can do if the talks are pre-recorded. | Thank you so much for all the work that went into making PLDI virtual! I wouldn't have been able to attend this year otherwise, regardless of COVID-19, and I really enjoyed the parts I could attend. |
| | I just want to say a big "Thanks!" to the committee on the work and thoughts that they put in in making a virtual PLDI 2020 a success! |
| Thank you very much for making it virtual and available free of cost. Otherwise, at least I could not have attended and benefited from it. Thanks again. | My responses here were fairly critical in nature because I do feel that virtual conferences will never be as useful as physical ones for some purposes. That said, I think the PLDI organizers did an *incredible* job, and that the conference was quite amazing all things considered. Fantastic job, keep up the good work. |
| Excellent event for this pandemic time, but would always prefer a physical conference. | |
| The video talks were so good that I think, even at in-person conferences, there should be an option to submit a YouTube video ahead of time and just play it instead of giving a live talk. This would also help people with visa issues or children at home give talks at conferences. I really think we still want (in non-pandemic days) several hundred people (at least) attending in person, because in person attendance is amazing, is just super fun and a great way to make lifelong connections. But adding some kind of online option to complement that seems like a strict victory. For future online conferences (really all | Thank you for organizing, this was an excellent conference! |
| | Thanks for organizing, it worked really well! Also the non-existent entrance fee made this event attendable for me, which I think was really helpful for me (currently a PhD student). Thank you! :) |
| | Information about going from paid to free Slack plan should have been shown in red |

everywhere way before the conference since it resulted in unavailability of search through the messages. Being a novice in Slack I didn't e.g. save the threads for two interesting papers to go through the post-session questions and now am unable to find them anymore

- Do demo day style pre-conference live streams to deal with timezones.  - Use surveys to decide which talks/sessions get a livestream during the conference and when. Everyone should regardless submit a video upload of their full talks  - Have small groups formed before the conference based on background diversity to rate (as a group, not individually) how cross functional the talks are. The group would necessarily have to discuss or communicate their own POV and experience resulting in more attendee intercommunication. The ratings them selves could be another award category, or could be meaningless and a necessary lie to get attendees to talk with their fellow attendees.

Very well done! We can improve, but the whole virtual conference idea still has a way to go in general; much to learn. The organizers did a great job for a first go-round at short notice.

This survey asks for ethnicity as part of the demographics questions. The classification of ethnicity is very US-centric. For an international, conference, this is not ideal.

It was a pleasure to be able to attend PLDI for the first time, given the circumstances. I think the live presentations were good and conducted to a high standard, with panels well organised with the use of Sldio for questions. The capability to watch presentations asynchronously, and rewind if necessary, is great to follow in-depth technical material. I wish physical conferences could also provide this option to attendees.

Thanks a lot for the awesome job in running PLDI online!

1) Maybe it's time to ditch the conference presentations (also in the physical conference). The virtual conference for just highlighted that they don't serve any use for me, and probably others.  2) Please don't roll over recording and editing responsibilities to authors. Authors end up spending so much time on questionably-useful activities. If we do go ahead with virtual, pre-recorded presentations, consider adding recording and video editing as services included in the conference fees. If you think doing so is prohibitively expensive, then re-think the previous paragraph: it is at least this cost that you are rolling over the community, who is definitely less qualified to deliver it than what organiser would be able to provide. If community members feel that these videos serve a real purpose for them, there's nothing stopping them from investing the time and effort involved in generating those videos, putting them up on a streaming platform, and linking to them from their own webpages. I just don't buy it that producing videos is something we should mandate.

Thanks for organizing it!

I think there were two key factors in the succeed of PLDI 2020 1) professionalism for part of the organizers 2) "charisma", "coolness", "engagement" (not really sure which word describe this better) of the organizers  The first point includes that I did not experience any technical difficulties during the conferences. The second point includes e.g. that the organizers thought about gather to mimic social interaction or the PC chair posting very often in slack like "come and join me for a drink" or "come and join in the mentoring channel".  While I think 1) has been and will be achieved by many conferences, none of the others two virtual conferences I attended so far achieved 2)

Job well done :)

A lot of confusion ensued every time I had to find the right YouTube link for a session. The AM/PM split was a bad idea given that it was attached to some timezone but not the one my program was shown in (which I set to mine). It also took me a bit to realize that AMAs had their own stream.  The slack channel topic should link to some document that collects *all* stream links, and it should always link to the stream(s) that are live right now. That way it is always clear where to get the current stream links.  > UTC+01:00 (Berlin, Rome, Paris, Madrid, Warsaw, Lagos, Kinshasa, Algiers, Casablanca)  Berlin is in UTC+2 in the summer (as in, right now), so this option is confusing.

Great conference! Thank you to the organizers for doing such an incredible job

It is overall a very good experience, thank you all for making this happen!

Thanks for all the effort that went into converting the conference into a virtual event!! Although I would have preferred to have been in London, it was definitely worthwhile. I also think that the pre-recorded talks and slack-channel record of answers to questions means that these artifacts are *better* than what we would have had from a physical meeting.

Great job PLDI 20!!

As a junior researcher, I am worried that physical conferences will become a thing of the past. I really hope that hybrid conferences can become a nice middle ground, giving younger researchers a chance to build relationships while still remaining accessible to many through the online component.

Thank you for much for organizing such a wonderful event in a difficult time.

Live Q&A where remote participants can type questions to be read by a session chair is definitely the way to go. Slack unfortunately doesn't encourage a way to sort questions by votes. It would be better if there was more encouragement to submit questions in advance and people had time to consider and upvote the best questions, and the session chairs were encouraged to priortize the best questions (combining popularity and their judgement on what questions would yield the most insightful discussion).

If the conference cannot be held physically, then just cancel the whole thing and let's go back to publishing in journals. There's no reason to have arbitrary deadlines without a physical conference.

Huge thanks, friends! I'm happy this happened (in a sense, I think COVID too!).

Please keep the AMA sessions in the future edition of PLDI. They were so great and useful!

Thank you so much for the great conference, it was my first time attending, i hope it will not be my last. I learned so much from researchers and experts.

Thank you so much for this great conference, it was my first attendance; I hope it won't be my last; I've learned so much from the researchers and the experts.

PLDI exceeded my expectation as a virtual conference, but in the long run I think only virtual conferences will weaken our community. We need to get together physically at least once or twice a year for all the informal social interactions that really make the community strong. There's no replacement for the experiences, mentoring, and community growth that are fostered by all the dinners, long evenings at the hotel bars, etc. with diverse mixes of new and established researchers that can happen when we physically meet.

I think that PLDI 2020 was organized fantastically! Especially given no prior experience with virtual conferences and the limited time to prepare. Thanks!

Virtual conference really reduces the barrier to entry - my area is only tangentially related, but had this been a costly (in terms of money and time) event, I would not have attended. But the flip side is that I'm also less committed, I can pick and choose events and intersperse things through my normal day, as opposed to doing nothing but conference activities.

This was the 3rd major virtual conference that I attended since the beginning of COVID. I attended EuroSys which used the DISCORD system for "meetings" and that was as much as disaster as PLDI. Didn't talk to a single new person, it was just groups of people who knew each other already doing Google Meet chats on the side. I also attended IEEE Security and Privacy and that worked A LOT better. They used a totally integrated solution with a paid Zoom account that allowed 1000 simultaneous attendees, and they had a nice web based platform that tied it all together. It also wasn't free - and IEEE apparently paid $65,000 for the three day technical setup. But the attendee experience was so many leaps and bounds above the two other conferences that it is probably worth it.

Thanks for all the hard work! Given the limited time you all had, it was very well executed, we all learned a lot.

Congratulations for making this happen! This was really impressive. Even if in principle ASPLOS slack-only mode with videos may seem to offer similar functionality, I believe that having scheduled video streams and questions makes a difference.

| | |
|---|---|
| Ally did a great job | I think the organizers did a great job reacting so well to this difficult situation! Some things could be improved, but as one of the first conferences to go fully virtual we had to learn these things, and others will benefit from it. In the end it was a great experience, made possible by the dedicated work of the organizers. Thank you! |
| Thanks a lot for the conference, it was a great opportunity for me :) | |
| It was hard to find earlier talks during an ongoing session, in order to catch up etc. | |
| As a speaker, if we go for live talks, it would be better to have meeting mode where participants turn video on. It is pretty hard to give a talk to a laptop with zero feedback from the audience. | I attended ECOOP at least twice, but it was not in the list :) Many of the other events only once. Many thanks to all the organizers! It was really good. I spent most of the time watching content, so did not have much time for social interactions, which is a bit sad. We should think about people who have hard time interacting online and watching talks without captions, but I think the accessibility of the virtual event was much better than usual. |
| Thank you for the conference! Despite some problems, it was good! | |
| Virtuality: "It's lonely out in space / On such a timeless flight." | |
| Thanks | |
| thanks a lot for the great effort! | |
| Virtual (and hybrid) are the future of conferences. PLDI 2020 was a first step to establish what this future looks like. I think some ideas worked well (the #mentoring channel and the AMA sessions) but the overall conference experience was - for me - significantly worse than any physical conference that I have attended so far. We must improve on this format going forward. | Thank you for organizing this wonderful conference during this hard time! |
| | It's the first year everyone tries moving online/virtual, so we should not set our expectations too high. PLDI 2020 is a huge success, regardless of all the shortcomings as a first trial! |
| | Thanks for organising it very well! Please have more musical aspects like #ThisIsPLDI |
| | Congratulations! It was very nice organized and I was vary happy to be able to attend. |
| I could not afford to attend, so this was a great opportunity. It would be great to have all the videos available after the conference (this one, or any future one, even if it is not held virtual). It is very difficult to work and follow the conference at the same time so I could not attend everything I wanted. I would like to watch some videos now, but these are not available. So it would be great if the availability of videos lasts longer for all the sessions. | I already filled out the survey, but forgot to say a few more things. I like that the talks were streamed/recorded, because I could go back and watch them later. Or I could pause and rewind if I missed something, or speed through parts I already know. |
| | Love the accessibility the technology provides to non-travellers. When I retire, will have lots more time to attend and interact. |
| I think the virtual mode is helpful. In this way we can attend sessions from multiple conferences. | I would have liked scheduled breaks, for socializing or actually taking a break. I have difficulty socializing at physical conferences, and actually found it a little easier at virtual PLDI. I am very impressed by how PLDI 2020 turned out, and look forward to seeing how future virtual conferences develop. Thank you!!! |
| I am amazed at the high number of aspects that the organizers got right. Big Thank You! | |
| Thanks for doing such a great job organising PLDI despite all the obstacles. Also, I think a good balance has been found between ensuring polite interactions and helping people feel welcome. I don't think there's any need to go any further in this direction. More science and less politics. | I said "slightly disagree" to "felt included", and thought I should elaborate. I'm an engineer in industry with an interest in research, but not (yet) participating in research myself. Looking through the #mentoring posts and PLMW |
| questions, I felt like the mentoring opportunities were very oriented towards people already/still in academia. The only people talking about their experience in industry seemed to be talking about it coming from academic research, not the other way around. (And I also think that's fine; not every event has to cater to my personal situation, but thought I'd mention it.) | that no fees were charged so that attendees without a budgets (e.g., from developing countries) were on equal footing with those who have the privilege of a travel and conference budget. |
| when a conference is virtual, going overtime is very disturbing | Please don't impose the use of proprietary software on the participants |
| It was an amazing effort to pull of a virtual conference in such a short time. I still feel like I got way less than a physical conference, but it was way, way better than I had originally anticipated. | While I thought this was well run overall, I felt documentation was either lacking or hard to find. I struggled to find the right video feed, saw something about /video but could never figure out what that was, am still not sure whether watching over the youtube link is what I was supposed to do, etc. Admittedly, I didn't watch the "how PLDI will work" video, so it's probably my fault. But I can't stand watching videos to find things like URLs and commands. |
| Thanks for organising this conference, especially during such difficult circumstances! | It was very well run! The videos prerecorded explaining the conference were fantastic! |
| The registration process is somewhat complicated. | The ability to pause, rewind, and then catch-up by playing at 1.5x in YouTube was incredibly useful for paper comprehension, and dangerously addictive as a way of watching... |
| We'll need to work on the tech to make it easier for the organizers, more seamless for the participants, and of more consistently high quality (for some of the talks, video was too fuzzy to read slides). | It was a very nice chance to explore the community of PLDI as someone from a different community. As I only had limited time available during the week I was glad that I could watch the talks on the go through Youtube. At the same time, there were several main track talks which I could not watch in my timezone. Otherwise: Thanks for making this possible! Actually, next time I would save more time in my schedule for a virtual PLDI |
| I would have appreciated a clearly labelled informal chat | |
| Congratulations on a job fantastically well done! | |
| Great job! | |
| Please add extended poster session for each presented paper. It is a great fit for virtual conference. It creates more interactions between the authors of a paper and other attendees of the conference. | Special mention for Jay Lorch, who changed costumes throughout his recorded presentation. It was hilarious. :) |
| Congratulations to all organisers! It's a massive effort to put a conference like that together, specially during this unprecedented time. | great event! a BIG thank you to all the organizers |
| I would like to thanks PLDI organizing committee for organize PLDI every year. for this we learned many things & enlarged our knowledge. | I think the community pulled it together in a positive way, and I think we should all be proud of this. But when we can resume in-person meetings we should do so. |
| Thank you! I thought it was great. I think it would be nice if future conferences had some small virtual component for people who otherwise wouldn't have attended. | I was a presenter at one of the PLDI co-located events, and the presentation was scheduled at 4:00 am in my timezone. Fortunately I was able to submit a pre-recorded presentation video, but it was a shame that I had to stay up late to attend the live Q&A session. It would have been better if the chairs had considered authors' time zone when scheduling the sessions. Still, I had a wonderful experience attending these |
| Please keep 1:1 Meeting and ask me anything! I really really enjoyed them! | |
| Thank you very much for organising an excellent virtual conference! I particularly liked | |
| events and I'm very grateful for all the people who made this conference possible in this hard time!! | I was not able to register a Zoom account in order to join the meeting and I did not know how to log into Slack properly; after all, I did not know to whom I can turn for help. |
| Thank you for all your work! And also for collecting this data and making sure other conferences can profit from the PLDI lessons learned :) | Thank you. It was amazing!! |
| | *** THANK YOU to Ally, James, Crista, Benjamin, Jens and all the others, you have worked wonders *** |
| I find almost all PLDI research track presentations too hard to follow; for most of them, I don't even get what problem that paper is trying to address. | Fantastic job by all the organizers, but very especially Ally!! |
| Congratulations to transitioning so efficiently to a remote conference | This is more a personal observation, but one challenge I had during the week was balancing PLDI with other commitments (work, family, etc).  It's not like a physical conference where it is easy to detach from everything else for a couple days.  More focussed times with clear break times may have helped.  Parallel tracks would probably have been fine to help with this.  Thanks for putting this on -- it was a huge effort on short notice and serves as a model of how we can continue to work as a community moving forward.  Thanks! |
| under normal circumstances (ie with child care!) I would have engaged much more with the virtual conference. it worked out amazingly well. congrats! | |
| Thanks for organizing virtual PLDI, which - given the circumstances - has been a great event! | |
| Given the situation, PLDI's organization was great! Thanks to Aly and everyone else involved | |
| Links to video streamings and other contents are not clearly exposed, so I've lost time trying to find the correct stream or I couldn't find it in slack channels. | Well done on this first online edition of PLDI. I was pleasantly surprised by how well it was organised. Don't keep this experience just within your group of people. Write up what you did, what considerations were made, etc. and publish it online! I'm sure very many other conferences would benefit from having an experience report / guide to online conferences to read. |
| A big THANKS to the organizers, it was great :) | |
| I'm a fan of various proposals knocking around to center SIGPLAN in-person events on one big event per year, with more virtual events throughout the year to complement. | |
| Thanks for virtual pldi, I love pldi | This was my first PLDI, if it were hosted in london I most likely wouldn't have been able to attend. I was really surprised with how caring people were to me as a student. If possible I will definitely try to attend future editions! |
| Thanks! I think this was very well executed. I didn't spend as much time as I would've liked to "at" the conference, since I was also existing in my own timezone and still had meetings/work/etc. But, I was really impressed with the amount of engagement. The mentoring channel was great, too. | |
| | I paid for proceedings at registration but haven't seen anything; this process seems broken. Also: slight condescending attitude in some sessions towards industry. |
| Virtual environments have a huge potential to be immersive in a way that streaming video, Slack and even Gather-style video chats are not. Twenty years ago I played networked Quake over a 56K modem, and while the latency was awful, it had an immediacy and sense of "sharing a space" that our virtual conferences would do well to capture using today's more capacious technology. Much hard work and hard funding needed of course.... | Thanks a lot for a superbly-organized conference! |
| | Your effort to take things virtual were absolutely excellent! |
| | Virtual conferences are good, useful, and help a lot of people. Except myself. I feel disadvantaged by them. Timing: I have a hard time justifying to myself attending a virtual event. Social interactions: I hate Zoom etc on a good day, and can't bring myself to interact |
| with people virtually. In a physical setting, I am forced to do it, which is much better for myself. So, yeah, I would like to have physical conference from time to time still. | (a) I would have liked a dedicated poster session/timeslot (b) I believe that the remote format hurt PhD students, who had the least opportunity for exposure to more "senior" people (as in physical conferences) |
| My research group would have been able to pay for me if we had to, but if PLDI stays virtual, I think it should be a priority to try and keep it also free, because everyone is not lucky to be able to pay several hundred dollars for that kind of events, and it's a great way to open up the PL community | Much of the mentoring was US/North America related, while I would have appreciated to know more about European academics (I don't plan on having a career in NA) |
| | Thanks to the organizers, you made a very good thing from a very difficult situation! |
| Due to my other responsibilities I could not watch everything I wanted to yet. I fully intend to watch more and ever rewatch some. I have. Thus, the amount of material I consumed is understated in this survey. I also shared this material with many of my students and colleagues. I think a great resource has emerged via this virtual conference. Much appreciated. | I felt that more information should be given out before the start of PLDI20, it felt that things were given out piecemeal at times which led to some confusion |
| | Thanks! |
| | This was obviously a huge amount of work to put on, and I am afraid I would have gotten a lot more value out of blocking out the same amount of time to simply read PLDI papers. |
| Great efforts, enormous thanks for putting on such a Great show! | Thanks a lot for all the time and effort you put into organizing virtual PLDI. Under the circumstances of a global pandemic, I think it was a great conference! |
| Amazing job. Went so smoothly that it was not until business meeting that I realised what you had all been up against.  Superheroes! | |
| I enjoyed tutorial on Spoofax | It felt like a great job for a virtual conference, and it was a lot better than I thought it would be. That being said, I still strongly prefer non-virtual conferences, though I see that with climate change we might not want to do these that often. Maybe a rotation model among the four big SIGPLAN conferences would be good, with two of them each year live (pandemic permitting) and the other two virtual (to save on CO2). As for one last problem with virtual conferences that I may have not been asked about because I didn't give a talk at PLDI, but personally, I feel the connection to the audience is missing both via Zoom and obviously when recording a video (also, recording a video is a lot more effort). |
| Many thanks to the organizers and all who helped this to happen. This is fantastic. | |
| It was not obvious that things were going so well and surely required a lot of hard work behind the scene, which often can be underestimated. Thank you for the really good organization of the virtual conference, I enjoyed it. | |
| Overall, great job given the time y'all had to prepare. In my view, a virtual conference does not have the same appeal as a in-person meeting. However, it has the advantage of being more inclusive than a physical meeting. | |
| Ally, John, Emina, and the student volunteers did a fantastic job organising this first-of-its kind virtual event. I think you already know it, but I feel like it couldn't be stressed enough. :) | Congratulations on holding a pretty good event in really adverse circumstances. |
| | A huge thank you to the organizers who are the heroes of PLDI. |
| I am planning on watching some of the records later (I hope they remain available), as the previous week has been a busy time for me at university. | Great Job Organizers - Hope you all get an award |
| | Thank you making it available to everyone. It was a nice learning experience attending PLDI 2020. The mentoring sessions in PLMW were |
| I thank organizers for all the efforts. | |
| outstanding. Thanks to all the researchers, participant and especially to the organizers. | Make it so AMA people can respond in text to individual questions later. I didn't see how to do that. |
| Overall I think this year PLDI is great and well organized. | Very good conference given the unfortunate circumstances. |
| the survey is way too longer than I expected, would you please make it shorter (next time)? | As a member of the ICFP OC, I learned a lot from this virtual PLDI. Huge thanks to the organizers, as well as the contributors of the PLDI song! |
| This was the first international conference I attended and it was amazing. I feel really inspired now! | Great Conference Installation |
| The goal of a virtual conference should not be to reproduce a physical one as closely as possible, this can only fail. Rather, the goal should be to identify what works *better* virtually than physically, and change the format to take advantage of this. | I really liked the ability to role back in the live feeds and to play those feeds at a faster pace to in some cases "catch up". Being able to pause was also quite helpful since we were at home and family interruptions are inevitable. I am super impressed overall with how well Ally and his crew handled the switch to online including giving keynote speakers the opportunity to opt for 2021, the AMA sessions, the incredible advertising of PLDI on twitter, the slack presence of organizers, and the clear information to authors. Well done!!!!! |
| Great job in a tight spot! | |
| Overall, PLDI 2020 has gone virtual successfully. | |
| Given how quickly this was put together, I think that this conference worked well. I think that youtube live streams are a good approach to broadcasting. I think something other than slack should be used (presumably most of the messages are going to disappear once someone stops paying for the premium account) | |
| | Thank you for all the organization, it was a complete success! |
| This was my first PLDI and found it amazing! The talks that I've seen live were incredible pieces of work. I hope that next PLDIs were on-line or have such streaming in order to allow people unable to attend due to financial reasons. | Great job, thanks! |
| | being able to replay talks with different playback speeds is very useful. Using @here in slack so often meant i just ignored the slack notifications most of the time |
| | Great job! |
| Thank you very much for giving me the opportunity to participate !!! | One advantage of virtual PLDI is that I can watch some video after the talk for more than one time. It would be great if PLDI could provide presentation videos online in the future. Thanks. |
| I look forward to seeing you in person | |
| Thanks everyone | |
| | It was a nice experience, and I hope virtual conferences continue. Thanks for setting it up! |